\documentclass[twocolumn]{aastex631}

\usepackage{graphicx}
\usepackage{graphbox}
\usepackage{hyperref}
\usepackage{amssymb,amsmath}
\usepackage{stmaryrd}
\usepackage{booktabs}
\usepackage{float}
\let\tablenum\relax
\usepackage[separate-uncertainty=true, multi-part-units=single]{siunitx}
\DeclareSIUnit \erg {erg}

\def\lea{\mathrel{<\kern-1.0em\lower0.9ex\hbox{$\sim$}}}
\def\gea{\mathrel{>\kern-1.0em\lower0.9ex\hbox{$\sim$}}}
\newcommand{\lta}{{\>\rlap{\raise2pt\hbox{$<$}}\lower3pt\hbox{$\sim$}\>}}
\newcommand{\gta}{{\>\rlap{\raise2pt\hbox{$>$}}\lower3pt\hbox{$\sim$}\>}}

\begin{document}
\title{
Population of X-ray Sources in the Intermediate-Age Cluster NGC 3532: a Test Bed for Machine-Learning Classification }

\shorttitle{NGC 3532}

\author{Steven Chen}
\affiliation{Department of Physics, The George Washington University, 725 21st St. NW, Washington, DC 20052}
\email{schen70@gwmail.gwu.edu}

\author{Oleg Kargaltsev}
\affiliation{Department of Physics, The George Washington University, 725 21st St. NW, Washington, DC 20052}

\author{Hui Yang}
\affiliation{Department of Physics, The George Washington University, 725 21st St. NW, Washington, DC 20052}

\author{Jeremy Hare}
\affiliation{NASA Goddard Space Flight Center, Greenbelt, MD, 20771}
\affiliation{NASA Postdoctoral Program Fellow}

\author{Igor Volkov}
\affiliation{Department of Physics, The George Washington University, 725 21st St. NW, Washington, DC 20052}

\author{Blagoy Rangelov}
\affiliation{Department of Physics, Texas State University, 601 University Drive, San Marcos, TX 78666}

\author{John Tomsick}
\affiliation{Space Sciences Laboratory, University of California Berkeley, CA 94720}

\begin{abstract}

Open clusters are thought to be the birth place of most stars in the Galaxy. Thus, they are excellent laboratories for investigating stellar evolution, and X-ray properties of various types of stars (including binary stars, evolved stars, and compact objects). In this work, we investigate the population of X-ray sources in the nearby 300-Myr-old open cluster NGC 3532 using Chandra X-ray Observatory and multi-wavelength data from several surveys. We apply a random-forest machine-learning pipeline (MUWCLASS) to classify all confidently detected X-ray sources (S/N$>5$) in the field of NGC 3532. We also perform a more detailed investigation of brighter sources, including their X-ray spectra and lightcurves. Most X-ray sources are confirmed as coronally-active low-mass stars, many of which are confidently identified by MUWCLASS. Several late B or early A-type \textbf{stars} are relatively bright in X-rays, most of which are likely binaries. We do not find any compact objects among X-ray sources reliably associated with NGC 3532, down to the limiting X-ray flux of $\sim\SI{2e-15}{\erg\per\s\per\square\cm}$, corresponding to $L_X\sim\SI{6e28}{\erg\per\s}$ at the cluster's distance. We also identify several Galactic sources beyond NGC 3532 that differ from typical coronally active stars, and were classified by MUWCLASS as potential compact objects. Detailed investigation reveals that these sources may indeed belong to rarer classes, and deserve follow up observations.

\end{abstract}


\section{Introduction}

Most stars are born in dense, gravitationally bound star clusters which are broadly classified into globular clusters (GC) and open clusters (OC). GCs are ancient ($\sim \SI{10}{Gyr}$), massive ($>\SI{e6}{M_\sun}$) and are typically located off the Galactic disk, while OCs tend to be young ($<\SI{1}{Gyr}$), less massive ($<\SI{e5}{M_\sun}$), and located within the Galactic disk \citep{larsen_young_2010}. Old (several Gyr) OCs are known to exist, but are rare, indicating that they tend to gravitationally dissolve on timescales of hundreds of Myrs. 

By the age of a few million years, gas which is not used in star formation is expelled from the cluster via several mechanisms, including ionization, stellar winds, supernovae, and radiation pressure \citep{larsen_young_2010, farias_difficult_2015}. At this age, the largest stars (O- and early B-type) have gone supernova, leaving behind compact objects (CO) in the form of neutron stars (NSs) and black holes (BHs).

The expulsion of gas reduces the cluster's gravitational binding energy, and may cause the dissolution of more than 90\% of OCs before 100 Myrs \citep{larsen_young_2010, lada_embedded_2003}. At that epoch, if the cluster survived gas expulsion, mass transfer in binaries becomes the prime factor for stellar evolution, while cluster evolution is primarily driven by stellar dynamics and external interactions. These clusters still undergo dissolution due to two-body relaxation, external shocks, and stellar evolution. Only clusters with total initial mass $>10^{4} M_{\odot}$ are likely to survive beyond 1 Gyr \citep{larsen_young_2010}.

In clusters that are a few hundred Myr old or younger, X-ray sources are typically represented by coronally active lower mass stars and various types of Young Stellar Objects (YSOs), Active Binaries \textbf{(e.g. RS CVn and BY Drac. systems)}, Cataclysmic Variables (CVs), and colliding-wind binaries (CWBs). Most NSs and BHs born in supernova (SN) explosions are expected to receive strong natal kicks and, hence, should escape the cluster quickly \citep{van_der_meij_confirming_2021}. However, some NSs and BHs could still remain bound to the cluster, e.g., NSs formed from electron capture SNe, especially if the SN explosion takes place in a binary system \citep{igoshev_combined_2021,stevenson_wide_2022,gessner_hydrodynamical_2018}.

With the exception of sources from a few special classes, (e.g., accreting NS with cyclotron lines in their spectra, AGN with redshifted broad iron lines, pulsating X-ray sources), little can be learned about the X-ray source nature \textit{solely} from X-ray data, especially if the source is not bright enough for a high resolution spectrum (e.g., detecting spectral lines helps to distinguish between thermal plasma and nonthermal emission). The vast majority of X-ray sources in clusters are relatively faint and their nature is largely unknown. Therefore, multi-wavelength analysis of these sources is crucial to discern their nature.

This paper, which is the first in a series of papers about the intermediate age clusters observed by the Chandra X-ray Observatory (CXO), presents the methodology and analysis of multiwavelength (MW) data for a well-known nearby cluster, NGC 3532, which has been studied in detail in the optical and near infrared (NIR). 

\subsection{NGC 3532}
\label{ngc3532}

NGC 3532 is located $484\substack{+35 \\ -30}$ pc away \citep{fritzewski_spectroscopic_2019} in the Carina region of the southern Milky Way. Its distance and Galactic coordinates ($l=289.6^\circ$, $b=1.3^\circ$) place it well within the Galactic plane. NGC 3532 has an accepted age of $\sim$300 Myr
\citep{fritzewski_spectroscopic_2019}. \cite{fernandez_photometric_1980} estimated the total cluster mass to be a moderate $\SI{2000}{M_\sun}$, with brighter stars covering a $14' \times 20'$ ($2\times 3 $ pc ) central region and fainter stars extending over $1^\circ \times 1^\circ$ (($8\times 8 $ pc, see Figure \ref{fig:ngc3532_cxo}). NGC 3532 exhibits a relatively low extinction $E(B-V)=0.034\pm0.012$ \citep{fritzewski_spectroscopic_2019} which allows for the detection of fainter and softer sources. 

NGC 3532 is covered by modern optical surveys, including the VST Photometric H$\alpha$ Survey of the Southern Galactic Plane and Bulge \citep[VPHAS+;][]{drew_vst_2014}, the DECam Plane Survey 2 \citep[DECaPS2;][]{saydjari_dark_2022}, and Gaia eDR3 \citep{brown_gaia_2021} and has also been the subject of dedicated spectroscopic \citep{fritzewski_spectroscopic_2019} and photometric studies \citep{clem_deep_2011}.

Temporal monitoring of NGC 3532 has been carried out with a 42-day long campaign with CTIO's Yale 1-m telescope \citep{fritzewski_rotation_2021}. Identifications of variable stars in the NGC 3532 field are also available from the catalog of large-amplitude variables in Gaia DR2 \citep{mowlavi_large-amplitude_2021}. 

Spectral classifications of optical stars in NGC 3532 have been performed by \cite{eggen_open_1981} and \cite{fritzewski_spectroscopic_2019}. \cite{fritzewski_spectroscopic_2019} confirmed 660 member stars within NGC 3532 using proper motion data from Gaia DR2, with the expectation that the cluster hosts over 1,000 stars in total, while \cite{clem_deep_2011} estimated over 2,000 stars in total when accounting for binaries.

Using a deep optical survey with the Cerro Tololo Inter-American Observatory, \cite{clem_deep_2011} derived a mass function power-law index of -2.54 for the higher mass star range ($>\SI{2}{M_\sun}$; assuming 40 stars $>\SI{2}{M_\sun}$ from Figure 21 of \citealt{clem_deep_2011}), which corresponds to $\sim 21$ stars with initial mass $>\SI{3}{M_\sun}$ that have died at the cluster age of 300 Myr, including $\sim 5$ stars $>\SI{8}{M_\sun}$ that could form NSs or BHs, leaving lower mass B8V-B9V stars as the heaviest remaining stars. \cite{clem_deep_2011} also estimated a binary fraction of $\sim 27\%$, based on the excess brightness, and listed 32 known and candidate WDs, with photometry and location on the CMD compatible with NGC 3532 membership. \cite{dobbie_further_2012} confirmed spectroscopically the cluster membership of a total of seven WDs in NGC 3532. They inferred the WD masses to be 0.76-1.00 $\SI{}{M_\sun}$ and corresponding progenitor masses to be 3.7-6.9 $\SI{}{M_\sun}$. \cite{raddi_search_2016} confirmed three more member WDs, with VPHAS J110358.0-583709.2 being one of the most massive WDs found in open clusters. This WD has a mass of $\SI{1.13}{M_\sun}$, and a modeled progenitor mass of 8.80 or 9.78 $\SI{}{M_\sun}$. This may be an Oxygen/Neon WD, or otherwise was formed from a binary merger \citep{raddi_search_2016}. No NSs or BHs have been reported in NGC 3532.

Dedicated analysis of X-ray sources in NGC 3532  dates back to the ROSAT era. \cite{franciosini_rosat_2000} analyzed ROSAT data for NGC 3532 observed from 1996-1997, discovering $\sim$50 X-ray sources, above $4\sigma$ detection significance level, fifteen of which have optical counterparts (belonging to the cluster) located within $10''$ from the corresponding X-ray source. 

Most ROSAT X-ray sources were matched to cluster F-type stars. Four A-type stars were also detected, with their X-ray emission suspected to be due to unseen companions. \cite{simon_x-ray_2000} analyzed the same ROSAT data, discovering 43 X-ray sources above $4\sigma$ detection significance level.

With 174 optical cluster stars selected by \cite{franciosini_rosat_2000} within $17'$ of the ROSAT pointing, the chance coincidence probability of one X-ray source to be matched \textbf{with at least one} cluster star, assuming the stars are uniformly distributed across the sky, is 1.7\%.\footnote{\textbf{T}he chance coincidence probability obeys a Poisson distribution, with $\lambda$ given by the average number of stars expected within the area of the X-ray source's positional uncertainty.}
However, with an updated list of cluster members from Gaia DR2 \citep{jaehnig_membership_2021}, $\sim 550$ probable cluster member sources are detected in the same $17'$ radius field. The chance coincidence probability is then 5.1\%. As NGC 3532 sits in the Galactic plane, there's also a large number of Galactic background stars. With $>48,000$ Gaia DR3 sources in the $12'$ field around the cluster center, the probability that an X-ray source is matched to at least one star is nearly 100\%. 

Thus, in both ROSAT studies, large positional uncertainties (PUs) of ROSAT sources prevented definitive determination of counterparts in most cases, and the authors did not discuss sources other than flaring low mass stars. This underscores the need for high-resolution X-ray images while studying X-ray sources in the densely populated galactic fields. Both ROSAT studies indicated the hydrogen column density toward NGC 3532 to be $n(H)=\SI{2e20}{cm^{-2}}$. 

The archival CXO data on NGC 3532 offer broader coverage in photon energies, better sensitivity, and sub-arcsecond angular resolution. The greatly improved positional accuracy and access to fainter X-ray source populations motivated us to carry out a detailed multi-wavelength study of NGC 3532, with a focus on classification of X-ray sources and identification of any unusual objects. For this purpose, we make use of our machine learning multi-wavelength classification pipeline, \textit{MUWCLASS}, described in detail in \cite{yang_classifying_2022}. In Section \ref{obs}, we describe the CXO observation of NGC 3532, the multi-wavelength catalogs, and the crossmatching procedure. In Section \ref{analysis}, we assess bulk properties of CXO sources using multi-wavelength plots, including color-magnitude diagrams (CMDs) and color-color Diagrams (CCDs). In Section \ref{classification}, we present Machine Learning (ML) classification results of X-ray sources in NGC 3532. \textbf{In Section \ref{detailed}, we follow up }with more detailed analysis of selected X-ray sources using their X-ray spectral and multi-wavelength properties in conjunction with the ML classification results, including a discussion of candidate compact objects. Finally, Section \ref{summary} summarizes our findings.

\section{Observations and Archival Data}
\label{obs}
\subsection{CXO data}

\begin{figure*}[hbt!]
    \centering
    \includegraphics[width=0.65\textwidth]{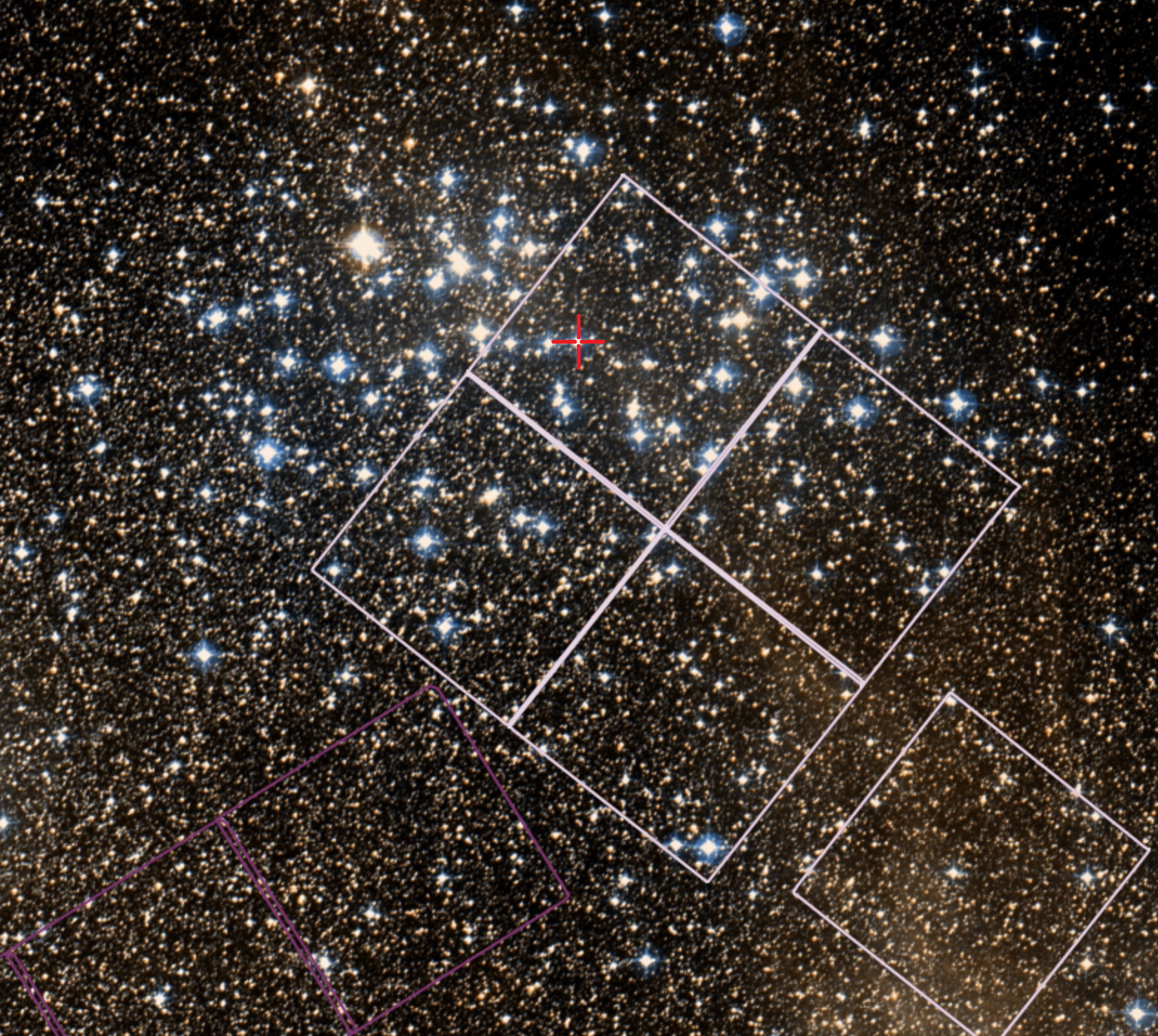}
    \begin{interactive}{animation}{plots/NGC3532_movie.mp4}
    \includegraphics[width=0.65\textwidth]{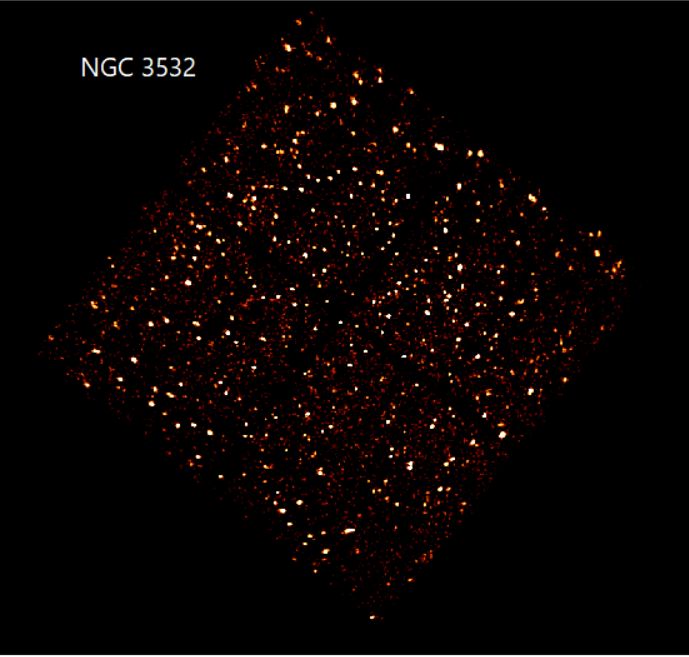}
    \end{interactive}
    \caption{The top panel shows the Digital Sky Survey image of NGC 3532 with the CXO ObsID 8941 (analyzed in this paper) ACIS field of view overlayed (white squares). The red cross shows the cluster center \citep{clem_deep_2011}. The bottom panel shows the ACIS-I image. An animated version of this image is available, showing 0.5 ks slices of the observation.}
    \label{fig:ngc3532_cxo}
\end{figure*}

{\sl CXO} conducted a single observation (ObsID 8941) of NGC 3532 with the Advanced CCD Imaging Spectrometer (ACIS; \cite{garmire_advanced_2003}) from 2008-10-23 to 2008-10-25 (MJD 54762-54764), for a total of 131,858 s ($\sim$36 hours). About half of the cluster (see Figure \ref{fig:ngc3532_cxo}; top panel) was imaged on the ACIS-I array operated in timed exposure mode (with time resolution of 3.2 s) using the Very Faint telemetry format (which provides a lower background). The CXO image is shown in the bottom panel of  Figure \ref{fig:ngc3532_cxo}. 
The Chandra Source Catalogue 2.0  \citep[hereafter CSC2;][]{2020AAS...23515405E}, released in 2020, contains detailed information (e.g., fluxes and variability measures) on a per-observation level, a stack-level, and a master-level. We use CSC2 to extract fluxes in three non-overlapping energy bands (hard band $h=$2.0-7.0 keV, medium band $m=$1.2-2.0 keV, soft band $s=$0.5-1.2 keV), as well as the broadband flux ($b=$0.5-7.0 keV). CSC2 provides the mode ($F_{\rm mode}$), as well as the lower and upper limits at 1-$\sigma$ confidence ($F_{\rm lo}$ and $F_{\rm hi}$) to the mode to characterize the flux distribution for each source in the catalog. We calculate the mean and the variance, using the same equation from \cite{yang_classifying_2022}, i.e.\ assuming the flux distribution to be the Fechner distribution with the equations from \citet{possolo_asymmetrical_2019}.

We only select sources with signal-to-noise ratio $>5$ and with off-axis angles $<10\arcmin$. We also require the X-ray sources to have valid flux measurements (that are not missing/null values) in at least one energy band for ML classification (see Section \ref{classification}). From an initial list of 300+ X-ray sources available in CSC2, 131 sources pass our selection criteria. The properties of these sources are compiled into a comprehensive machine-readable master table available online (a subset of this large table is shown in Table \ref{table:detailed}). Each source in the master table is assigned a unique identification number which is used throughout the rest of this paper.

We construct three hardness ratios (HRs) from the three CSC2 fluxes:

\begin{subequations}
\begin{align}
{\rm HR}_{ms} & = \frac{f_m-f_s}{f_m+f_s},  \\
{\rm HR}_{hm} & = \frac{f_h-f_m}{f_h+f_m}, \\
{\rm HR}_{h(ms)} & = \frac{f_h-(f_m+f_s)}{f_h+f_m+f_s}. 
\end{align}
\label{eq:CSC-1}
\end{subequations} 

CSC2 does not apply any astrometric corrections to their X-ray coordinates, which is accounted for with a systematic error of 0.71$\arcsec$ \textbf{(95\% confidence)} to account for this. Rather than using these PUs with uniformly added systematic uncertainty, we calculate the X-ray PUs using the empirical equation 12 from \cite{kim_chandra_2007}.

Then, we apply our own astrometric corrections. We use the CIAO \texttt{wcs\_match} algorithm to align the coordinates of CXO sources to the Gaia eDR3 catalog (see Appendix \ref{astrometric}). We find an astrometric correction of $\Delta {\rm RA} \cos ({\rm DEC})=0.23\,\arcsec$ and $\Delta {\rm DEC} = 0.15\,\arcsec$ with a 1-$\sigma$ alignment uncertainty of $0.092\,\arcsec$, which is then added to the X-ray PUs in quadrature. 

Several metrics for detecting intra-observation source variability are available from CSC2, including P-values based on Kolmogorov-Smirnov and Kuiper's statistics. We decided to use Kuiper's statistics, as it is more robust.\footnote{For additional details, see \url{https://cxc.harvard.edu/csc/why/ks_test.html}} In this work, sources with Kuiper variability probability above 99\% \textbf{($\approx 2.6\sigma$)} are taken as variable.

\subsection{Gaia}

The Gaia eDR3 catalog was released on December 2020 \citep{brown_gaia_2021}. It contains positions, photometry, parallax, and proper motion data for 1.8 billion sources. Typical PUs range from $\sim 0.02$ mas at $G<15$ to $\sim 1.0$ mas at $G=20$. 

Gaia's photometric information is provided in the broad G band (330-1050 nm) and two narrower BP (330-680 nm) and RP (640-1050 nm) bands. 
The Gaia G band is sensitive to about $G=21$, with a magnitude uncertainty of 0.3 mmag at $G<13$, rising to 6 mmag at $G=20$ \citep{brown_gaia_2021}.
The BP band overestimates the flux of faint red sources, leading to these sources appearing bluer than they should be. BP uncertainties increase from 0.9 mmag at $G<13$ to 108 mmag at $G=20$.\footnote{$\sim 60$ mmag at $BP=20$ for the field of NGC 3532} \citep{brown_gaia_2021}

From Gaia eDR3, distances to 1.3 billion objects were estimated from parallax data by  \cite{bailer-jones_estimating_2021}. These distances, $r_{\rm geo}$, are purely geometric, i.e., they do not rely on photometry. The accuracy of these distances depends heavily on the reliability of the parallax measurement, so only distances inferred from positive parallax measurements, with $\pi/\sigma_\pi >= 2$ are used in our ML classification (see Section \ref{classification}). A large peak is seen in the distribution of source distances around 475 pc, consistent with the NGC 3532  cluster distance of 484 pc derived from Gaia DR2 \citep{fritzewski_spectroscopic_2019}. 

Shortly before the submission of this work, Gaia DR3 was released. While the release did not include new astrometry or photometry, many derived astrophysical parameters for millions of sources were made available, including distance, mass, age, temperature, spectral type, and emission lines \citep{gaia_collaboration_gaia_2022}. These parameters were derived using the Apsis Pipeline, which includes multiple, independent analysis modules. Although the quality of any one parameter should be taken with caution, when the stellar parameters from independent modules are consistent, these parameters should be more reliable. Therefore, we supplement our analysis of NGC 3532 with Gaia DR3 astrophysical parameters, when they are consistent \textbf{between} Gaia modules \sout{and applicable}. We primarily used the ESP-ELS module for the classification of spectral types, the FLAME module for mass and age, and the GSP-Phot Aeneas module for temperature. While multiple modules provide distances, \cite{gaia_collaboration_gaia_2022} suggested that they may not be reliable, so we continued to use the Gaia eDR3 distances from \cite{bailer-jones_estimating_2021}. 

\subsection{2MASS}

The Two Micron All-Sky Survey (2MASS) is a near-infrared (NIR) all sky survey conducted between 1997-2001 \citep{skrutskie_two_2006}. 2MASS conducted observations in the near-infrared J (1.25 $\mu$m), H (1.65 $\mu$m), and K (2.16 $\mu$m) bands, with 10$\sigma$ point source detection levels at 15.8, 15.1, and 14.3 mag respectively. For sources with magnitudes in the K band between 8.5-13 mag, the photometric uncertainty is about 0.03 mag. The astrometric accuracy ranges from $<100$ mas for brighter sources to $>200$ mas for fainter sources above 16 mag. 

\subsection{WISE}

The WISE telescope is an infrared (IR) all-sky survey mission launched in 2009. WISE conducts observations in 4 infrared bands, W1 (3.4 $\mu$m), W2 (4.6 $\mu$m), W3 (12 $\mu$m), and W4 (22 $\mu$m), with a full width at half maximum (FWHM) of $6''$, translating to a typical sub-arcsecond level angular resolution. The 5$\sigma$ point source detection levels for the 4 bands occur at the equivalent of 16.5, 15.5, 11.2, and 7.9 Vega mags respectively, with a uncertainty of 0.185 mag \citep{wright_wide-field_2010}. The AllWISE catalog, released in 2013, combines WISE data from the primary mission phase, as well as the NEOWISE mission phase \citep{cutri_vizier_2021}.

The UnWISE \citep{schlafly_unwise_2019} and CatWISE2020 \citep{marocco_catwise2020_2021} catalogs combine previous catalog data with more recent NEOWISE observations to increase sensitivity beyond AllWISE.
In particular, UnWISE has 5 times, and CatWISE2020 has 6 times longer exposure times compared to AllWISE.
UnWISE 50\% completeness limits are W1 = 17.93 mag and W2 = 16.72 mag. CatWISE2020 S/N=5 limits are W1 = 17.43 mag and W2 = 16.47 mag. UnWISE and CatWISE2020 do not offer W3 or W4 data. 

In this work, observations from all three catalogs are used for plotting and ML classification (Section \ref{classification}). UnWISE fluxes in the W1 and W2 bands were converted to magnitudes. AllWISE sources and magnitudes are preferred over CatWISE2020 sources when both are available to maintain consistency with the use of W3 magnitudes from AllWISE, while both are preferred over UnWISE sources.

\subsection{DECaPS2 and VPHAS+}

To complement the above all-sky, but relatively shallow surveys, we used the deeper DECam Plane Survey 2 \citep[DECaPS2;][]{saydjari_dark_2022}. DECaPS2 is an optical and NIR survey conducted with the Dark Energy Camera at the Cerro Tololo Inter-American Observatory in Chile. It reaches a typical single-exposure depth of 23.7, 22.7, 22.2, 21.7, and 20.9 mag\footnote{This is the photometric depth corresponding to 50\% source recovery rate\citep{saydjari_dark_2022}. } in the optical and NIR $g$, $r$, $i$, $z$, $Y$ bands, with a typical seeing of $1''$. 

DECaPS2 magnitudes were converted into Gaia magnitudes using a linear model fit for $\sim 40,000$ sources with both Gaia and DECaPS2 magnitudes in the field of NGC 3532. The $g$, $r$, $i$, $z$ bands were fit to Gaia G band; $g$, $r$, bands to RP band; and $r$, $i$, $z$ bands to BP band. Since DECaPS2 extends significantly deeper than the surveys used in the \textbf{training dataset}, this survey was not used to classify sources in the ML pipeline as it may introduce biases. The standard deviation of converted magnitudes at Gaia $G=21$ is $\sim 0.2$ mag for $G$, and $\sim 0.5$ mag for $G_{BP}$ and $G_{RP}$. Extrapolation of converted DECaPS2 magnitudes to fainter ranges than Gaia reaches may result in larger errors. However, for the purposes of this work, having precise magnitudes is not essential. 

We also analyzed the VST Photometric H$\alpha$ Survey of the Southern Galactic Plane and Bulge \citep[VPHAS+;][]{drew_vst_2014} data of NGC 3532. However, only 1 CXO source (Source 77) had VPHAS+ counterparts without Gaia counterparts, and this source was detected in more bands in DECaPS2. 

\subsection{Crossmatching}
\label{sec:crossmatching}

CXO sources in NGC 3532 were crossmatched to optical and infrared counterparts to enable multi-wavelength analysis, plotting, and ML classification. After the astrometric correction (see Appendix \ref{astrometric}), CXO sources were first cross-matched to Gaia eDR3 sources using the combined $2\sigma$ PUs by adding (in quadrature) the X-ray and Gaia PUs. Source positions at the Gaia eDR3 epoch (2016) are propagated to the epoch of the CXO observation (2008) using proper motions, when available. 

The CXO PU is calculated by combining the empirical PU using equation 12 from \cite{kim_chandra_2007} and the alignment uncertainty measured from the astrometric correction (see Appendix \ref{astrometric}) in quadrature. Gaia PUs include the Gaia coordinate uncertainty, uncertainty in proper motions, parallaxes and their uncertainties, and astrometric excess noise. 
The CXO PUs for sources in the NGC 3532 field range from $0.25 ''$ to $2.4 ''$ with a median value of $0.79 ''$. 

2MASS and ALLWISE counterparts were then identified using the Gaia eDR3 pre-computed cross-matched sources, using the ``best neighbor" source \citep{2021gdr3.reptE...9M}. For multi-wavelength counterparts from other catalogs (DECaPS2, CatWISE2020, UnWISE) that do not have pre-computed cross-matches, or the 2MASS and ALLWISE counterparts of sources that do not have Gaia counterparts (such that pre-computed cross-matches are not available), the counterparts were matched using the PUs of the multi-wavelength and X-ray catalogs added in quadrature. For all multiwavelength catalogs but Gaia eDR3, we multiply the Gaia eDR3 proper motion by the catalog reference epoch difference, and add it to the total PU.

The recalculated CXO source PUs are significantly smaller than the PUs in CSC2, and we suspect they may be underestimated (e.g., several soft X-ray sources were $<1''$ away from fairly bright optical stars). Therefore, we increased the combined CXO and multiwavelength catalog PUs by a factor of 1.5. As a result, 6 additional sources previously lacking any counterparts are matched to a counterpart, while 31 additional counterparts are added in total.\footnote{\textbf{A CXO source that only has one counterpart, may be matched to counterparts in other catalogs after the expansion of the combined PU.}} Given that the CXO PUs are $2\sigma$ uncertainties, these 6 additional matches are expected. \textbf{Assuming a median of $1.2''$ for the expanded CXO PU, the chance coincidence probability for a CXO source to be matched with at least one cluster member, assuming an average density of $\sim1,000$ cluster members in a $20'$ radius field that covers the CXO field (see Section \ref{membership}), is $\sim 0.1\%$, while the probability to be matched with any Gaia source (including background sources), assuming an average density of $\sim48,000$ Gaia sources in the $12'$ radius field directly surrounding the CXO field, is $\sim 12.5\%$. We emphasize this mostly affects sources near the edge of the CXO field with large PUs that were not already matched to Gaia counterparts (which in most cases are well within CXO PUs), and we discuss some of these sources in Section \ref{detailed}.}

Of the 131 CXO sources in the field of NGC 3532 that pass our selection criteria, 109 have Gaia counterparts; 15 have DECaPS2+ counterparts but not Gaia; 95 have 2MASS counterparts; 82 have WISE counterparts, of which 47 were from AllWISE, 25 were from CatWISE2020, and 10 were from UnWISE.

\section{Cluster Analysis}
\label{analysis}

We summarize various multiwavelength properties of CXO sources in the field of NGC 3532 with several plots, including luminosity function plots, color-magnitude diagrams (CMDs), color-color diagrams (CCDs), and a hardness ratio diagram (HRD). 

\subsection{Cluster membership}
\label{membership}
Cluster membership is determined by a set of distance and proper motion cuts using Gaia eDR3 data \citep{brown_gaia_2021, bailer-jones_estimating_2021}. About 134,000 Gaia sources within $20'$ from the center of the ACIS-I array field-of-view (see Figure \ref{fig:ngc3532_cxo}) were included in the analysis. First, we apply a preliminary cut by excluding sources outside $\pm 33\%$ pc and $\pm 5$ mas/yr of the mean cluster distance of 484 pc, and proper motion of $\mu_\alpha = -10.37$ mas/yr, $\mu_\delta = 5.18$ mas/yr \citep{fritzewski_spectroscopic_2019}. Then, the sources within one standard deviation of the median value of all three parameters are taken as cluster members. This process produces a membership list of 916 stars which is relatively pure. Compared to a list of 660 members produced by \cite{fritzewski_spectroscopic_2019} from radial velocity data and Gaia DR2, our list is larger, but may be less pure. Within our $20'$ radius field, \cite{fritzewski_spectroscopic_2019} select 356 members, from which we also select 344 as members. However, we have close to three times the total number of members. Compared to another list of 1,300 members produced from Gaia DR2 parallax and proper motions using Gaussian mixture models \citep{jaehnig_membership_2021}, our list is less complete, because we restricted our selection of sources to $r<20'$, but it is more pure, having less contaminants with obviously wrong proper motions and distances. The number of CXO sources crossmatched to cluster members also increases to 57 compared to 40 from \cite{jaehnig_membership_2021}

\subsection{Variability}

Using the definition of variability discussed in Section \ref{obs}, we find that 37 X-ray sources out of 131  (i.e.\ 28\%) are significantly variable. Of these, 34 have Gaia, 30 have 2MASS, 24 have WISE, and 2 have DECaPS2 counterparts.  About 20 variable sources are likely to be cluster members, and 18 display flares. 
For the 16 flaring sources having Gaia distances, their average flare luminosities\footnote{All flare luminosities we provide hereafter are average flare luminosities.} are in the range $\SI{7e29} - \SI{9e31}{\erg\per\s\per\square\cm}$. The largest flare from a cluster member is the flare of Source 29 at $\SI{3.4e30}{\erg\per\s\per\square\cm}$.

\subsection{Luminosity Function}

\begin{figure*}[hbt!]
    \centering
    \includegraphics[width=0.75\textwidth]{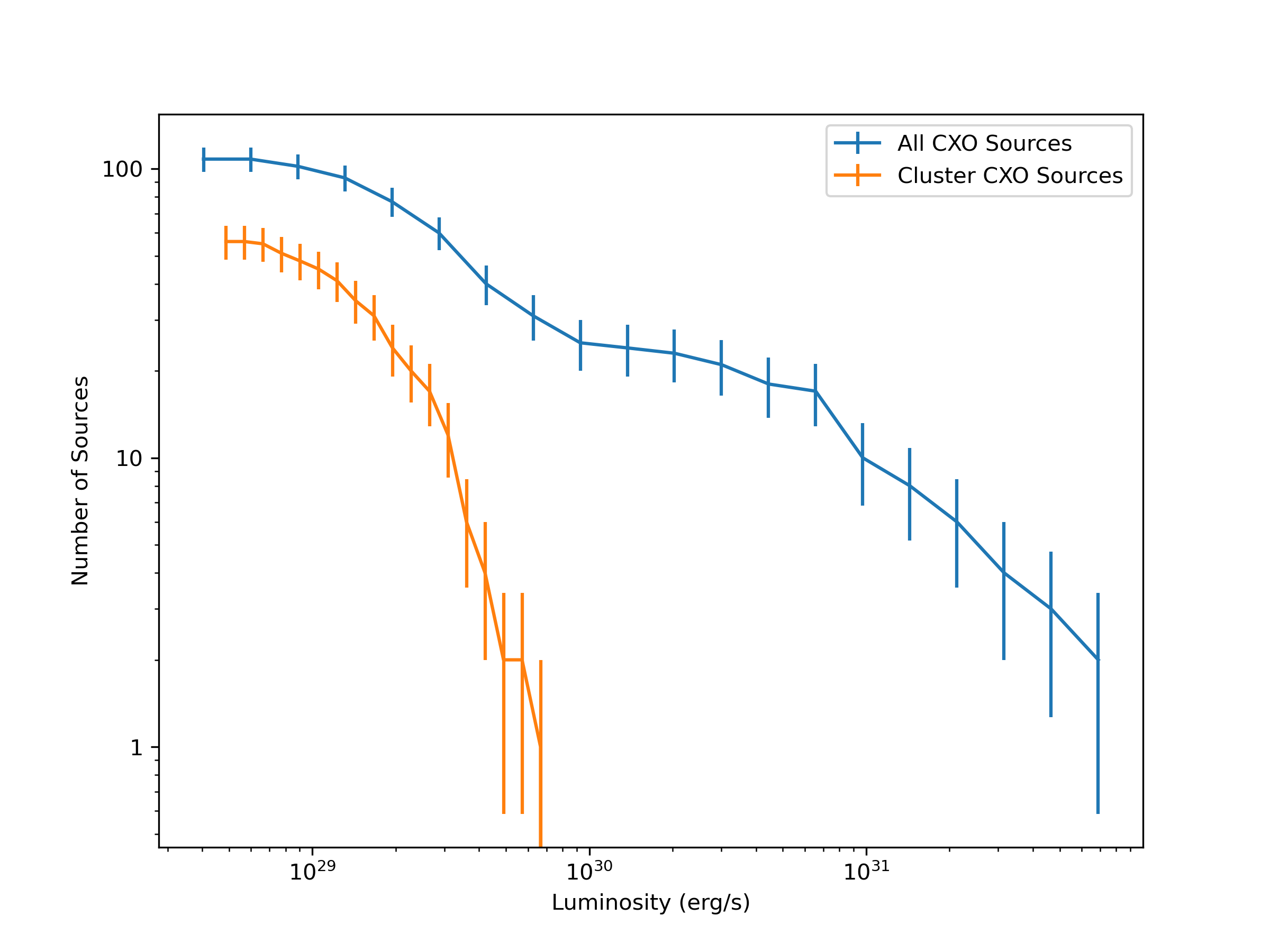}
    \caption{Cumulative luminosity function of CXO sources in the field of NGC 3532. Top: 108 sources in field with a distance measurement. Bottom: 60 CXO sources crossmatched to cluster members.}
    \label{fig:ngc3532_lf}
\end{figure*}

The cumulative luminosity function of CXO sources in the field of NGC 3532 is shown in Figure \ref{fig:ngc3532_lf}. Luminosity is calculated from the CXO broadband (0.5-7 keV) flux using Gaia distances \citep{bailer-jones_estimating_2021} for sources with a Gaia counterpart. Sources without Gaia counterpart are not shown. The top curve  shows the 108 CXO sources with a distance measurement, while the bottom curve   shows the 60 cluster members. 

All sources brighter than $\SI{1e31}{\erg\per\s}$ are not cluster members. At higher luminosities the cluster luminosity function may be approximated by a power-law, while at lower luminosities it comes to a plateau. While the plateauing can be explained by the limiting sensitivity of the observation, below which objects are not detected, the apparent break near $L_X\approx \SI{3e29}{\erg\per\s}$ should not be related to the sensitivity limit of $\sim\SI{5e28}{\erg\per\s}$.  

\subsection{Color-Magnitude Diagrams}
\label{cmd}

\begin{figure*}
    \centering
    \includegraphics[width=\textwidth]{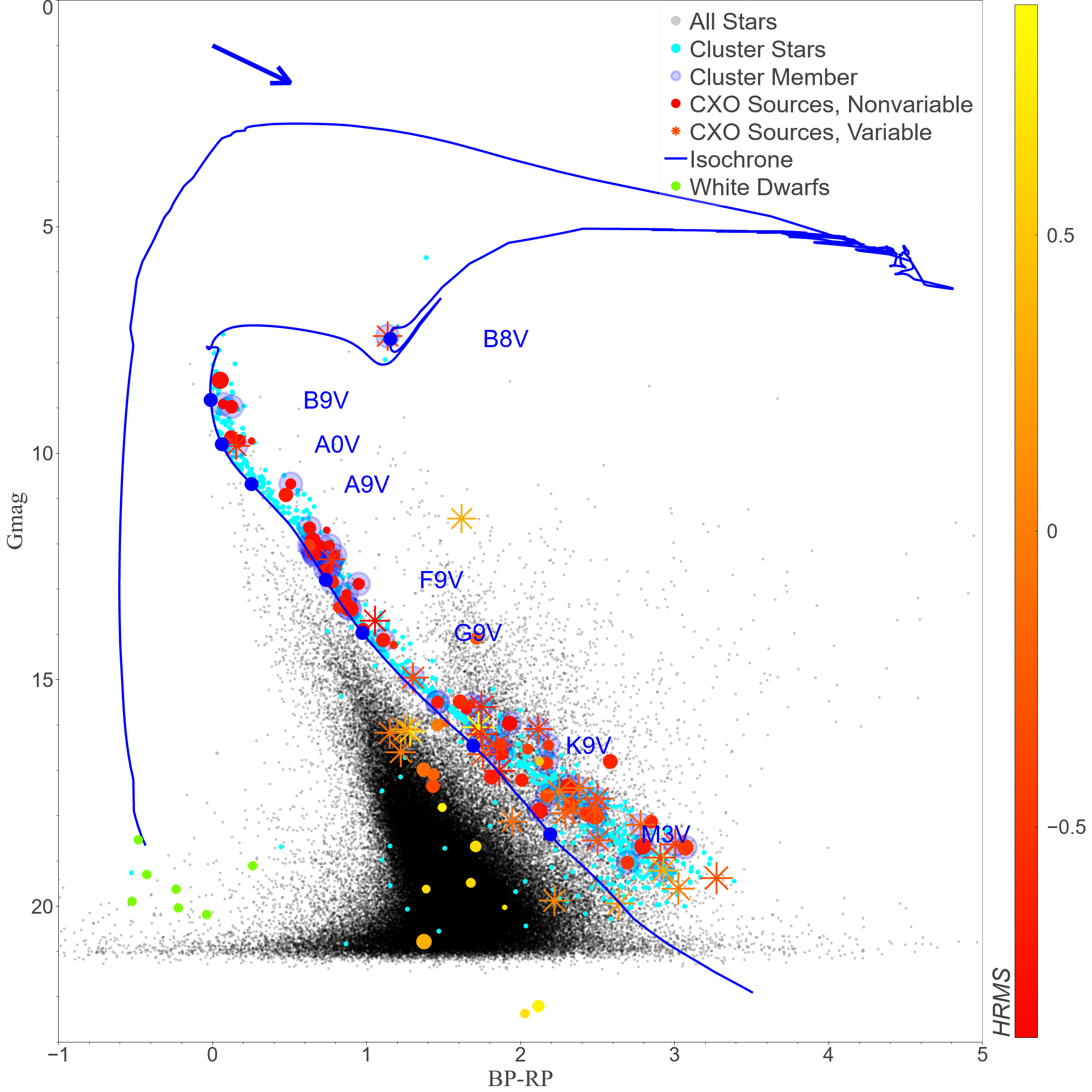}
    \caption{Color Magnitude Diagram (CMD) of NGC 3532. All Gaia sources located within $20'$ from the center of CXO observation are shown in black. Cluster members are shown in cyan. Cluster WDs with Gaia counterparts are shown in green. Sources with X-ray counterparts are shown in red-yellow color scale, with color indicating the CXO hardness ratio $\mathrm{HR}_{ms}$ (redder color corresponds to softer spectrum), and size proportional to the logarithm of the broadband flux ($F_b$). Variable X-ray sources marked with asterisks. An extinction vector corresponding to $A_V=1$ is shown in blue, while the total Galactic $A_V$ in this direction is $\approx4$. An isochrone corresponding to the age of 300 Myrs, $d=484$ pc, and extinction $E(B-V)=0.034$ is also plotted. Reference labels for several spectral types based on isochrone masses are shown. }
    \label{fig:ngc3532_mw_bp-rp_gmag}
\end{figure*}

A color-magnitude diagram (CMD) of NGC 3532 constructed from Gaia and DECaPS2 data is shown in Figure \ref{fig:ngc3532_mw_bp-rp_gmag}. All Gaia eDR3 sources within the 12$'$-radius around the center of ACIS-I field of view are shown in black. Cluster members are shown in cyan. Gaia sources with CXO counterparts are shown with a red-yellow color scale, with color indicating the value of the medium-soft hardness ratio, $\mathrm{HR}_{ms}$. The sizes of the markers for these sources scale with the logarithm of the CXO broad-band flux, $\log(F_b)$. Variable X-ray sources are marked with asterisks. Several known WDs in NGC 3532 crossmatched to Gaia sources are shown in green, and appear below the main sequence.\footnote{See also Table \ref{table:wd}.} An isochrone for the age of 300 Myr, distance 484 pc, solar metallicity, and extinction E(B-V)=0.034 (discussed in Section \ref{ngc3532}) is also shown.\footnote{Isochrones are constructed with Python \texttt{Isochrones} package, using MIST stellar evolution models \citep{morton_isochrones_2015}.}

The cluster members form a clear main sequence. A few evolved cluster stars are well-fitted by the isochrone (except for one). The isochrone appears to be slightly offset to  the left of the main sequence, with the deviation more apparent in the lower mass range. This deviation is due to an issue with how isochrone models transform colors, and is also present in isochrones in \cite{fritzewski_spectroscopic_2019}. One cluster member appears near the known white dwarfs, but is not classified as a white dwarf by Gaia DR3 DSC-Combmod \citep{fouesneau_gaia_2022}. There are a few sources that passed our fairly strict cut for cluster membership (Section \ref{membership}) but are still located below the main sequence. The origin of these sources is unclear. Since none of these outliers coincides with CXO sources, we do not investigate them further. 

Many sources with X-ray counterparts are located near the isochrones, indicating their cluster membership. Given the optical properties and the relative X-ray softness (see colormap), these sources are probably stars with active coronae (this conclusion is confirmed later in Section \ref{classification} with ML classification and in Section  \ref{detailed} with spectral analysis). Most variable X-ray sources appear at the fainter part of the NGC 3532 main sequence populated by low-mass stars.

There are two additional structures that are visible in the CMD plot, one above and one below the main sequence. These structures were also noticed by \cite{clem_deep_2011}. The structure below the main sequence are contaminating field stars withing the plane of the Galaxy beyond NGC 3532. A number of counterparts of harder X-ray sources fall within this region. Their hardness can be attributed to the additional absorption through the plane, and/or to the intrinsically harder spectra. The plume of sources above the main sequence (mostly field giant stars according to \citealt{clem_deep_2011}) merges with the main sequence at fainter magnitudes, but branches off at brighter magnitudes. The two CXO sources with DECaPS converted magnitudes at $G>22$ are discussed in \ref{classification}.

Similarly constructed NIR and IR CMDs are shown in Figure \ref{fig:ngc3532_ir_cmd}. In the NIR CMD, the same three structure as in the optical CMD are visible. Most X-ray sources still appear on the main sequence, with a number of NIR-faint sources with harder X-ray spectra clustering toward the bottom of the main sequence. Many of these sources are variable in X-rays. These are likely to represent a mix of flaring low-mass stars in the cluster, or beyond it. 

The structures seen in the optical and NIR CMDs are not apparent in the IR CMD. The main sequence is still visible, but non-cluster sources now appear close to the main sequence at brighter magnitudes. Most variable sources are clustered at the fainter end of the CMD similarly to the optical and NIR CMDs.

\subsection{Hardness Ratio Diagram}
\label{hardness ratio}

\begin{figure*}[hbt!]
    \centering
    \includegraphics[width=\textwidth]{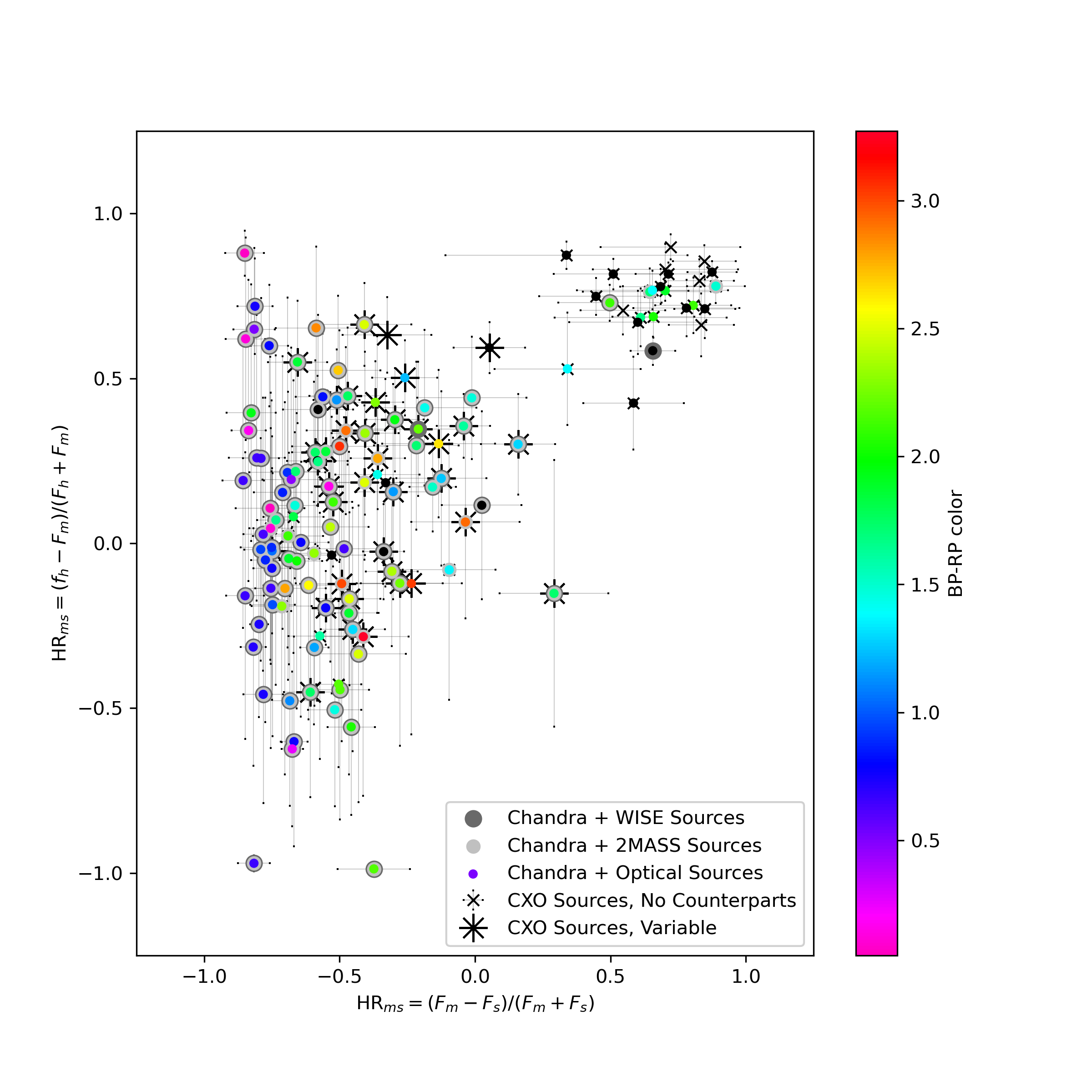}
    \caption{HR diagram for CXO sources. Variable sources are marked with asterisk. Sources with WISE counterpart marked with dark gray circle; 2MASS counterpart with light gray circle; optical counterparts have colormap corresponding to Gaia BP-RP color, with sources missing BP-RP color shown in black. Sources with multiple counterparts have overlapping markers.\textbf{ Non-variable sources without counterparts are marked with small 'x' crosses. }}
    \label{fig:ngc3532_mw_hrms_hrhm}
\end{figure*}

\begin{figure*}
    \centering
    \includegraphics[width=0.45\textwidth]{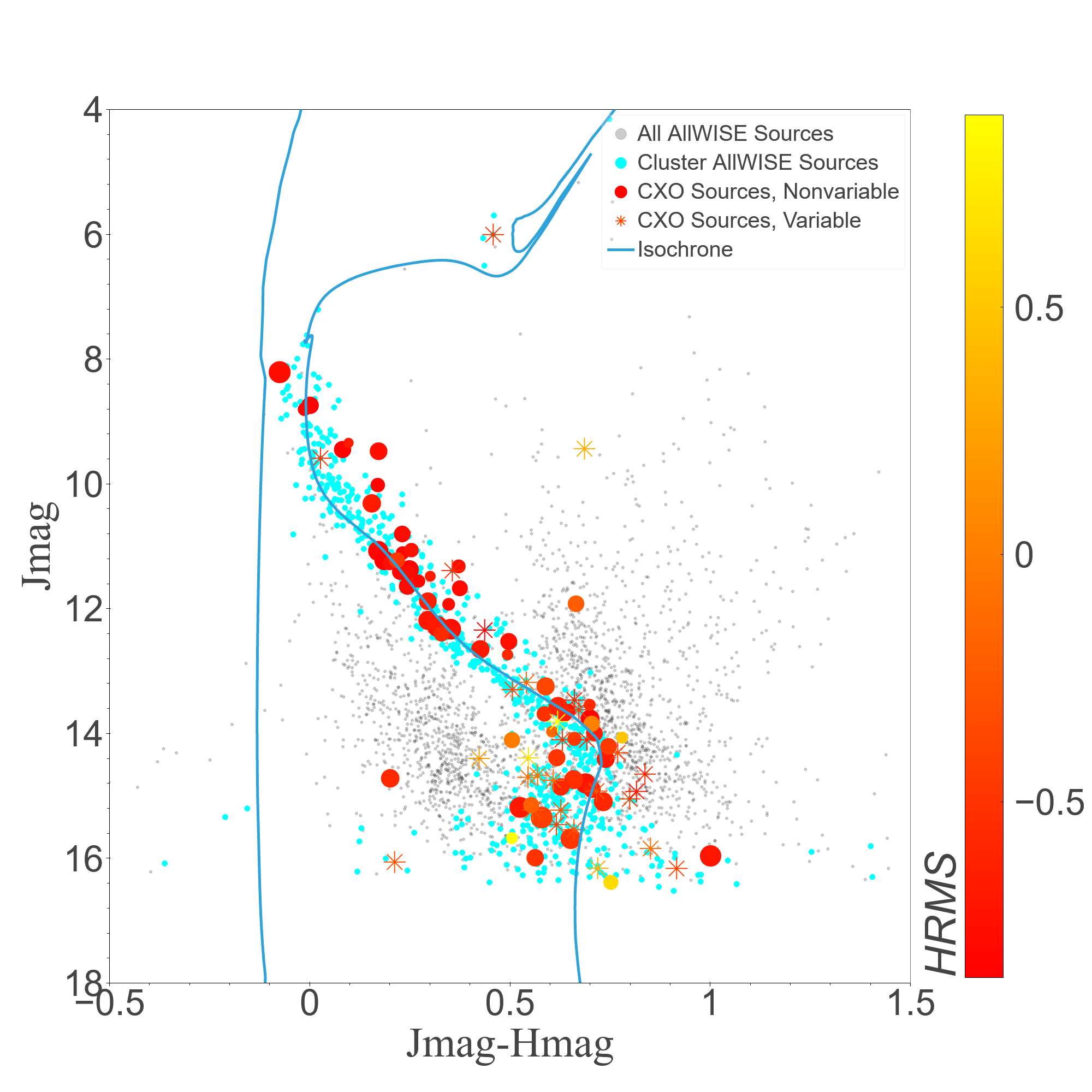}
    \includegraphics[width=0.45\textwidth]{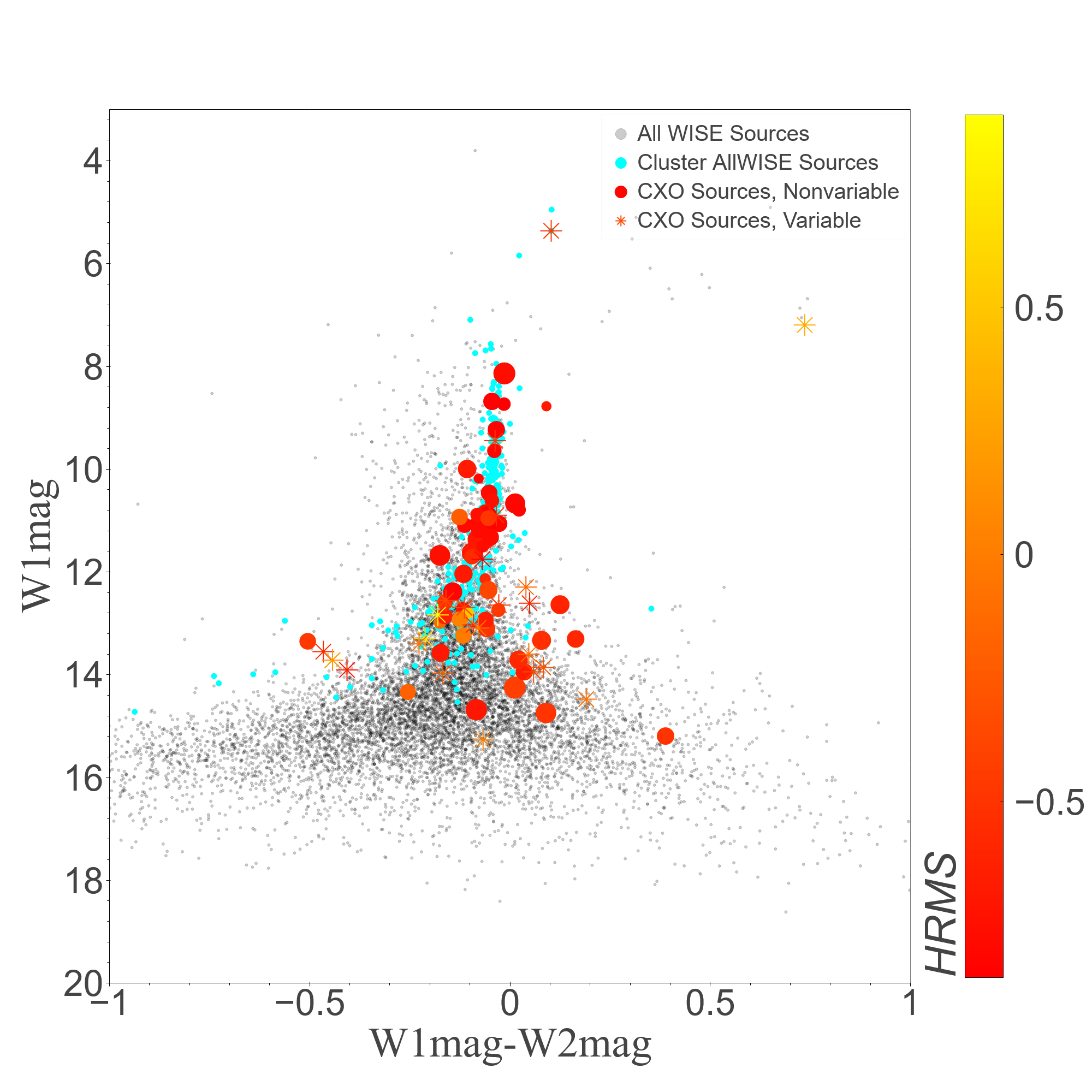}
    \caption{Color-Magnitude Diagrams of NGC 3532 in near-IR (left) and IR (right). The left panel also shows the isochrone corresponding to the age of 300 Myrs, $d=484$ pc, and extinction $E(B-V)=0.034$. Cluster members shown in cyan, field sources shown in black, sources with X-ray counterparts in red-yellow color scale (with color indicating medium-soft hardness ratio ($\mathrm{HR}_{ms}$), and size proportional to the logarithm of the broadband flux ($F_b$). Variable X-ray sources marked with asterisks. \textbf{For the left panel, the AllWISE catalog, which cross-matches to 2MASS sources, are used for background sources. For the right panel, AllWISE+UnWISE+CatWISE2020 sources are used for background sources.}}
    \label{fig:ngc3532_ir_cmd}
\end{figure*}

A hardness ratio plot for all X-ray sources in the field of NGC 3532 is shown in Figure \ref{fig:ngc3532_mw_hrms_hrhm}. Any counterparts are indicated by overlapping markers, see plot legend. 

The numerous sources with soft ($\mathrm{HR}_{ms}<-0.7$) spectra and fairly blue optical counterparts ($0<$BP-RP$<1.5$) are main sequence stars (cf.\ Figure \ref{fig:ngc3532_mw_bp-rp_gmag}) belonging to NGC 3552. The soft X-ray emission can be attributed to active stellar coronae with typical temperatures of a few million degrees (a fraction of a keV). As we show in Section \ref{classification}, the MUWCLASS pipeline indeed classifies these sources as low-mass stars. The softest and bluest of these sources (lying solely on the main sequence) are virtually all non-variable, implying that the 130 ks CXO observation was too short to catch any flares. Their X-ray luminosities correspond to a steady level of coronal activity at $\sim \SI{1e29}{\erg\per\s}$. For comparison, the Sun's quiescent X-ray luminosity ranges from $ \SI{1e27}{\erg\per\s}$ to $ \SI{1e28}{\erg\per\s}$ \citep{judge_estimate_2008}, significantly lower than the luminosities of these cluster stars. This is consistent with the expectation that younger stars are more coronally active \citep{gudel_x-ray_2009, davenport_evolution_2019}. 

The redder sources ($1.5<$BP-RP$<3$) mostly correspond to the bottom part of the cluster's main sequence (see Figure \ref{fig:ngc3532_mw_bp-rp_gmag}) with most of these sources exhibiting somewhat harder X-ray spectra. The central part of the HR diagram contains a number of these redder variable sources, which could be active binaries or flaring coronae of more active solitary stars. Finally, there are several soft X-ray sources that lack optical and NIR/IR counterparts, or with only faint DECaPS2 counterparts. Their properties are discussed in more detail in Section \ref{background_softer}.

The upper right region of the HR diagram features strongly absorbed sources with relatively hard (either due to strong absorption or intrinsically hard) X-ray spectra. Twenty of these sources have optical counterparts, 13 of which have only faint ones in DECaPS2. As discussed in Section \ref{classification} and Section \ref{background_hard}, many of these sources are likely AGNs, while the ones for which we can exclude an extragalactic origin may be Galactic CO systems.

\subsection{Color-Color Diagrams}

\begin{figure*}
    \centering
    \includegraphics[width=0.45\textwidth]{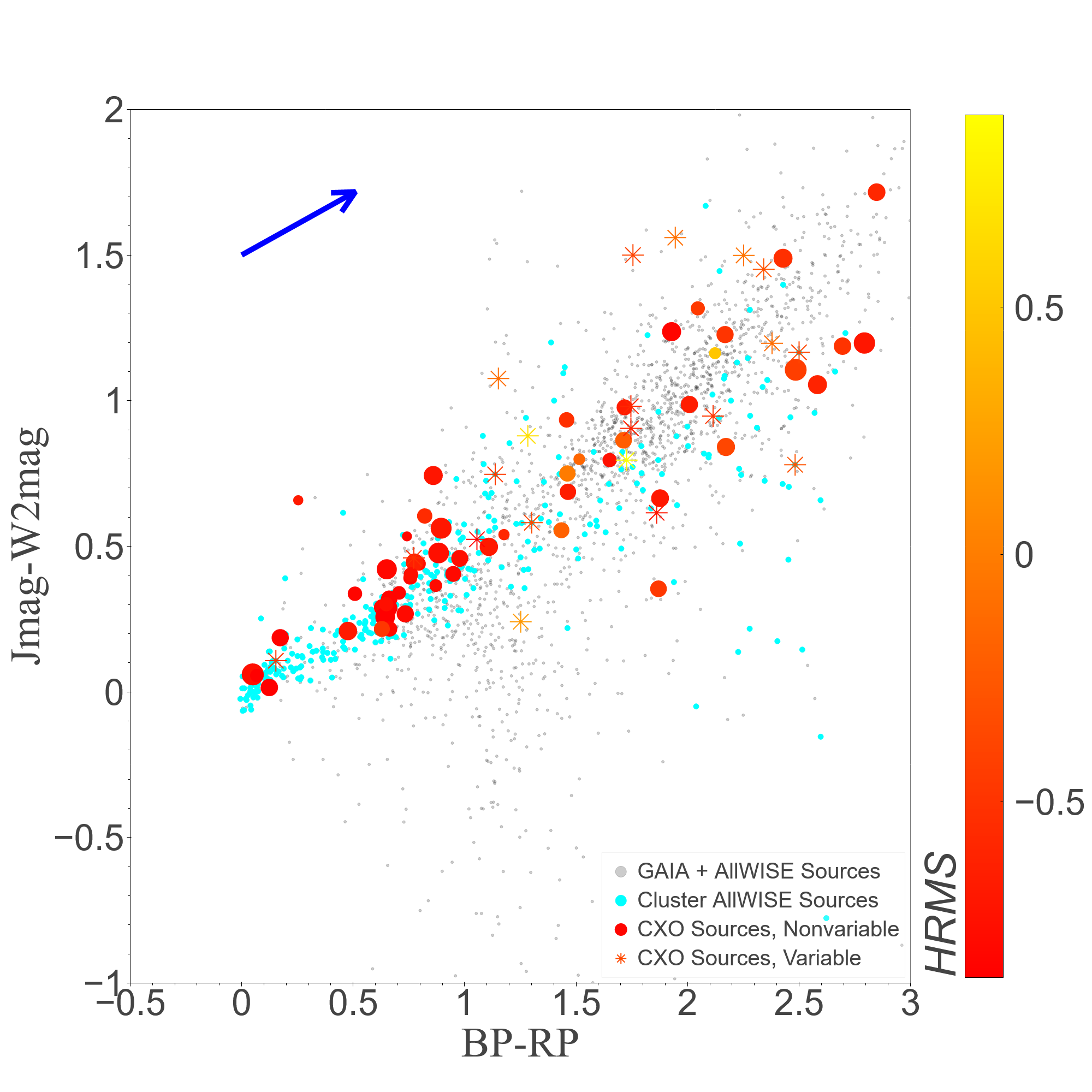}
    \includegraphics[width=0.45\textwidth]{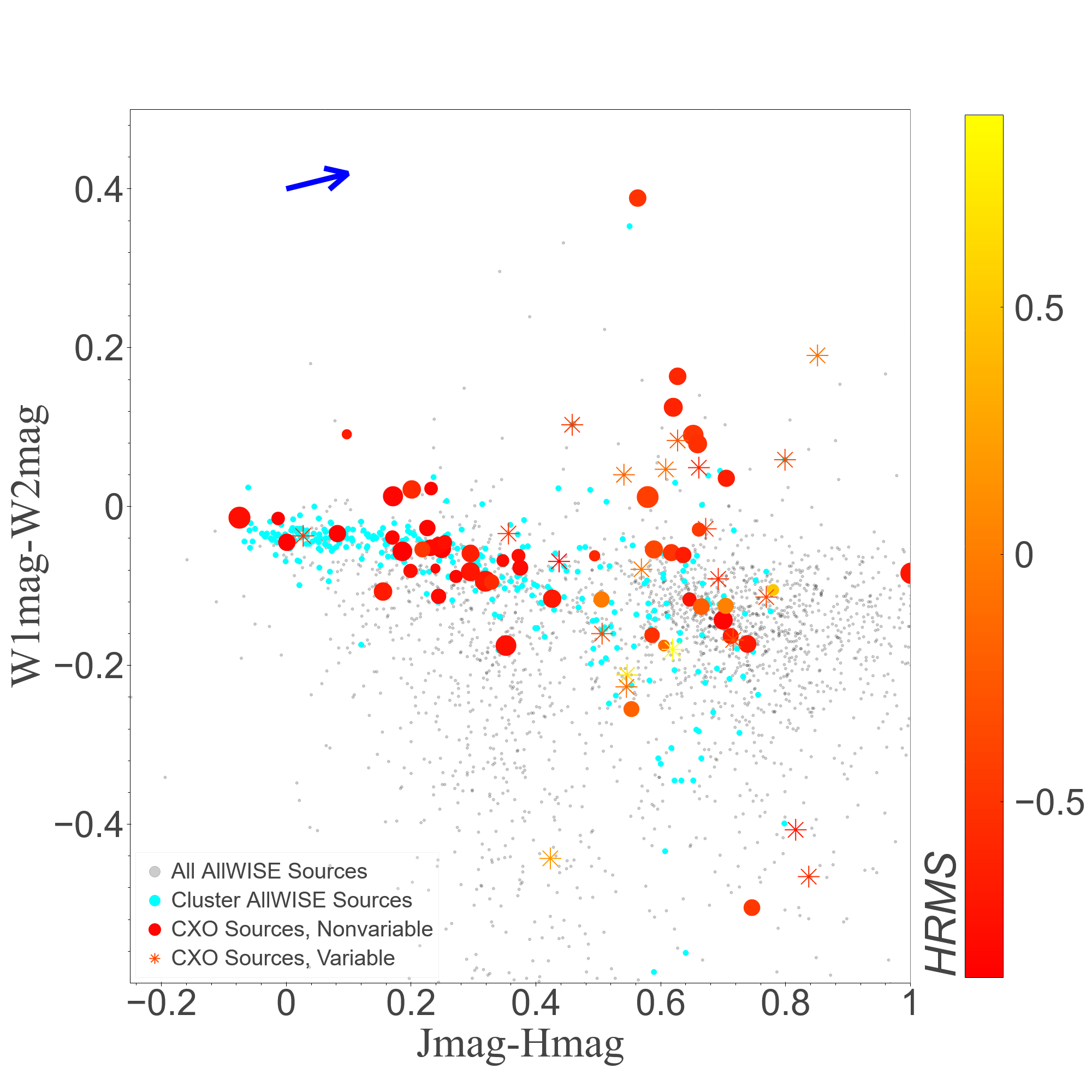}
    \caption{Optical and Infrared CCDs of NGC 3532 \textbf{, constructed similar to \ref{fig:ngc3532_ir_cmd} are used.} An extinction vector corresponding to $A_V=1$ is shown in blue. }
    \label{fig:ngc3532_cc}
\end{figure*}

Color-Color Diagrams of NGC 3532 constructed from Gaia, 2MASS, and WISE data are shown in Figure \ref{fig:ngc3532_cc}, with the same color scheme as in Figure \ref{fig:ngc3532_mw_bp-rp_gmag}. 
The sources along the diagonal locus of points are mostly stellar, while the outliers are more likely to be binaries or non-stellar sources. The harder sources are typically associated with redder sources in BP-RP and J-W2 colors, suggesting that both X-ray HRs and colors are affected (at least partly) by the extinction (see extinction vectors). The W2 band is too red to be affected by the extinction, and must be more representative of the intrinsic spectrum of the source. 

\section{Machine Learning Classification}
\label{classification}

We supplement our analysis with automated classification of X-ray sources using a multiwavelength machine-learning classification (MUWCLASS) pipeline described in detail by \cite{yang_classifying_2022}. The pipeline makes use of a training dataset (TD; see also \cite{yang_classifying_2022}) with $\sim3,000$ X-ray sources of known classes and 33 multiwavelength features from CSC2, Gaia, 2MASS, and three WISE catalogs, including fluxes, magnitudes, colors, X-ray variability characterization, distances, and luminosities.\footnote{Note that the pipeline described in \cite{yang_classifying_2022} did not use distances and luminosities. We added these features in this work.}

MUWCLASS uses a Random Forest algorithm to classify X-ray sources into eight classes: low-mass stars (LM-STARs\textbf{,up to late B-type}), high-mass stars (HM-STARs\textbf{, OB and Wolf-Rayet}), AGNs, Young Stellar Objects (YSOs\textbf{, protostars and pre-main sequence stars}), Low-Mass X-ray Binaries (LMXBs\textbf{, including binaries in quiescence  , and spider-type systems}), High Mass X-ray Binaries (HMXBs\textbf{, including gamma-ray binaries}), Cataclysmic Variables (CVs), and Neutron Stars (NSs\textbf{, only isolated ones are included}). \textbf{For additional details of which types of sources and catalogs comprise each class, please refer to Section 2.1 of \cite{yang_classifying_2022}.}

Since NGC 3532 is near the Galactic plane, and nearly all AGNs included in the TD are located outside of the plane, the reddening through the Galactic plane in the direction of NGC 3532 corresponding to $E(B-V)=1.3$ \citep{ruiz_ruizcagdpyc_2018}, as well as photoelectric absorption corresponding to $n_{H}=\SI{9e21}{\cm^{-2}}$ \citep{guver_relation_2009} has been applied to all TD AGNs in the optical-NIR-IR and X-rays, respectively (see \citealt{yang_classifying_2022} for details).

For each feature of each source (in both the TD and the field data to be classified), MUWCLASS creates a probability distribution function of the feature values based on the measurement uncertainties. We run MUWCLASS 1,000 times, each time sampling features from their probability distribution functions, and each time producing classification probabilities for each class, based on the percent of trees in the random forest that predict that class.\footnote{For example, a source to be classified may have Gaia feature $G=15$ mag with uncertainty $e_G=0.05$ mag. For the 1,000 samplings, one sampling may produce $G=14.99$ mag, while another could give $G=15.02$ mag. As a result, for one sampling, 80\% of trees in the random forest may classify the source as a LM-STAR, 20\% a YSO, while for another sampling the probability outcomes will be slightly different.}
After 1,000 samplings, the mean probability ($P_{\rm class}$) of a source belonging to each class, and its standard deviation ($\Delta P_{\rm class}$; hereafter the classification probability uncertainty) which characterizes the width of the $P_{\rm class}$ distribution, can thus be calculated by incorporating uncertainty information for each feature (see \citep{yang_classifying_2022} for further details).

Confidently classified CXO sources are selected using a classification confidence threshold defined as:

\begin{equation}
{\rm CT} = \min_{\rm class}(\frac{P_{\rm predicted\,class}-P_{\rm class}}{ \Delta P_{\rm predicted\,class}+ \Delta P_{\rm class}}),
\label{eq:CT}
\end{equation}

where the class index runs through the classes that are different from the predicted class. We define confidently classified sources as those with CT$\ge2$.

Unlike \cite{yang_classifying_2022}, we use distance measurements, $r_\mathrm{geo}$, from the Gaia eDR3 distance catalog \citep{bailer-jones_estimating_2021} to the list of features. This allows for the incorporation of NIR J-band, optical G-band and broadband X-ray luminosities for sources with reliable distances, defined by a cut on the Gaia eDR3 parallax measurements $\pi/\sigma_\pi >= 2$. This cut removes the distances of most sources in the TD where a real parallax measurement is not expected, e.g., AGNs. About one third of all CXO sources in the TD, and in the field of NGC 3532, have distances after the cut. About 95\% of CXO sources with Gaia counterparts in the field of NGC 3532 have distances, which is expected due to the proximity of the cluster, and its location in the Galactic plane.

Due to the inclusion of the additional features, we re-evaluate the performance of the MUWCLASS pipeline, which is summarized by the confusion matrices in Appendix \ref{cm}. Overall, the addition of these distance-dependent features slightly improves the performance of the pipeline. Similar to the unmodified pipeline, the best performing classes are AGNs, LM-STARs, and YSOs, which are the best represented classes in the TD. Since sources that include stellar COs are both diverse in nature, and lower in number in the TD, the classification performance of CO classes (LMXBs, HMXBs, CVs, NSs) tend to be worse, and classifications tend to be confused among these classes. 

Therefore, to more efficiently search for CO candidates in NGC 3532, we combined LMXBs, HMXBs, CVs and NSs into a candidate CO class, with the classification probability calculated as the sum of the probabilities to belong to each of the four classes, and the corresponding classification probability uncertainties combined in quadrature. After merging the four classes into one, the previous 8-class scheme turns into a 5-class scheme which includes AGNs, HM-STARs, LM-STARs, YSOs and candidate COs. The same confidence threshold in equation \ref{eq:CT} was recalculated to evaluate the confident classifications in the 5-class scheme. The performance evaluation of the pipeline using the 5-class scheme is shown in the lower panel in Fig. \ref{fig:cm} in Appendix \ref{cm}. 

\subsection{Classification Summary}

Among the 131 X-ray sources in the NGC 3532 field, 70 have already been classified in \cite{yang_classifying_2022} while others were dropped either because they have large PUs or have confused and extended CSC2 flags raised. Of these 70 sources, 31 are confidently classified in this work, with their classification mostly consistent with the results of \cite{yang_classifying_2022}.\footnote{Among the confidently classified in this work sources, only 3 sources classified as LM-STARs were classified as HM-STARs in \cite{yang_classifying_2022} albeit at lower confidence. } These include 19 LM-STARs, 6 AGNs, 4 YSOs, 1 HM-STAR, and 1 LMXB.

\begin{figure*}
    \centering
    \includegraphics[width=450pt]{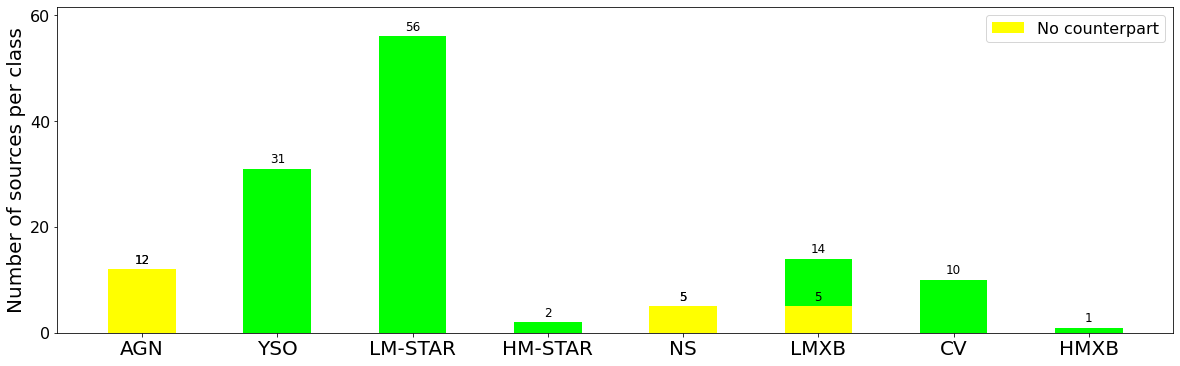}
    \includegraphics[width=450pt]{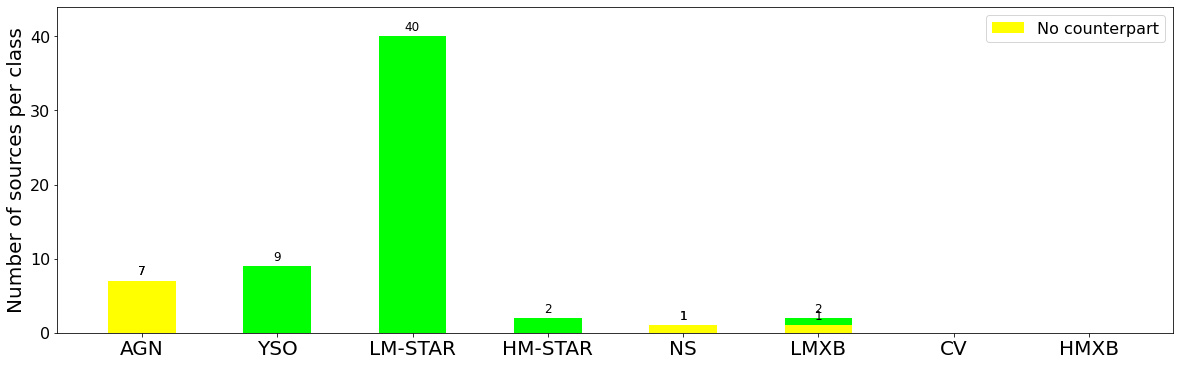}
    \includegraphics[width=450pt]{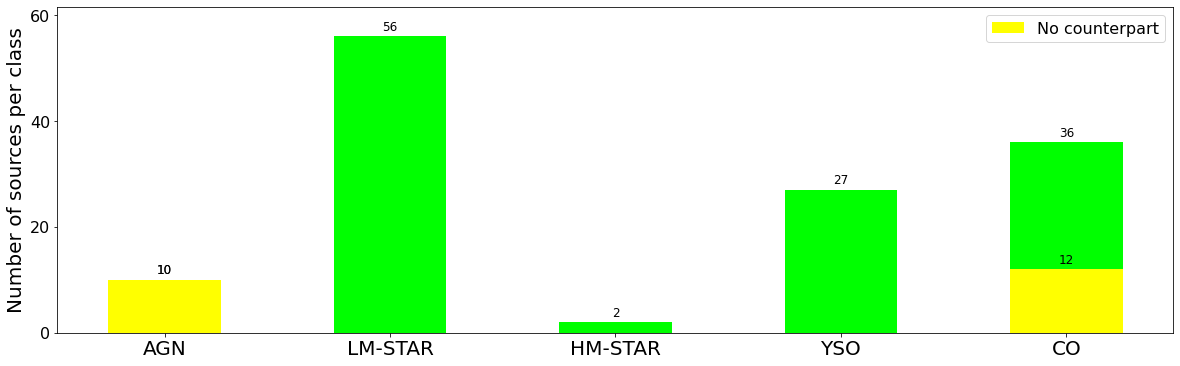}
    \includegraphics[width=450pt]{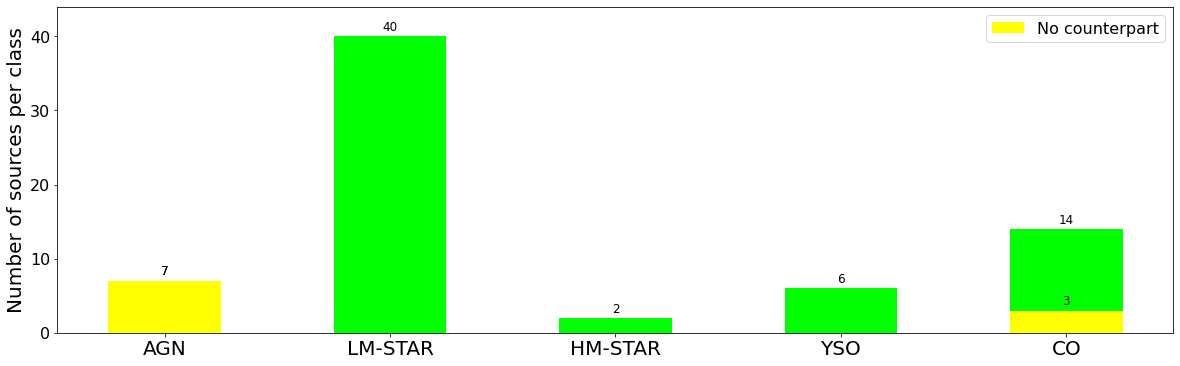}
    \caption{Summary of the classification outcomes for X-ray sources in NGC 3532. The green histograms show the classification distributions of all sources per class while the yellow histograms show the subsets without multiwavelength counterparts (DECaPS2 counterparts, which were not used for classification, are not counted here). The bins are labeled with the number of source belong to each class. The first panel shows the distributions for all classifications using the 8-class scheme. The second panel shows the distributions for confident classifications (CT$>2$) using the 8-class scheme. The third and forth panels show the same but for the 5-class scheme.}
    \label{fig:NGC3532_classification_summary}
\end{figure*}

The classification breakdown of the 131 X-ray sources in this work is shown in Figure \ref{fig:NGC3532_classification_summary}, with the 8-class scheme results shown in the first two panels, and the 5-class scheme results shown in the last two panels. The second and fourth panels show the sources that passed the confidence cut at CT=2 for their respective class schemes.

In the 8-class scheme, only 3 out of 31 sources classified as one of the CO classes pass the confidence cut. After combining the 4 classes into a single CO class (the 5-class scheme), 14 sources out of 37 classified as a candidate CO pass the confidence cut. None of the candidate COs were crossmatched to a cluster member. \textbf{Two of the 14 only have DECaPS counterparts, which were not used in ML classification, while one of the 14 have no MW counterparts in any catalog.}

In both schemes, MUWCLASS confidently classify 40 LM-STARs, 7 AGNs, and 2 HM-STARs, while the 5-class scheme confidently classify three less YSOs due to differences in the candidate CO class uncertainties between the two schemes. As the goal of the 5-class scheme is to identify candidate COs, for the purposes of plotting we use the 8-class scheme, and overlay candidate COs on top.

All confidently classified stellar objects (including LM-STARs, HM-STARs and YSOs) have multi-wavelength counterparts, while all confidently classified AGNs do not, except for faint ($>20$ mag) DECaPS counterparts, which may be caused by the substantial extinction ($E(B-V)=1.3$ or $A_V\approx4$) through the Galactic plane in the direction of NGC 3532. 

\subsection{Diagrams with Classification Results}

\begin{figure*}
    \begin{interactive}{js}{plots/NGC3532_BP-RP_Gmag_classification_rgeo_lum.html}
    \centering
    \includegraphics[width=\textwidth]{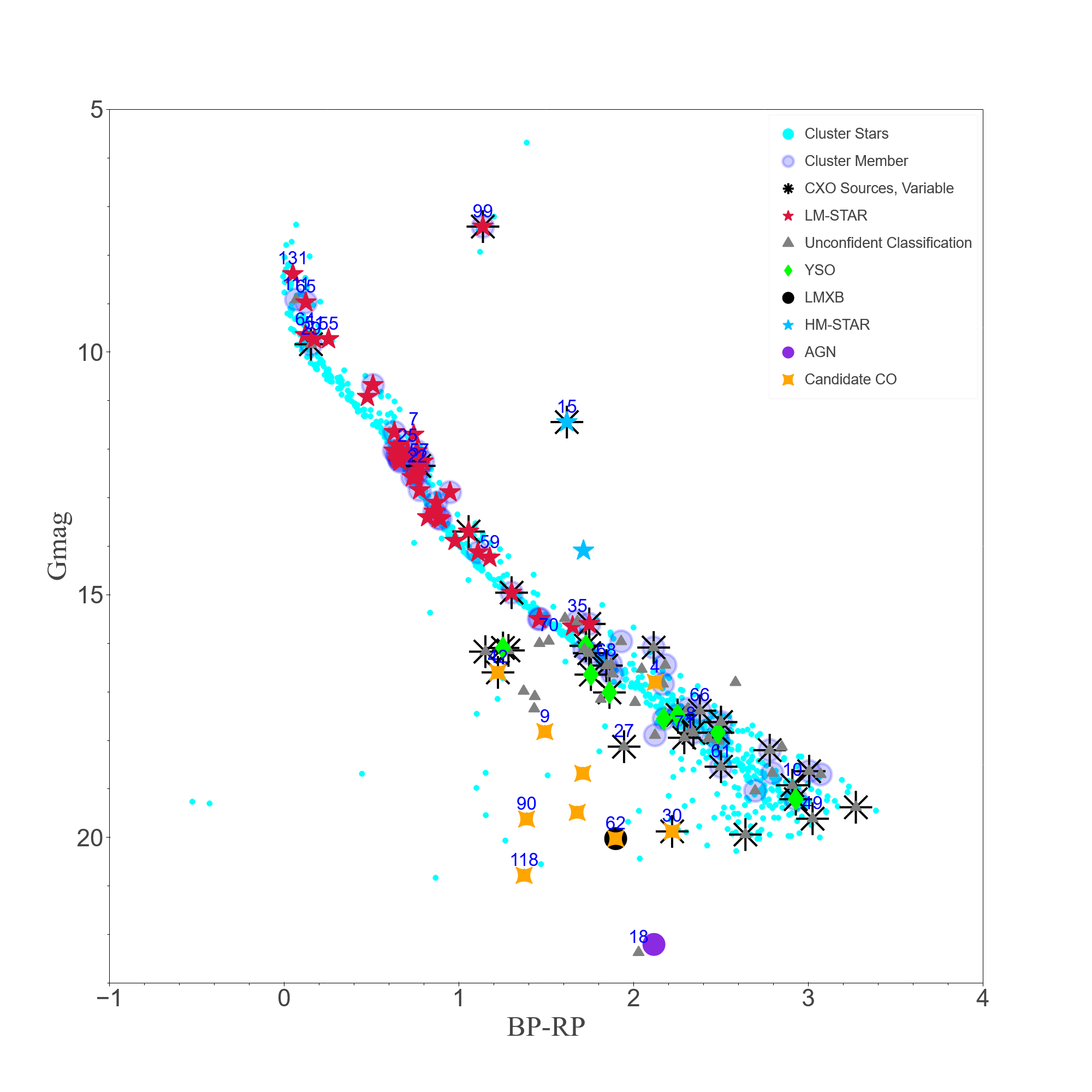}
    \end{interactive}
    \caption{CMD of NGC 3532. Gaia cluster members shown in cyan. Classifications of CXO sources with optical counterparts are labeled according to legend. Candidate COs marked with orange stars. Sources discussed in Section \ref{detailed} are labeled. This figure is available online as an interactive figure, with the ability to zoom, pan, and display detailed information for each source. }
    \label{fig:NGC3532_mw_BP-RP_gmag_classification}
\end{figure*}

Figure \ref{fig:NGC3532_mw_BP-RP_gmag_classification} shows the CMD with confidently classified sources marked by various symbols. Most sources classified as LM-STARs and YSOs are located on the main sequence. LM-STARs appear to be brighter in the G-band and are redder in color. One LMXB, along with other candidate COs appear below the main sequence. \textbf{The two CXO sources with DECaPS converted magnitudes at $G>22$ have highest AGN probabilities, with one passing the confidence threshold}. This is consistent with the reddening procedure described in Section \ref{classification}, which results in almost all AGNs in the TD being reddened to BP-RP$>2$, $G>18$.

\begin{figure*}
    \begin{interactive}{js}{plots/NGC3532_HRMS_HRHM_classification_rgeo_lum.html}
    \centering
    \includegraphics[width=\textwidth]{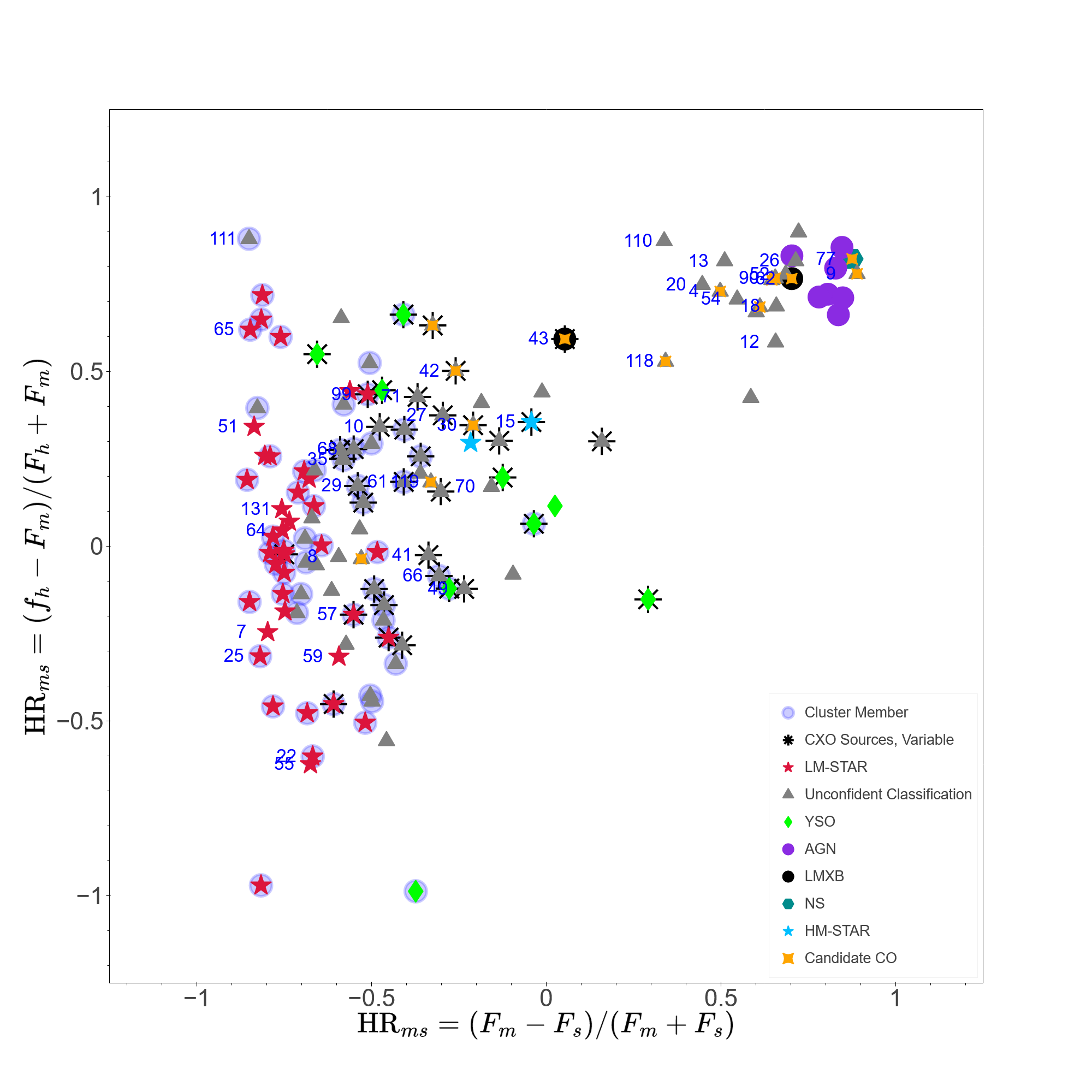}
    \end{interactive}
    \caption{HR diagram of CXO sources with classifications labeled according to legend. Candidate COs marked with orange stars. Sources discussed in Section \ref{detailed} are labeled with numbers. This figure is available online as an interactive figure, with the ability to zoom, pan, and display detailed information for each source. }
    \label{fig:NGC3532_HRMS_HRHM_classification}
\end{figure*}

Classified sources lacking optical colors do not appear on the CMD plot. Therefore, we also plot a HR diagram with classification results in Figure \ref{fig:NGC3532_HRMS_HRHM_classification}. A clear segregation of source classes along the medium-soft HR scale is seen: LM-STARs are soft; many unconfidently classified sources, including a majority of variable sources are slightly harder; YSOs, LMXBs, and some candidate COs are closer to the middle; other candidate COs are harder on both scales; and AGNs appear as the hardest class.

The larger HRs for classified AGNs are consistent with the expected high X-ray absorption of AGNs through the entire Galactic plane, as well as their intrinsically harder spectra compared to stars. Note that the uncertainties \textbf{on} HRs (not shown in the figure to reduce clutter; see Fig. \ref{fig:ngc3532_mw_hrms_hrhm}) can be very large for fainter sources, and their actual location may be significantly different than the observed location.

\begin{figure*}
    \begin{interactive}{js}{plots/NGC3532_flux_G_flux_aper90_ave_b_classification_rgeo_lum.html}
    \centering
    \includegraphics[width=\textwidth]{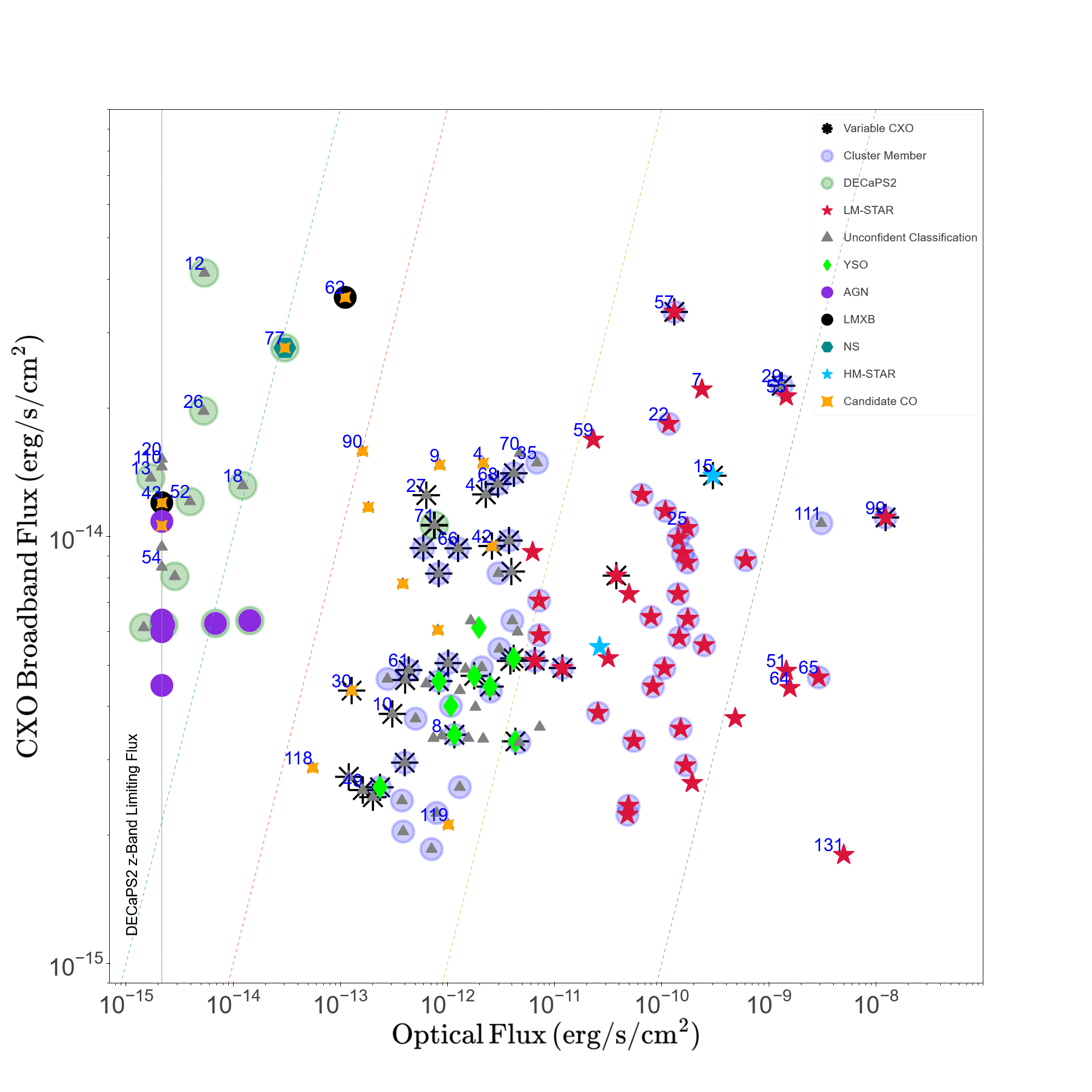}
    \end{interactive}
    \caption{X-ray and optical fluxes for CXO sources in the field of NGC 3532. X-ray source classifications labeled according to legend. Lines of constant X-ray to optical flux ratios are shown. CXO sources without optical counterparts are shown to the left, on a line corresponding to DECaPS2 $z=21.7$ \citep[photometric depth at which 50\% sources are recovered;][]{saydjari_dark_2022}. Sources discussed in Section \ref{detailed} are labeled. This figure is available online as an interactive figure, with the ability to zoom, pan, and display detailed information for each source. }
    \label{fig:NGC3532_flux_G_flux_aper90_ave_b_classification_rgeo_lum}
\end{figure*}

Figure \ref{fig:NGC3532_flux_G_flux_aper90_ave_b_classification_rgeo_lum} shows a diagram of X-ray versus optical fluxes with classification results. \textbf{Optical fluxes are calculated with}

\begin{equation}
    F_G = \Delta \nu \mathrm{ZP}_{\nu} 10^{M_G/2.5}
\end{equation}

\textbf{where $\Delta \nu$ is the frequency range, and $\mathrm{ZP}_\nu$ is the zero point of the G-band.}.\footnote{\textbf{Values taken from \url{http://svo2.cab.inta-csic.es/svo/theory/fps3/index.php?mode=browse&gname=GAIA&asttype=}}} CXO sources lacking an optical \textbf{counterpart} are shown on a line corresponding to DECaPS2 $z=21.7$ \citep[photometric depth at which 50\% sources are recovered;][]{saydjari_dark_2022}. Confidently classified LM-STARs are seen to the right of the $(F_{\rm X}/F_{\rm O})=10^{-3}$ line, while unconfidently classified variable X-ray sources, as well as candidate COs, are relatively brighter in X-rays and located to the left of this line. 

\begin{figure*}
    \begin{interactive}{js}{plots/NGC3532_lum_G_lum_aper90_ave_b_classification_rgeo_lum.html}
    \centering
    \includegraphics[width=\textwidth]{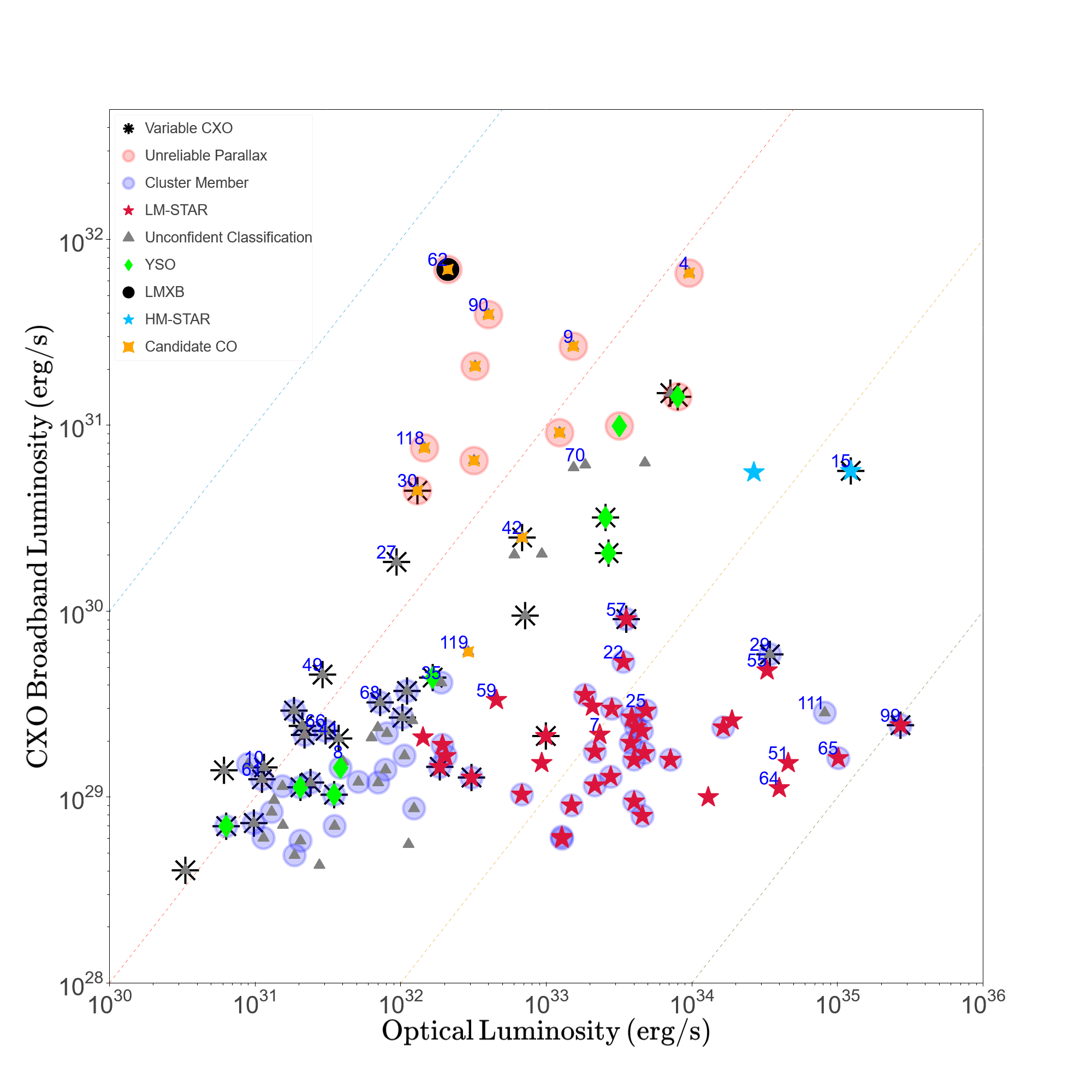}
    \end{interactive}
    \caption{X-ray and optical luminosities for CXO sources with Gaia counterparts. Arrows extending from mean luminosity to flare luminosity for flaring sources are shown. Lines of constant X-ray to optical luminosity ratios are shown. Sources discussed in Section \ref{detailed} are labeled. This figure is available online as an interactive figure, with the ability to zoom, pan, and display detailed information for each source. }
    \label{fig:NGC3532_lum_G_lum_aper90_ave_b_classification_rgeo_lum}
\end{figure*}

We also plot X-ray versus optical luminosities in Figure \ref{fig:NGC3532_lum_G_lum_aper90_ave_b_classification_rgeo_lum}. For elucidation, all available Gaia distances are used, but sources with $\pi/\sigma_\pi < 4$ (stricter than the cut used for ML classification) are marked as having unreliable parallaxes. For sources showing flares in their lightcurves, the flare luminosities are indicated by arrows pointing from the mean source luminosity to the flare luminosity (see Section \ref{detailed} for details). This plot confirms that sources classified as YSOs, HM-STARs, and candidate COs are more luminous in the X-ray compared to LM-STARs. 

The majority of variable sources have fairly low mean X-ray luminosities, as well as low optical luminosities consistent with M-dwarfs. As we discuss in Section \ref{detailed}, most of these are likely coronally flaring cluster LM-STARs.

\subsection{X-ray Sources without Counterparts}
\label{counterparts}

Since a lack of MW counterparts may be an indication of an unusual (non-stellar) nature of X-ray emission, we compiled the 7 CXO sources without Gaia, 2MASS, WISE, and DECaPS2 counterparts in 
Table \ref{table:Chandra only sources}.

The X-ray fluxes of these sources span from $\SI{4.5e-15}{}$ to $\SI{1e-14}{\erg \per \s \per \square \cm}$, similar to sources with counterparts. The X-ray to optical flux ratio limit for these sources ranges from 0.15 to 0.4, while most X-ray sources with MW counterparts are significantly brighter in the optical than in the X-ray. These source cluster on the hard-hard region in Figure \ref{fig:ngc3532_mw_hrms_hrhm}, and some of them are confidently classified as AGNs.

\begin{deluxetable*}{clllllllll}

\tablehead{ 
\colhead{Source} & \colhead{2CXO Name} & \colhead{Det. Signif.} & \colhead{Class} & \colhead{$P_\mathrm{Class}$} & \colhead{Can. CO} & \colhead{$F_b$} & \colhead{HR$_{MS}$} & \colhead{HR$_{HM}$} & \colhead{$P_\mathrm{var}$
}
}

\startdata
54&J110453.3-584900&7.8&NS?&$0.46\pm0.27$&N&$0.85\pm0.13$&$0.55\pm0.16$&$0.71\pm0.07$&0.52
\\
53&J110434.8-584908&6.8&AGN&$0.93\pm0.09$&N&$0.60\pm0.12$&$0.84\pm0.12$&$0.66\pm0.10$&0.5
\\
60&J110525.5-584727&6.4&AGN?&$0.68\pm0.18$&N&$0.95\pm0.16$&$0.72\pm0.26$&$0.90\pm0.04$&0.52
\\
36&J110458.3-585053&6.4&AGN&$0.93\pm0.07$&N&$0.64\pm0.11$&$0.83\pm0.13$&$0.80\pm0.06$&0.013
\\
17&J110538.0-585419&6.1&AGN&$0.80\pm0.14$&N&$1.08\pm0.18$&$0.85\pm0.12$&$0.86\pm0.05$&0.39
\\
23&J110526.1-584225&5.8&LMXB?&$0.56\pm0.11$&Y&$1.06\pm0.21$&$-0.33\pm0.16$&$0.63\pm0.11$& 1
\\
92&J110445.4-584807& 5&AGN&$0.93\pm0.14$&N&$0.45\pm0.10$&$0.70\pm0.27$&$0.83\pm0.08$&0.83
\\
\enddata

\label{table:Chandra only sources}
\caption{CXO sources without optical or NIR/IR counterparts. Columns include detection significance, most probable ML classification and its probability, candidate CO status in 5\textbf{(if CT$>2$ for CO class probability, see Equation \ref{eq:CT})}, broadband (0.5-7 keV) flux in units of $\SI{1e-14}{\erg\per\s\per\square\cm}$, hardness ratios, and variability probability. Unconfident classifications (as determined by Eq. \ref{eq:CT}) are marked with "?". 
}

\end{deluxetable*}

\section{Detailed Analysis of Selected Sources}
\label{detailed}

Beyond summarizing the bulk properties of the X-ray sources in the field of NGC 3532 above, we perform a more detailed analysis of these sources to draw further conclusions about X-ray source populations in and beyond the cluster, and to check the accuracy of our ML classifications. 

Spectra for 107 CXO sources with more than 50 net counts and S/N$>5$ in CSC2 were extracted using the \texttt{wavdetect} and \texttt{specextract} functions in CIAO tools version 4.14, and fitted using the \textsl{Sherpa} package \citep{2006SPIE.6270E..1VF}. Spectra for two additional sources (\# 118, 119), with slightly lower number of counts were also extracted because of their classifications as candidate COs. The extracted spectra were fit with the thermal plasma emission model ({\tt mekal}) and the powerlaw {\tt xspowerlaw} (PL) models modified by the interstellar photoelectric absorption according to {\tt xsphabs} (phabs) model \citep{wilms_absorption_2000}. For sources that were not well fit by either model, we attempted fits with a two-component thermal plasma ({\tt mekal}) model. We also tried fits with a blackbody model \texttt{bbodyrad}, but it did not fit any source significantly better than other models. The \texttt{wstat} statistic was used in all of the fits performed with {\sl Sherpa}.\footnote{see https://cxc.cfa.harvard.edu/sherpa/}

Additionally, we extracted lightcurves for the same sources using the \texttt{dmextract} function in CIAO tools, with 500 s bins. To extract flare spectra for flaring sources, we determine the flare time interval from their lightcurves with the following procedure: The lightcurve is split into 50 bins. The starting point of the flare is set at the bin with $4.5\sigma$ probability that it did not have counts above the median count rate by chance, and the ending point of the flare is set at the next bin where this probability drops below 99\% \textbf{($2.56\sigma$)}. 

We discuss some \textbf{groups of} sources below, selected based on their X-ray brightness ($>100$ net counts), presence of flares in their lightcurves, their ML classifications as a candidate CO, or if the optical counterparts are higher mass stars (A or earlier type; see Figure \ref{fig:ngc3532_mw_bp-rp_gmag}). These sources are categorized into \textbf{lower mass} cluster members, higher mass cluster members, background sources in the Galaxy, hard background sources \textbf{with counterparts}, sources with only DECaPS counterparts, and sources without any counterparts. \textbf{The most interesting sources of each group are presented here, while additional sources are presented in Section \ref{detailed_appendix} of the Appendix.} For convenience, variable sources are labeled with an asterisk next to the source number. The properties of these sources, including classification results and best-fit spectral model parameters are shown in Table \ref{table:detailed}. The full table of all CXO sources detected with S/N$>5$ is available electronically. 

Potential binary sources were identified using Gaia's Renormalised Unit Weight Error (\texttt{RUWE}) parameter, which measures goodness of fit of the astrometric data to a single star model. A value significantly greater than 1 (around 1.4) indicates binarity, or potential problems with the astrometric solution \citep{brown_gaia_2021}. Since NGC 3532 has a well-defined binary sequence visible above the solitary star main sequence in Figure \ref{fig:ngc3532_mw_bp-rp_gmag}, an offset from the main sequence can also indicate binarity.

\subsection{Cluster Lower Mass Stars}

Sources 7, 10$^{\ast}$, 22$^{\ast}$, 25, 35, 57$^{\ast}$, 61$^{\ast}$, 66$^{\ast}$, and 68$^{\ast}$ have Gaia counterparts that are low-mass members of NGC 3532. Their spectra and lightcurves are shown in Figures \ref{fig:NGC3532_cluster_members1} and \ref{fig:NGC3532_cluster_members2}. 

During the CXO observation, the average luminosities of these sources range from $\SIrange{1e29}{9e29}{\erg\per\s}$, with the most luminous source being source 57$^{\ast}$. Their X-ray to optical flux ratios range from $10^{-4}$ to $10^{-2}$. Several sources are variable, and three display flares. Source 22$^{\ast}$ is borderline variable by Kuiper's statistics (variability probability 0.987), but visibly shows a minor flare. The X-ray spectra of all these sources are soft or relatively soft (with $-0.8<$HR$_{\rm ms}<-0.3$ and $-0.6<$HR$_{\rm hm}<0.4$). Most can be fitted with an absorbed PL with $\Gamma\approx 2.4-3.7$ or {\tt mekal} with $kT=0.4-1.0$ keV. Sources 35, 57$^{\ast}$, \textbf{66$^{\ast}$}, and 68$^{\ast}$ are not well fit by either simple model, while a two-temperature {\tt mekal} model fits well with $kT_1=0.2-0.4$ keV and  $kT_2=1.2-2.5$ keV. The lightcurves of sources 10$^{\ast}$ and 61$^{\ast}$ show flares with a sharp-rise and slow-decay profile typical for stellar (coronal) flares \citep{pye_survey_2015}. The profiles of the flares of sources 22$^{\ast}$ and 66$^{\ast}$ appear more symmetric, possibly due to noisier data. 

The optical colors of these sources are consistent with being low-mass stars on the cluster's main sequence or the binary track right above it. Source 22$^{\ast}$ has Gaia eDR3 \texttt{RUWE} of 1.3, possibly indicating binarity, which is consistent with its location on the binary track. Sources 7, 25, 35, 57$^{\ast}$, and \textbf{66$^{\ast}$} are visibly above the main sequence in the binary track, but do not have high \texttt{RUWE} values. The two-temperature spectra of the latter three sources could be explained if they are systems of coronally active binary stars \citep{mcgale_rosat_1996}.

All of these sources are confidently classified as LM-STAR, or otherwise have high combined LM-STAR/YSO probabilities, consistent with their soft spectra, and probable coronal X-ray emission.

Based on the above analysis we conclude that the X-ray emission of most CXO sources matched to a cluster member have a coronal origin, although some of these sources may be active binaries rather than solitary stars.\footnote{We currently do not have an active binary class in our TD, so these systems may be classified as another class, such as YSOs.} We find that the ML classifications of these sources are mostly accurate (see main sequence on Figure \ref{fig:NGC3532_mw_BP-RP_gmag_classification}), \textbf{but we note that four K/M-type stars are classified as YSOs. (The other YSOs are not cluster members.)} 

The large number of unconfidently classified variable sources at the fainter end of the CMD (Figure \ref{fig:NGC3532_mw_BP-RP_gmag_classification}) correspond to the variable sources in the middle of the HR diagram (Figure \ref{fig:NGC3532_HRMS_HRHM_classification}), and the variable sources with low X-ray and optical luminosity in Figure \ref{fig:NGC3532_lum_G_lum_aper90_ave_b_classification_rgeo_lum}. Many of these sources are cluster members on the main sequence, and their multi-wavelength properties make them likely to be coronally active LM-STARs. 

\textbf{These unconfident classifications (which have high combined LM-STAR/YSO probabilities), as well as the four YSO classifications, are likely due to the large number of YSOs with properties similar to those of underrepresented K/M stars in the TD ($>1000$ YSOs, compared to $\sim 40$ K/M stars ). Additionally, the pre-main sequence stage of lower-mass stars ($>0.5 M_\sun$) can last tens or hundreds of Myrs, during large portions of which they evolve slowly through the optical and infrared feature spaces close to the main sequence \citep{amard_first_2019}. Therefore, at the cluster age of 300 Myr, some M-type stars may still be in their pre-main sequence stage, while other LM-STARs may be easily confused for YSOs}

\begin{figure*}[hbt!]
    \centering
    \includegraphics[width=\textwidth]{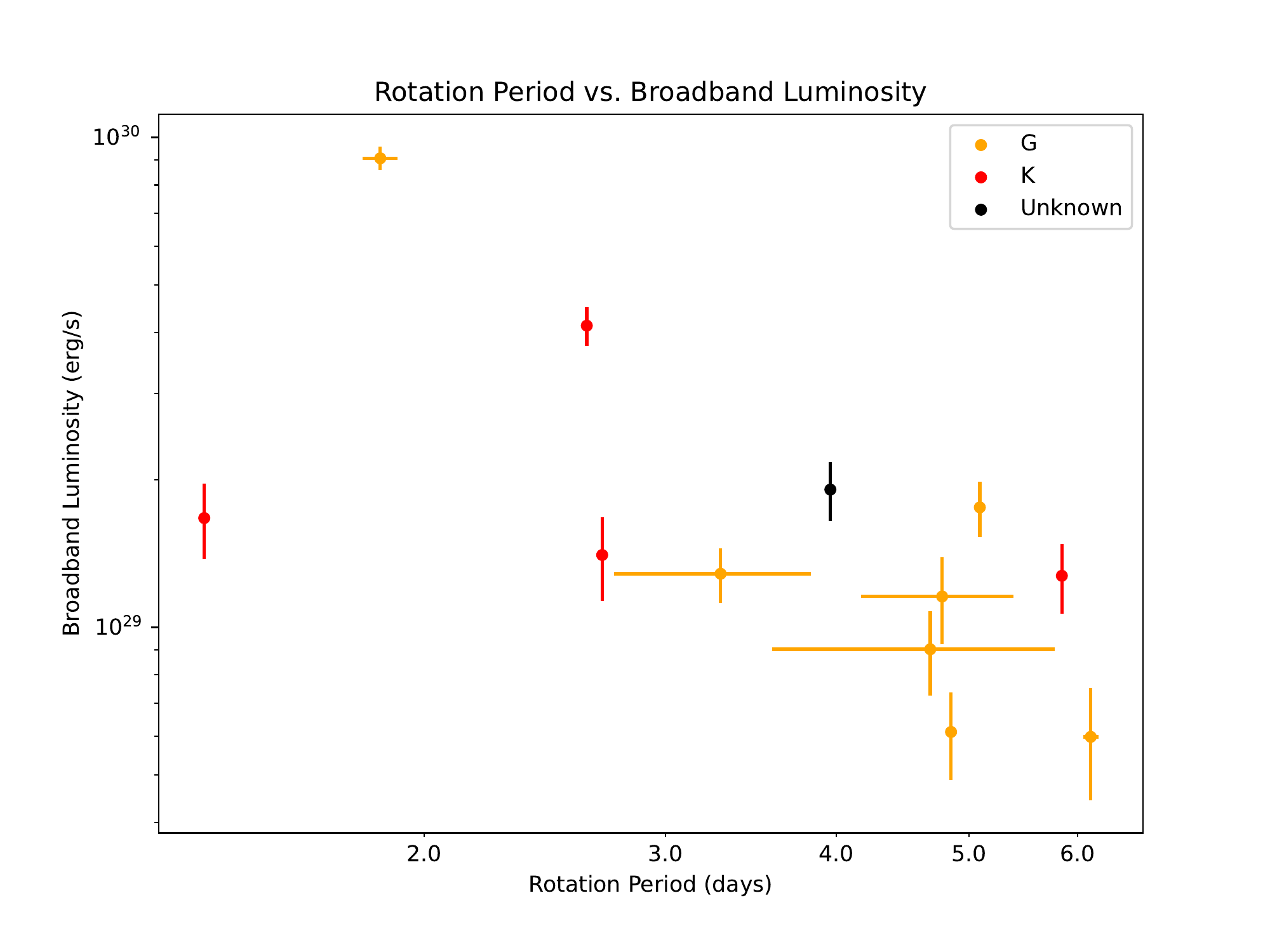}
    \caption{Rotation period of cluster stars from \cite{fritzewski_rotation_2021} vs. CXO broadband luminosity for crossmatched sources. Colormap shows Gaia DR3 spectral types \citep{fouesneau_gaia_2022}.}
    \label{fig:NGC3532_rotation_lum}
\end{figure*}

\begin{figure*}
    \centering
    \includegraphics[width=0.37\textwidth,align=t]{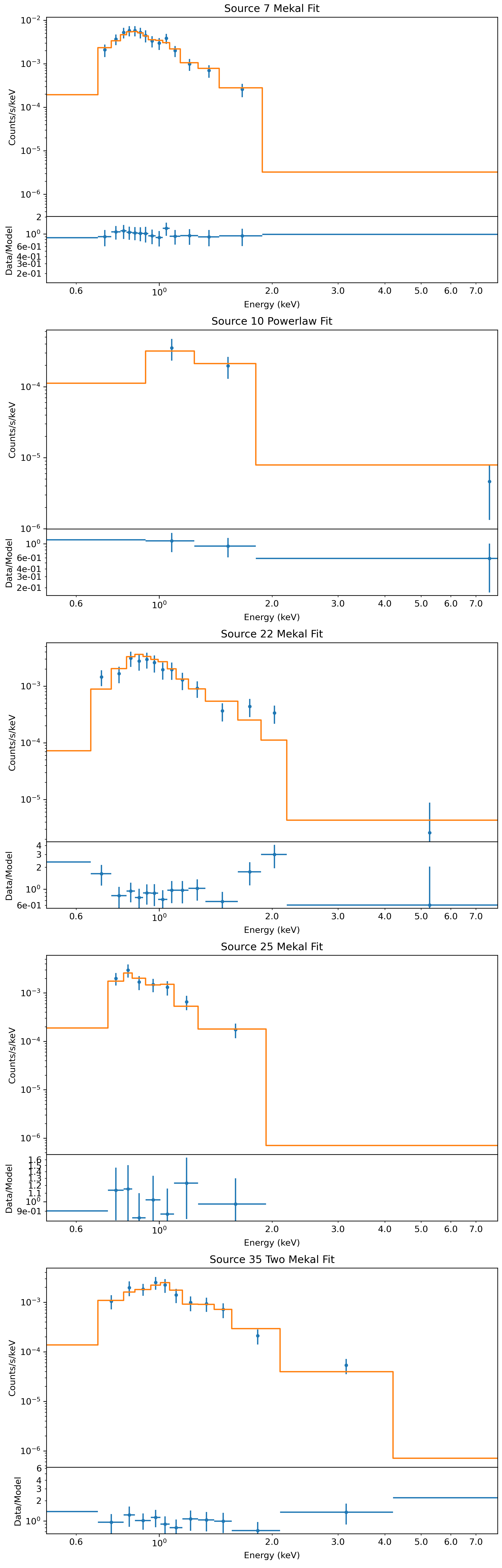}
    \includegraphics[width=0.37\textwidth,align=t]{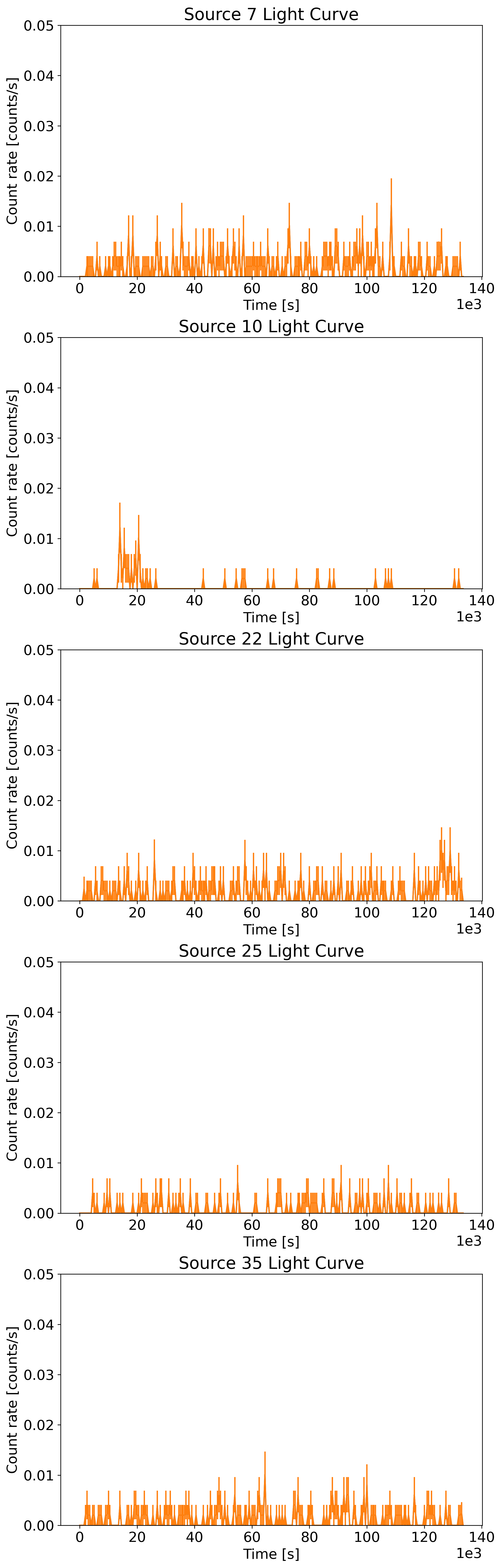}
    \caption{Spectra and lightcurves for selected cluster CXO sources. \textbf{Spectral model} fitted to each source shown in plot title. }
    \label{fig:NGC3532_cluster_members1}
\end{figure*}

\begin{figure*}
    \centering
    \includegraphics[width=0.37\textwidth,align=t]{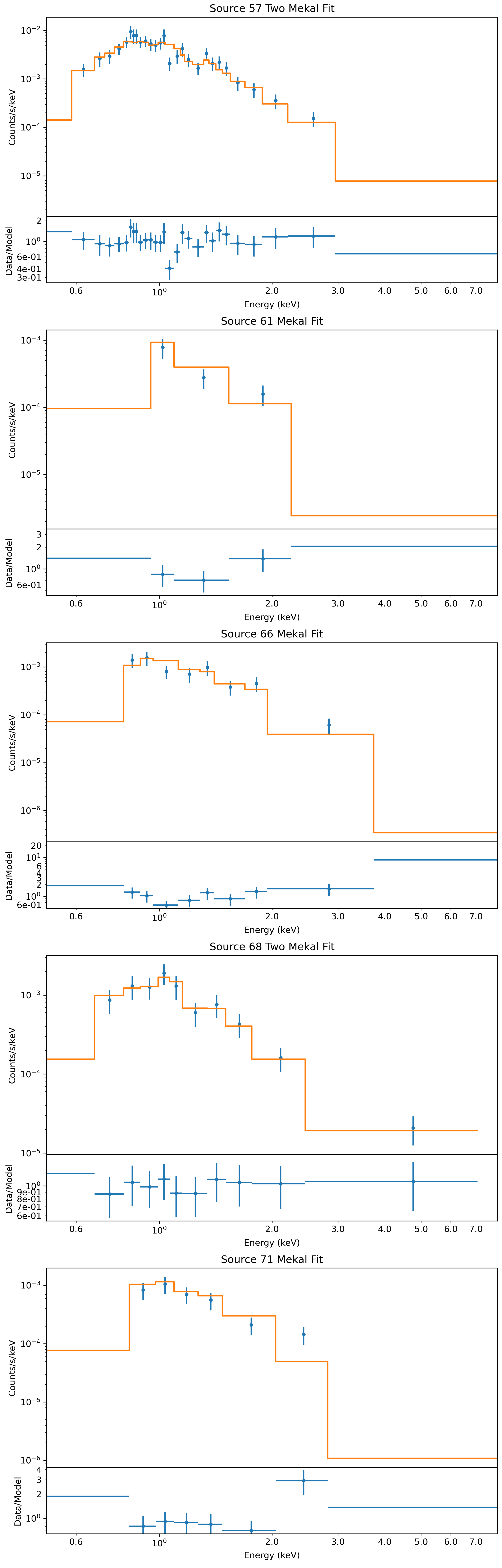}
    \includegraphics[width=0.37\textwidth,align=t]{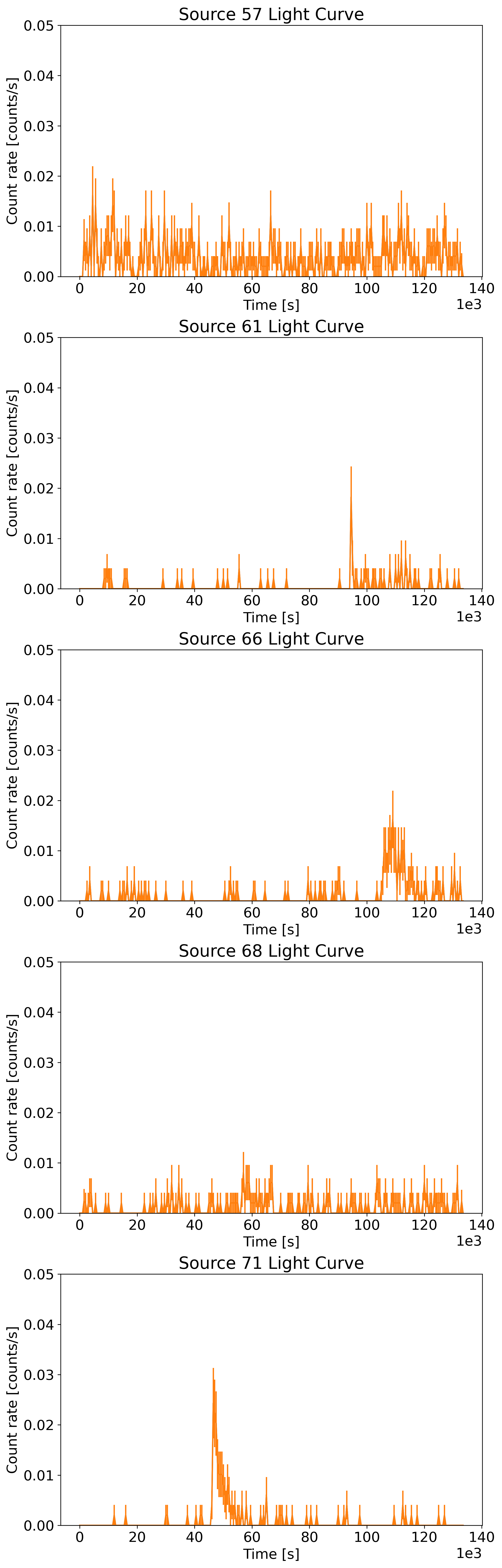}
    \caption{Spectra and lightcurves for selected cluster CXO sources. }
    \label{fig:NGC3532_cluster_members2}
\end{figure*}

The coronal activity of low-mass stars is known to be correlated with the star's rotation rate \citep{pizzocaro_activity_2019,notsu_kepler_2019,fritzewski_rotation_2021}. We crossmatched CXO sources to stars with rotation periods derived in \cite{fritzewski_rotation_2021}. An X-ray luminosity vs. rotation period plot is shown in Fig. \ref{fig:NGC3532_rotation_lum}. As expected, there is an inverse correlation between the stellar rotation period and X-ray luminosity. However, it shows substantial scatter (which is also seen in  Fig.~11 of \citealt{pizzocaro_activity_2019}) suggesting that factors other than rotation period, such as the presence of a close companion, may be important. Somewhat surprisingly, only two of these sources are  variable in X-rays, and none exhibit significant flares. This may be because the more frequently flaring stars tend to be less massive, and therefore fainter, and less likely to have their rotation periods measured.

\subsection{Cluster A-Type and B-Type Stars}

\begin{figure*}[hbt!]
    \centering
    \includegraphics[width=0.37\textwidth,align=t]{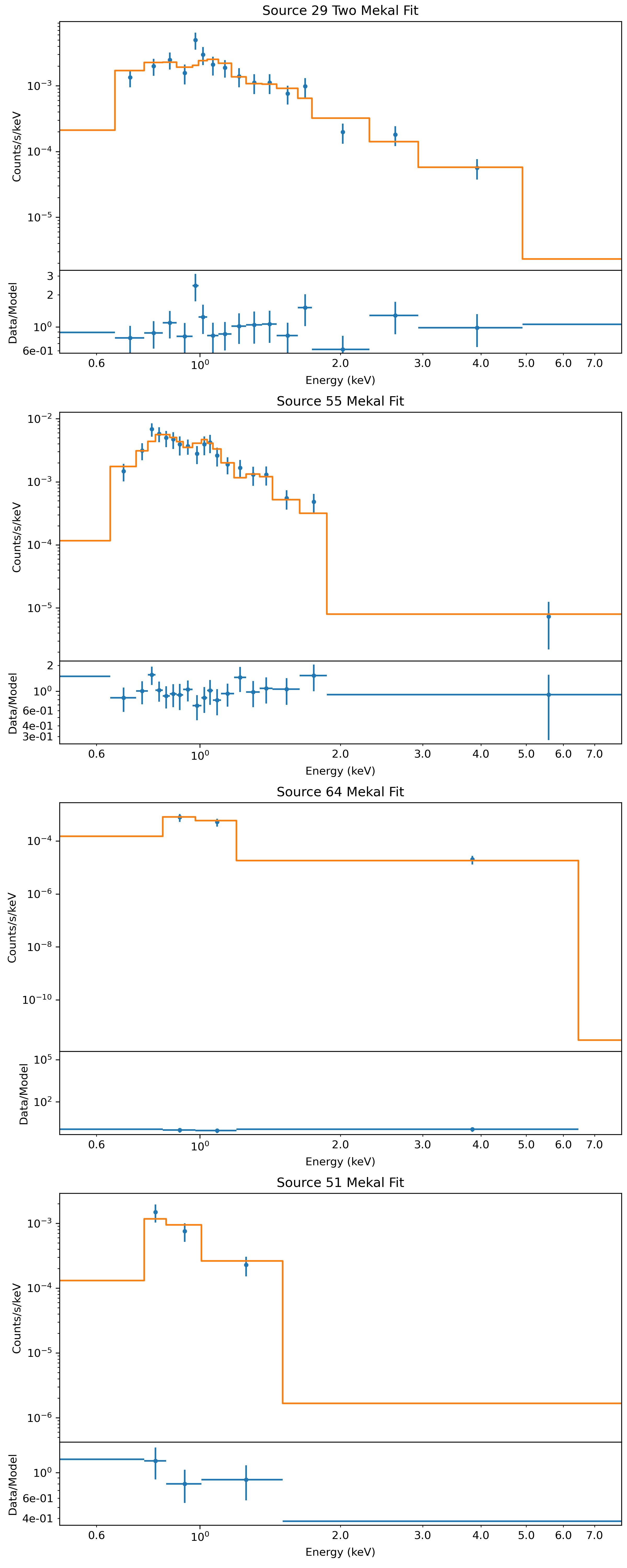}
    \includegraphics[width=0.37\textwidth,align=t]{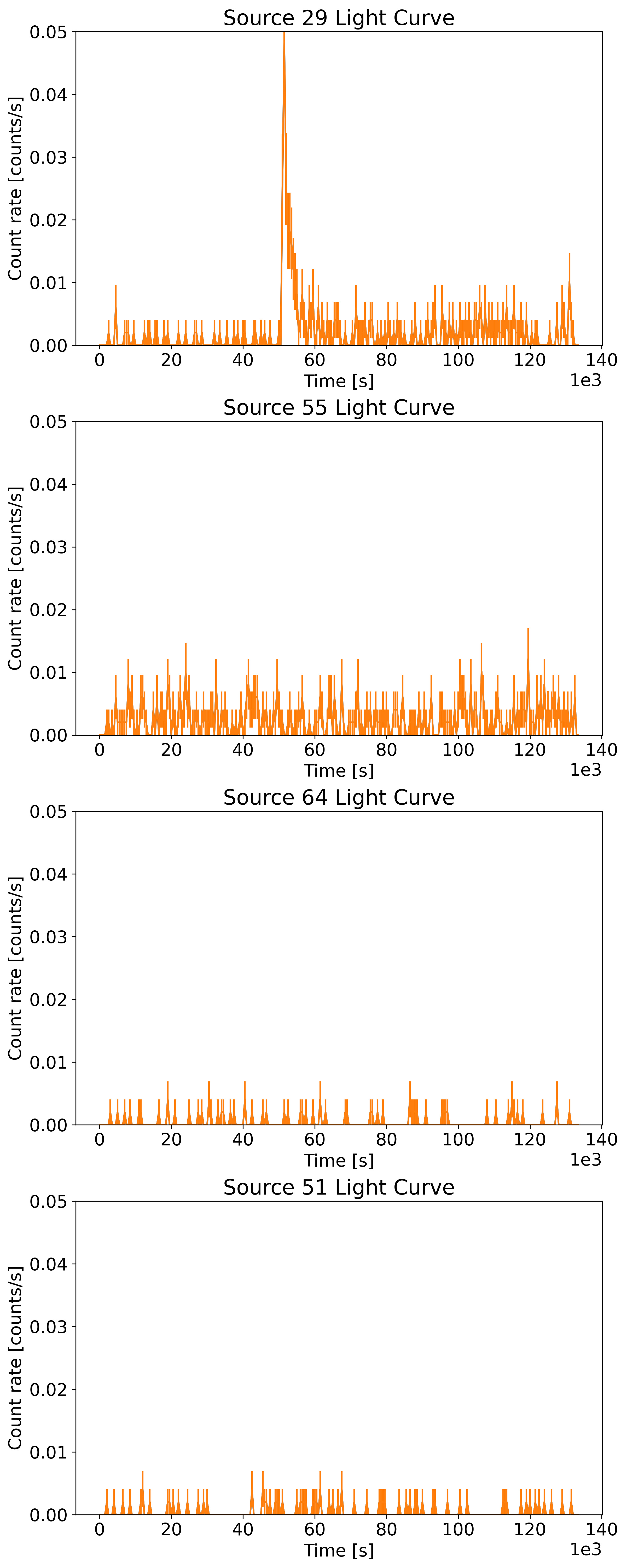}
    \caption{Spectra and lightcurves for CXO sources matched to cluster A-type stars. }
    \label{fig:ngc3532_A_stars1}
\end{figure*}

\begin{figure*}[hbt!]
    \centering
    \includegraphics[width=0.37\textwidth,align=t]{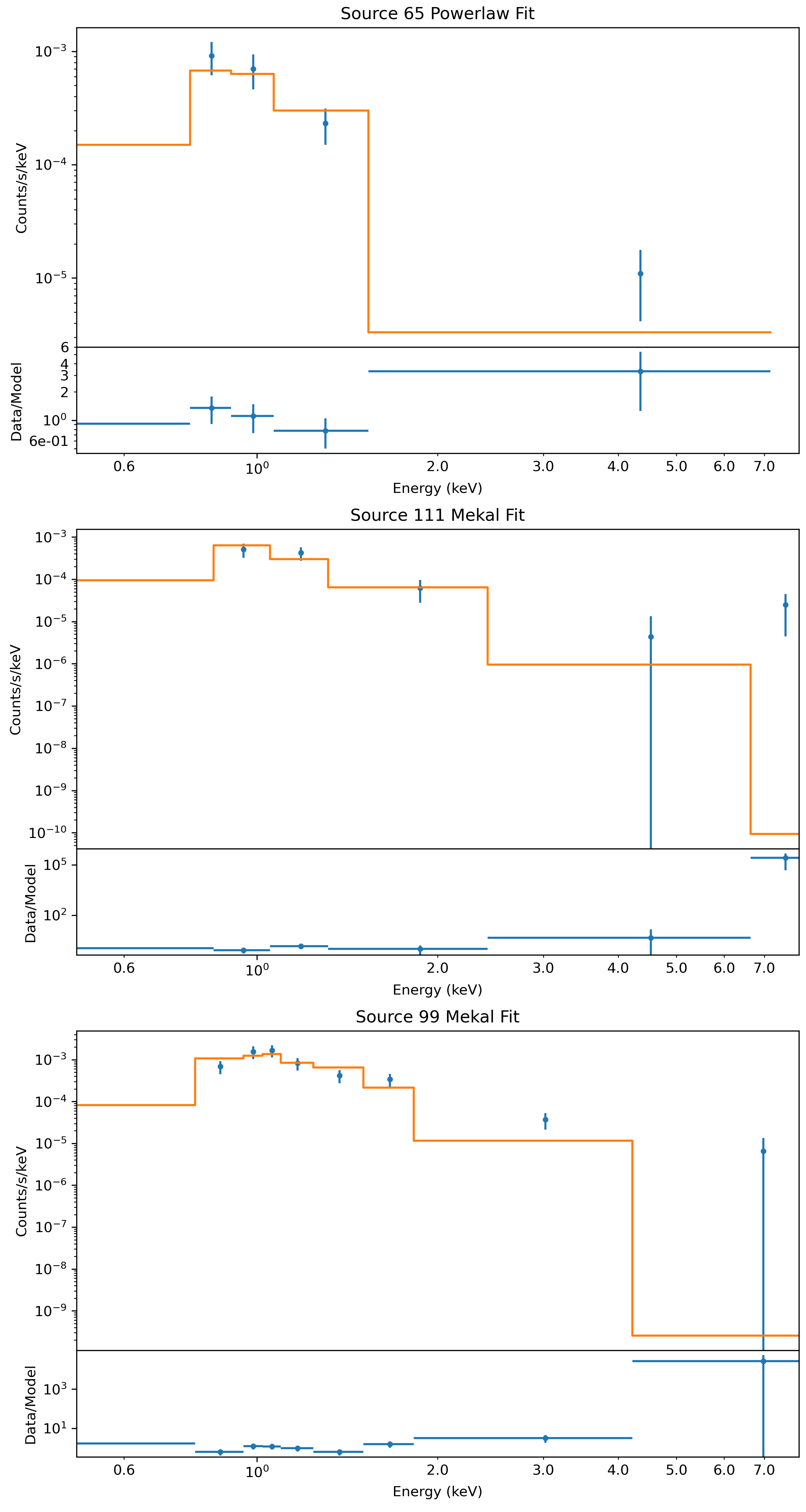}
    \includegraphics[width=0.37\textwidth,align=t]{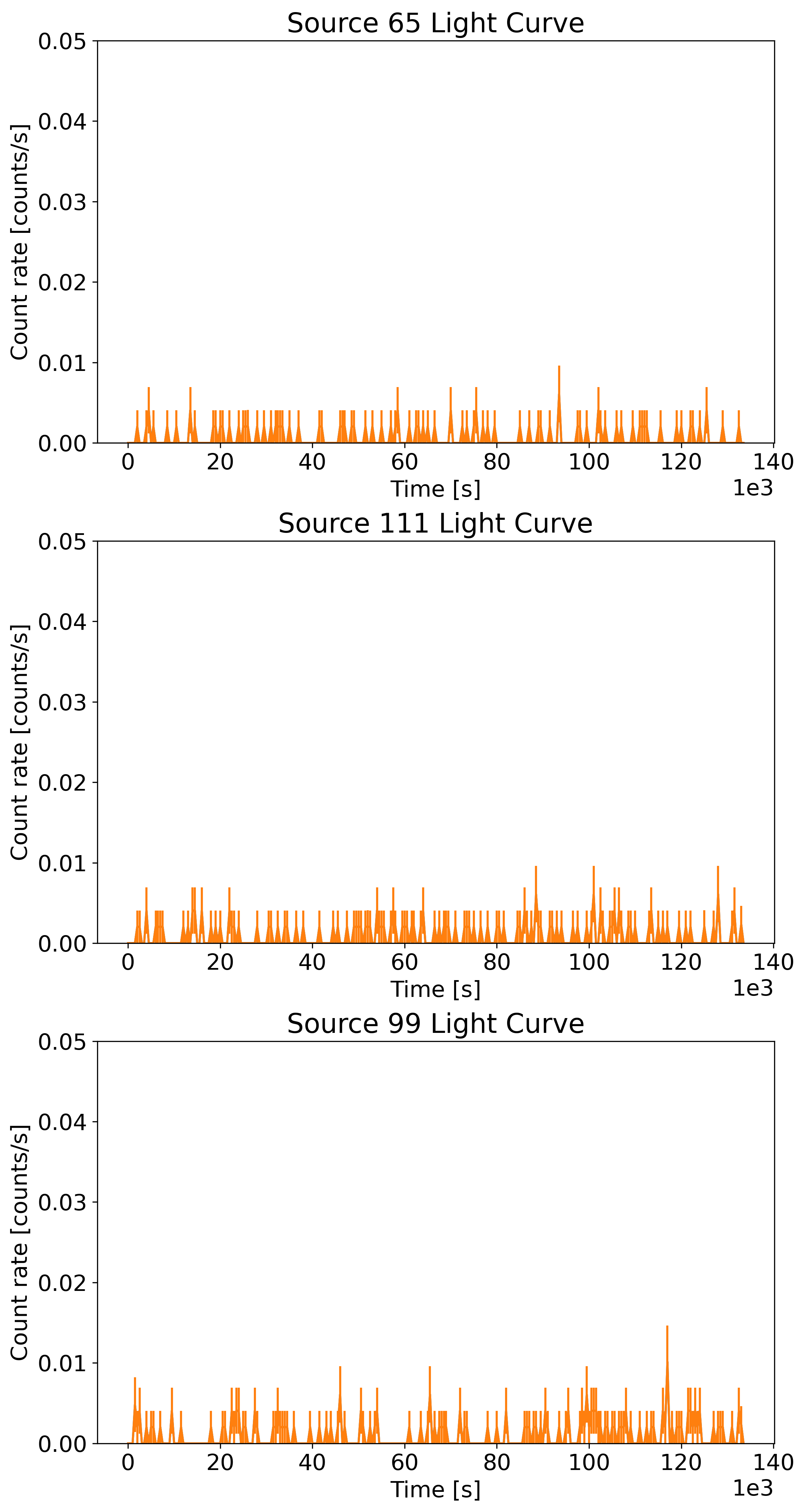}
    \caption{Spectra and lightcurves for CXO sources matched to cluster B-type stars. }
    \label{fig:ngc3532_A_stars2}
\end{figure*}

Several A-type (Sources 29$^{\ast}$, 55, 64) and B-type (Sources 51, 65, 111, and 131) stars belonging to the cluster are also coincident with X-ray sources. Their spectra and lightcurves are shown in Figures \ref{fig:ngc3532_A_stars1} and \ref{fig:ngc3532_A_stars2}. 

Sources 51, 65, 111, and 131 (identified as \object{HD 96192} - A, \object{CPD-58 3069} - A1V, \object{V* GV Car} - A0, \object{HD 96246} - A0V, respectively) have similarly high \texttt{RUWE} values, positions above the solitary star track of the main sequence, non-variability, and X-ray luminosities $\sim\SI{e29}{\erg\per\s}$. Source 131 has too few counts to extract a spectrum, while the other sources have soft spectra with $kT\approx0.4$ to 0.5 keV. Their literature A-type classifications conflict somewhat with the Gaia DR3 classifications as B-type stars. Source 51, in particular, appears slightly lower than sources 55 and 64 on the main sequence. Isochrone fitting suggests their masses to be between 2-3 $M_\sun$, broadly consistent with late-B or early-A classes.  
Source 111 is not confidently classified as a LM-STAR by MUWCLASS, because its X-ray spectrum shows a hard excess (above 6 keV) in its otherwise typical stellar spectrum. Given the \texttt{RUWE} value of 1.3, it's possible that interactions with a companion star is responsible for the hard excess. The nature of the companion could be constrained by a radial velocity study.

Source 29$^{\ast}$ (\object{HD 96157}) is identified as an A0 star in SIMBAD. It is strongly variable, exhibiting the largest flare among all CXO sources detected in NGC 3532, with a sharp rise, slow decay, and a duration of $\sim5$ ks. The average flare luminosity is $\SI{4.1e30}{\erg\per\s}$, a factor of $\sim10$ larger than the average quiescent luminosity of the source. The average spectrum can be described by {\tt mekal} with $kT\simeq1.4$ keV, but shows a soft excess that's better described by a two-temperature {\tt mekal} model with $kT_1=0.37$ keV and $kT_1=2.5$ keV. The \texttt{RUWE} of 0.84 does not indicate binarity, but it has a slightly elevated position on the solitary star track of the main sequence. The source is classified as 60\% LM-STAR, and 33\% as HM-STAR. (Note, that in our TD HM-STAR class consists of OB type stars and WR stars, which do not extend down into A-type stars.) 

It is commonly accepted that solitary A stars should be very faint in X-rays, since they have fairly small convective zones (compared to late type stars) and lack strong winds (compared to OB stars) \citep{gunther_coronal_2022}. Therefore, the detection of X-ray bright solitary A-type stars is unexpected. Since most of these sources are likely to be binaries, the detected X-ray emission may be attributed to a lower mass companion. However, there is \textit{only weak} evidence of binarity for the strongly flaring Source 29$^{\ast}$. Sensitive optical spectroscopy is needed to perform an additional search for a low-mass companion. If it is indeed a binary system, then the companion may be very low mass, which would be consistent with the strong flare.

\subsection{AGNs}
\label{AGN}

All seven confidently classified AGNs appear in the hard-hard region of the HR diagram (see Figure \ref{fig:NGC3532_HRMS_HRHM_classification}), have corresponding hard spectra ($\Gamma < 1.5$, except for source 49 with $\Gamma=3$), are non-variable, and have relatively few counts ($\sim 60$). Three of these sources have faint (magnitude $>20$) counterparts in the DECaPS2 survey, and none have any other counterparts. Based on these properties, we consider the AGN classifications to be reliable.

\subsection{Background Sources with Gaia Counterparts}
\label{background_softer}

\begin{figure*}[hbt!]
    \centering
    \includegraphics[width=0.37\textwidth,align=t]{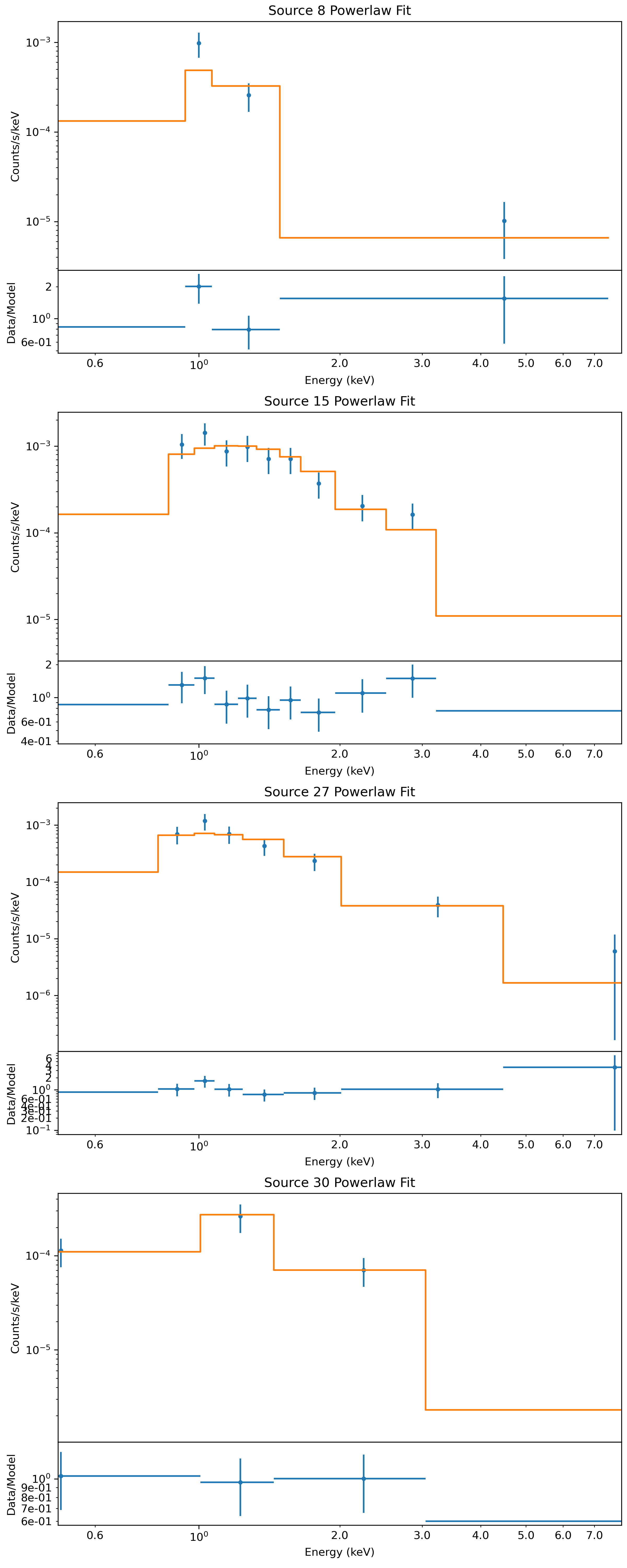}
    \includegraphics[width=0.37\textwidth,align=t]{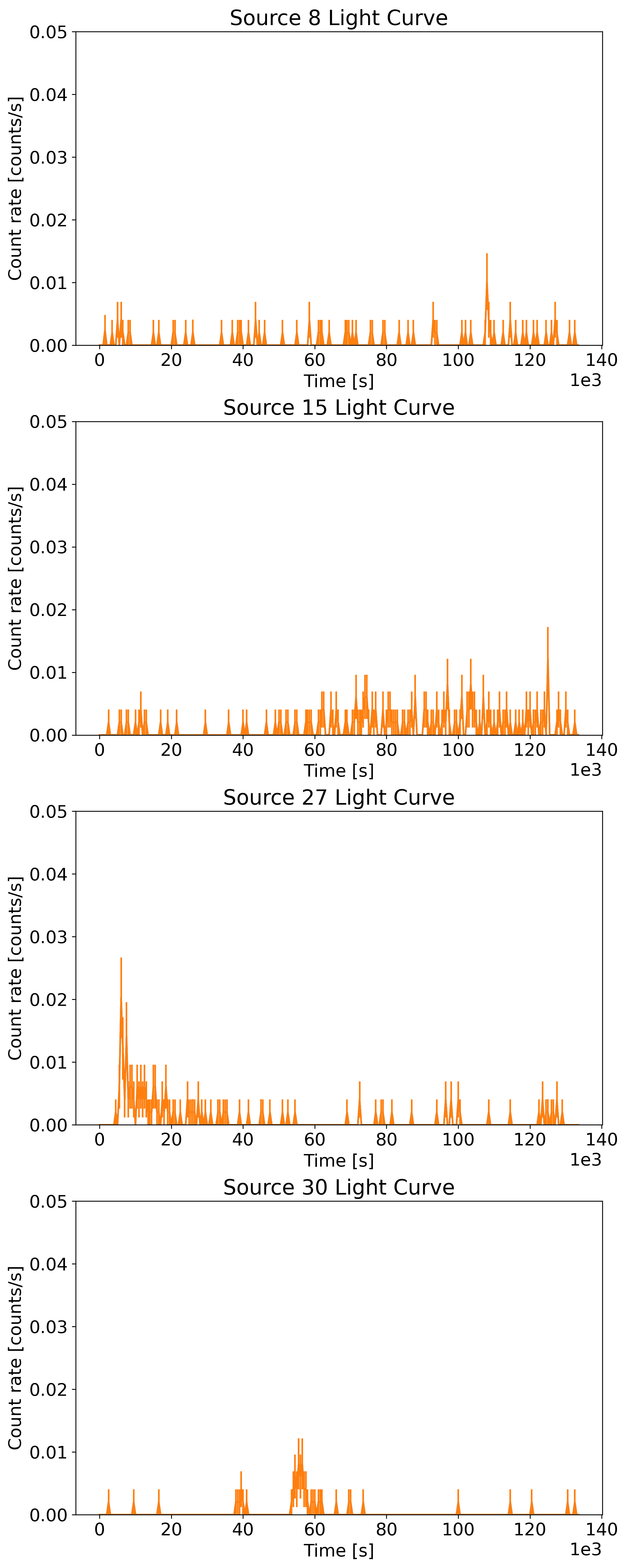}
    \caption{Spectra and lightcurves for CXO sources matched to cluster background sources. }
    \label{fig:ngc3532_background1}
\end{figure*}

\begin{figure*}[hbt!]
    \centering
    \includegraphics[width=0.37\textwidth,align=t]{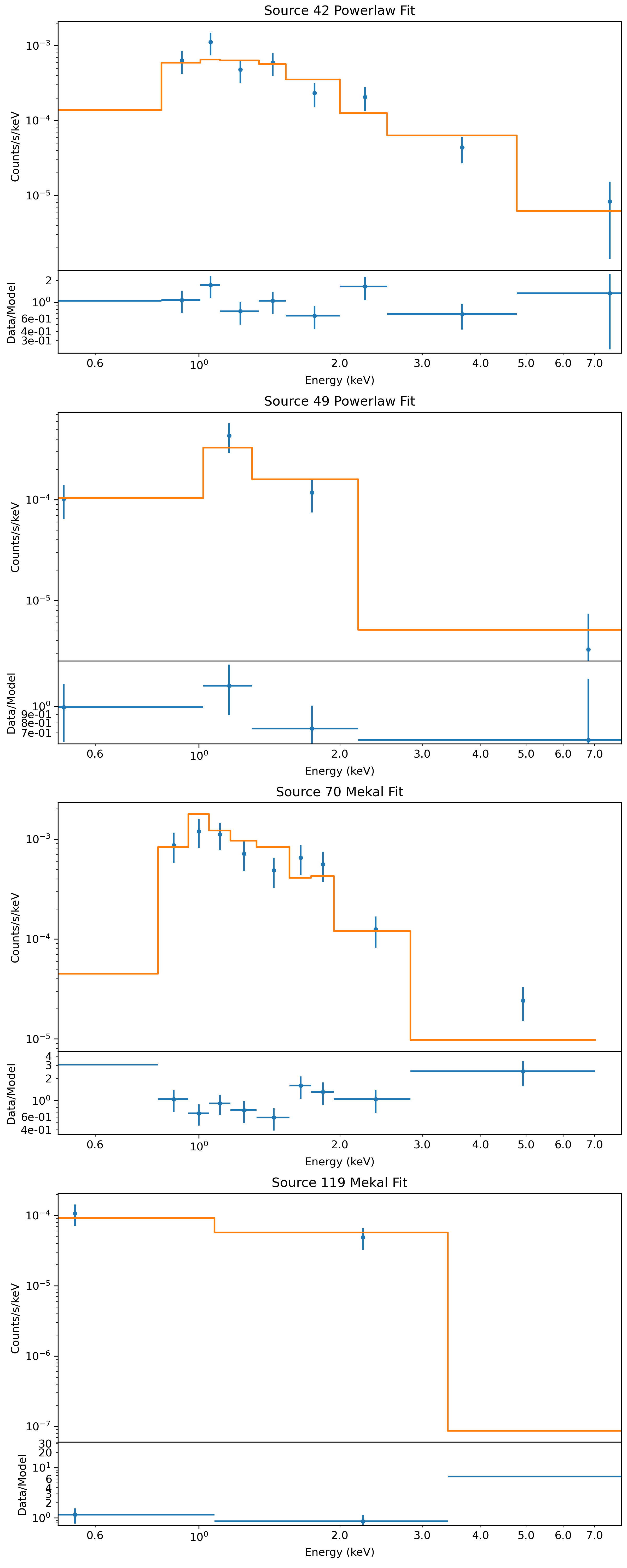}
    \includegraphics[width=0.37\textwidth,align=t]{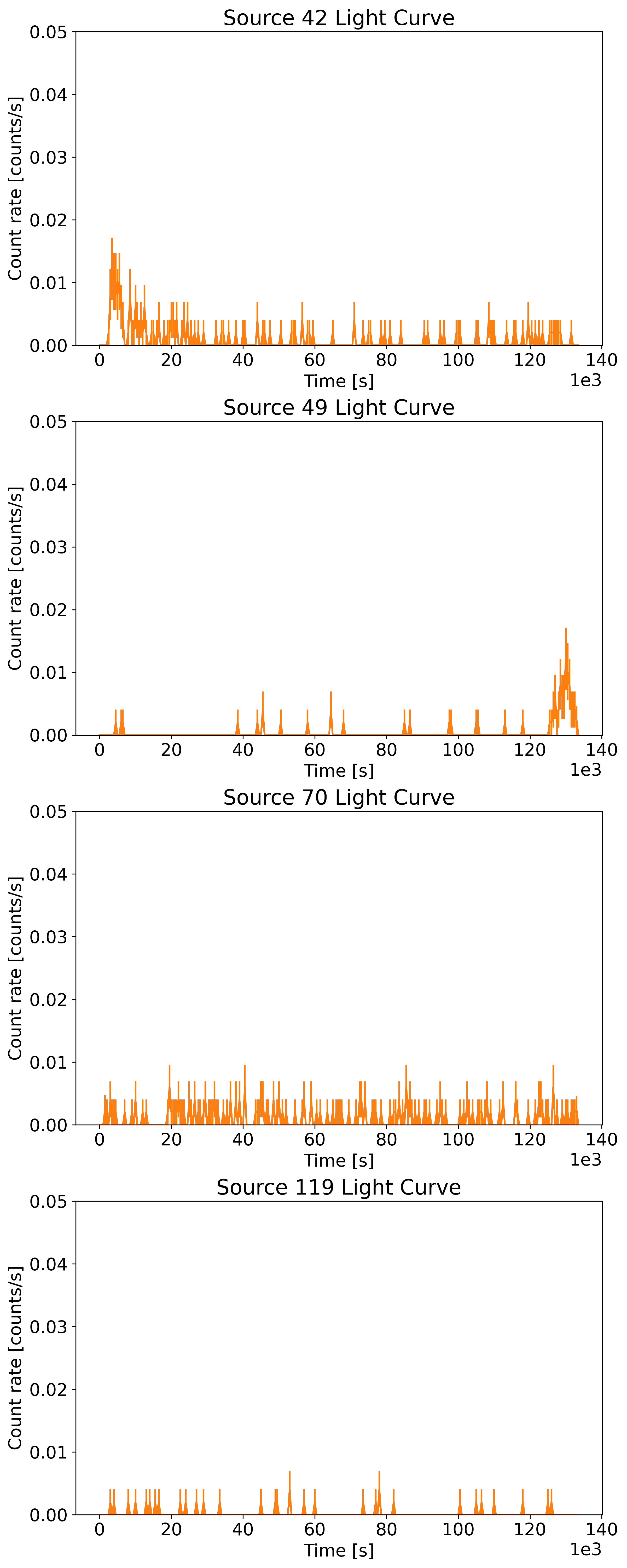}
    \caption{Spectra and lightcurves for CXO sources matched to cluster background sources. }
    \label{fig:ngc3532_background2}
\end{figure*}

Sources 8, 15, 27, 30, 42, 49, 70, and 119 have medium hardness ratios in Figure \ref{fig:NGC3532_HRMS_HRHM_classification}, and are bright enough for more detailed analysis. They have Gaia counterparts with distances beyond the cluster which are well-constrained, except for sources 30 and 49, which still have significant proper motions that exclude an extragalactic nature. Their spectra and lightcurves are shown in Figures \ref{fig:ngc3532_background1} and \ref{fig:ngc3532_background2}. 

Sources 30$^{\ast}$, 42$^{\ast}$, and 49$^{\ast}$ ($d\approx3$, 1.5, 1.2 kpc, respectively) show similar flares with symmetric profiles (unlike the sharp-rise slow-decay flares \textbf{common for LM-STARs} discussed above) and relatively hard spectra with $\mathrm{HR}_{ms}\approx 0.2$ and $\mathrm{HR}_{hm}=$ 0.3, 0.5, -0.1 respectively. Their spectra can be described by an absorbed PL model with $\Gamma=2.0-3.1$, and show some evidence of hardening during the flares. The X-ray flare luminosities for these sources are $\approx \SI{1e31}{\erg\per\s}$, while their quiescent emission is much fainter. The preferred ML classification for these sources is LMXB, but at fairly low confidence, with other possibilities being YSO or CV. Sources 30$^{\ast}$ and 42$^{\ast}$ are classified as candidate COs, which is supported by the atypical flare profiles and higher luminosities.

Source 70, located at $d\approx1.8$ kpc, is similar to these three sources in all respects (including the classifications) except that it does not exhibit a flare during the {\sl CXO} observation. Its highest classification probability is YSO at 57\%. 

\subsection{Hard Sources with MW Counterparts}
\label{background_hard}

\begin{figure*}[hbt!]
    \centering
    \includegraphics[width=0.37\textwidth,align=t]{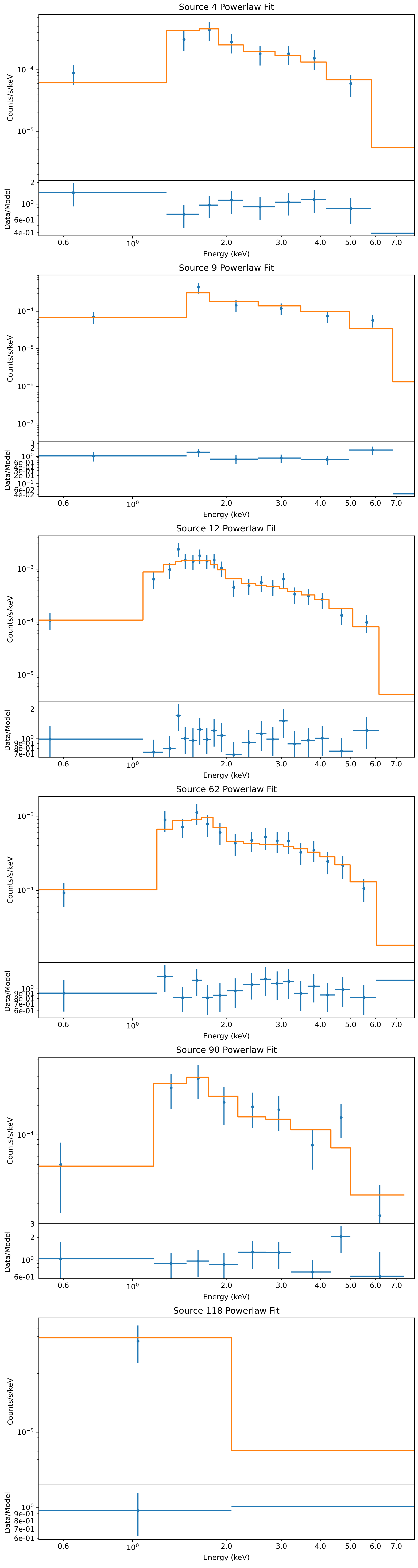}
    \includegraphics[width=0.37\textwidth,align=t]{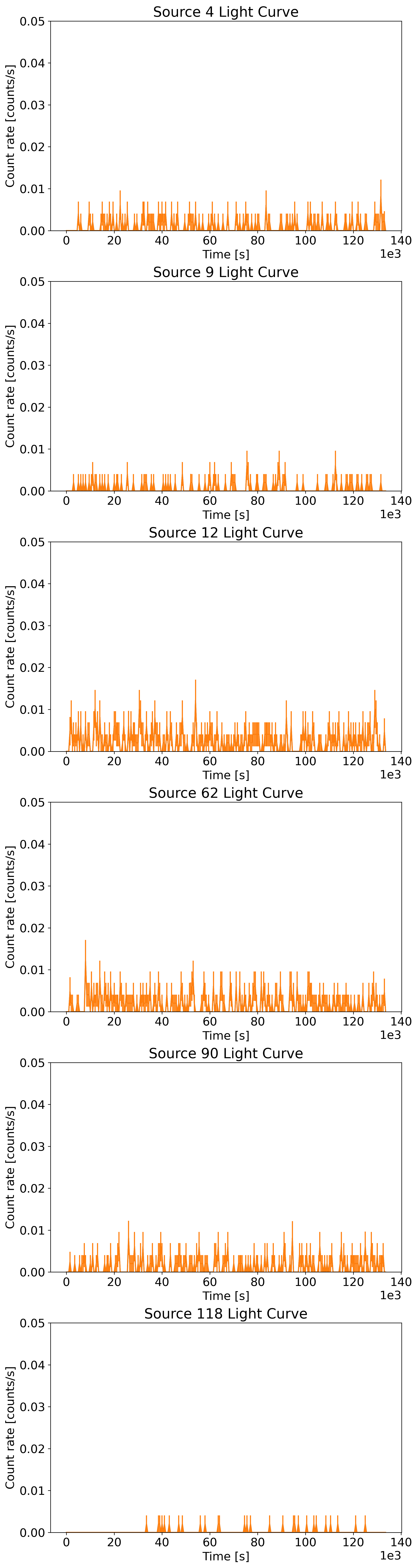}
    \caption{Spectra and lightcurves for hard CXO sources matched to background sources.}
    \label{fig:ngc3532_background_hard}
\end{figure*}

Sources 4, 9, 12, 62, and 90 have at least one multiwavelength counterpart in Gaia, 2MASS, or WISE surveys. Of these, sources 4, 12, and 90 are located at the edge of ACIS field of view, and thus have particularly large PUs that increases the chance coincidence probability. They appear in the hard-hard (upper right) corner on the HR diagram, being slightly softer than confidently classified AGNs (see Figure \ref{fig:NGC3532_HRMS_HRHM_classification}). Their spectra resemble those of AGNs (see Figure \ref{fig:ngc3532_background_hard}), and are well fit by both models, with PL photon indices $\Gamma\approx2.0,\ 1.6$ and {\tt mekal} $kT\approx 5.4,\ 6.5$ keV. The lightcurves are not variable. The distances (when present) of the Gaia counterparts have large uncertainties (in excess of $1,000$ pc), and most of the parallaxes do not pass the $\pi/\sigma_\pi >= 2$ cut that determines whether their distances are used in ML classification. However, these sources still have highly significant proper motions, and their BP-RP colors (when present) are bluer than the color of any AGN in the TD after applying extinction through the plane. These factors exclude an extragalactic origin. At their fiducial distances, the X-ray luminosities $\sim\SI{e30}-\SI{e31}{\erg\per\s}$ are at the high end for coronally active stars and at the low end for X-ray binaries. The \texttt{RUWE} values of $\sim 1$ do not indicate binarity, but this could be due to the large distances and optical faintness. The ML pipeline classifies some of these sources as candidate COs in the 5-class scheme, which is supported by the hard spectra and fairly high X-ray luminosities. 

Source 9 has a Gaia counterpart with a large \texttt{RUWE} value of 1.9, which suggests a background Galactic binary system. Its extremely large proper motion of $\SI{18.3\pm 0.2}{mas/yr}$ translates to a large tangential velocity of $\sim\SI{300}{\km\per\s}$ at its fiducial distance of $\SI{4\pm 2}{kpc}$, which after accounting for differential Galactic rotation, is still in excess of $\SI{100}{\km\per\s}$. Its hard spectrum ($\Gamma=1.6\pm0.5$), combined with its inferred large velocity and \texttt{RUWE} may indicate a binary system containing a non-accreting pulsar responsible for the hard emission \citep{jennings_binary_2018}.

Source 12 only has an UnWISE counterpart in the W2 band, and a faint DECaPS2 counterpart in the $i$ and $z$ bands ($>21$ mag). It has the highest X-ray flux among detected sources with $F_X = \SI{4.1e-14}{\erg\per\s\per\square\cm}$. The absorbed PL fit indicates $n_H=\SI{1 \pm 0.2 e22}{\per\square\cm}$ which is compatible with an extragalactic origin (based on the total $n_H\approx9\times10^{21}$ cm$^{-2}$ expected for $A_V\approx4$; \cite{guver_relation_2009}), unless the source is intrinsically obscured. This source also has the highest limiting flux ratio $L_X/L_O \gtrsim 1.5$ of all sources (see Figure \ref{fig:NGC3532_flux_G_flux_aper90_ave_b_classification_rgeo_lum}). 

Source 62 has a highly significant proper motion ($\SI{7.5\pm 0.6}{mas/yr}$) which implies a Galactic nature. Given its faintness in the optical/NIR, and the very high X-ray to optical flux ratio (see Figure \ref{fig:NGC3532_flux_G_flux_aper90_ave_b_classification_rgeo_lum}) it could be an LMXB, in agreement with its ML classification.

\subsection{Sources with Only DeCAPS2 Counterparts}
\label{DeCAPS2_only}

\begin{figure*}[hbt!]
    \centering
    \includegraphics[width=0.37\textwidth,align=t]{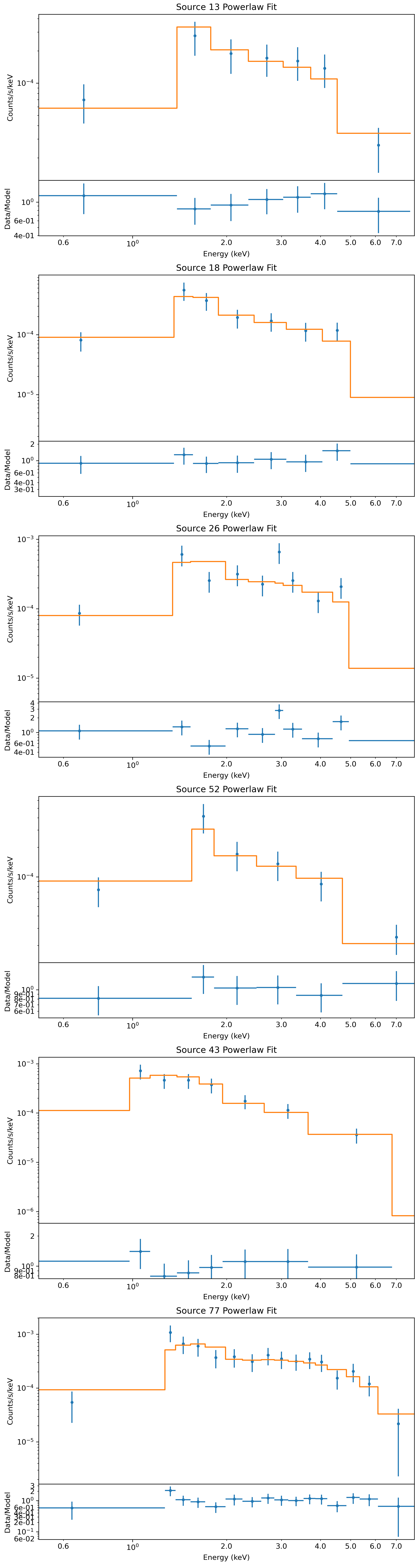}
    \includegraphics[width=0.37\textwidth,align=t]{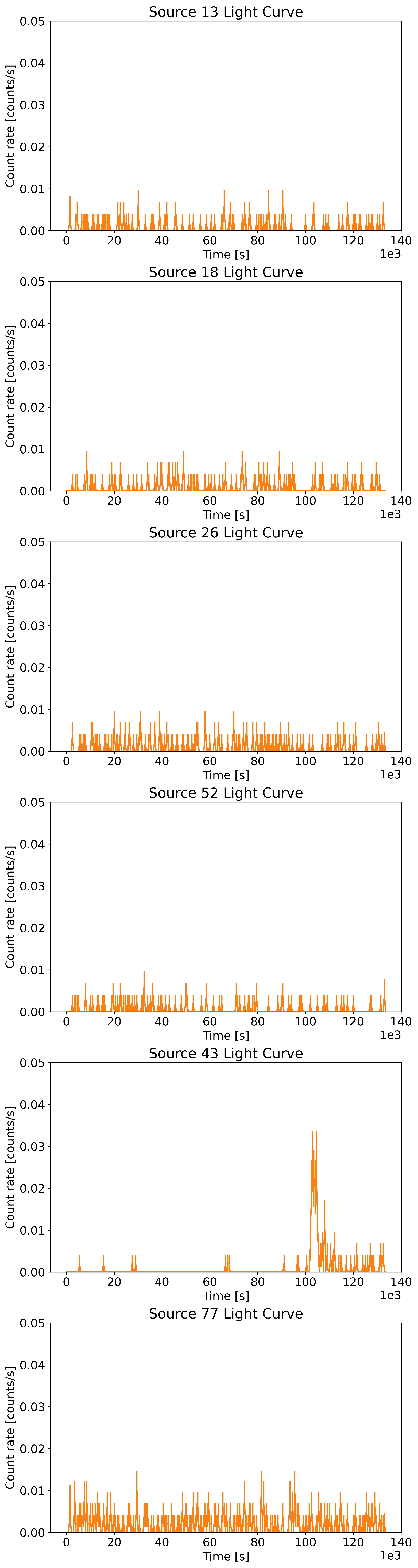}
    \caption{Spectra and lightcurves for CXO sources with only DECaPS2 counterparts. }
    \label{fig:ngc3532_DECaPS2_only}
\end{figure*}

Sources 13, 18, 26, 52, 43$^{\ast}$, and 77 do not have counterparts, except for faint counterparts in DECaPS2. Because of this, it is difficult to confirm or exclude these sources as AGNs, except for source 43. Their spectra and lightcurves are shown in Figure \ref{fig:ngc3532_DECaPS2_only}. 

Source 43$^{\ast}$ is variable (most counts are seen during the $\sim$5-ks long flare), with a hard spectrum which is fit by the absorbed PL or {\tt mekal} models, with $\Gamma\approx1.9$ or $kT\approx4.3$ keV, respectively. On the hardness ratio diagram, this source appears near the middle of the medium-hard scale, harder than LM-STARs, and away from confidently classified AGNs and other hard sources on the top right. The flare itself reaches peak luminosity in $\sim 1$ ks, and plateaus for $\sim 4$ ks. During the flare the spectrum is quite hard with the absorbed PL fit having $\Gamma\approx 1.8$. This behavior is distinct from typical coronal flares. The source is classified as 70\% LMXB, and consequently, is identified as a candidate CO. This source is only $1.5''$ away from a bright ($G=12.6$) background A-type star. Although the star is likely too offset to be the counterpart of the X-ray source, its brightness may be precluding the detection of a fainter counterpart to the X-ray source. In fact, in the DECaPS2 survey, this source has 2 counterparts within a $1''$ radius in the $Y$-band, $Y=17.7$ and 18.6 respectively. However, the reliability is uncertain, given the proximity of the bright star. If the source does have an optical counterpart, its classification is likely to change. 

Source 77 lacks counterparts, except for a faint counterpart in VPHAS+ and DECaPS2, with VPHAS+ $i=20$ and DECaPS2 from $r$=21.7 mag to $Y$ = 19.4. Being near the edge of the CXO observation field, the source has a large PU ($1.08''$) and a higher chance coincidence probability. This source appears on the top right corner of the HR diagram, close to confidently classified AGNs. It shows a very hard spectrum that's well fit by the PL model with $\Gamma\approx1.3$. Significant classification probabilities are 81\% NS, and 17\% AGN. The source is probably not a member of NGC 3532, because of substantial absorption in the X-ray spectrum ($n_H=\SI{0.9 \pm 0.3}{\per\square\cm})$. \textbf{If the DECaPS2 counterpart is a true match, then this source would not be classified as a NS.}

Sources 13, 18, 26, and 52 are similar to source 77, except that they have fainter DECaPS2 counterparts ($>22$ mag within $0.5''$ of CXO positions). They have high AGN and NS classification probabilities, but the presence of faint IR counterparts makes them more likely to be AGNs. This underscores the importance of having deep NIR survey coverage to discriminate between AGNs and possible CO classes.

\subsection{Sources without MW Counterparts}
\label{no_cp}

\begin{figure*}[hbt!]
    \centering
    \includegraphics[width=0.37\textwidth,align=t]{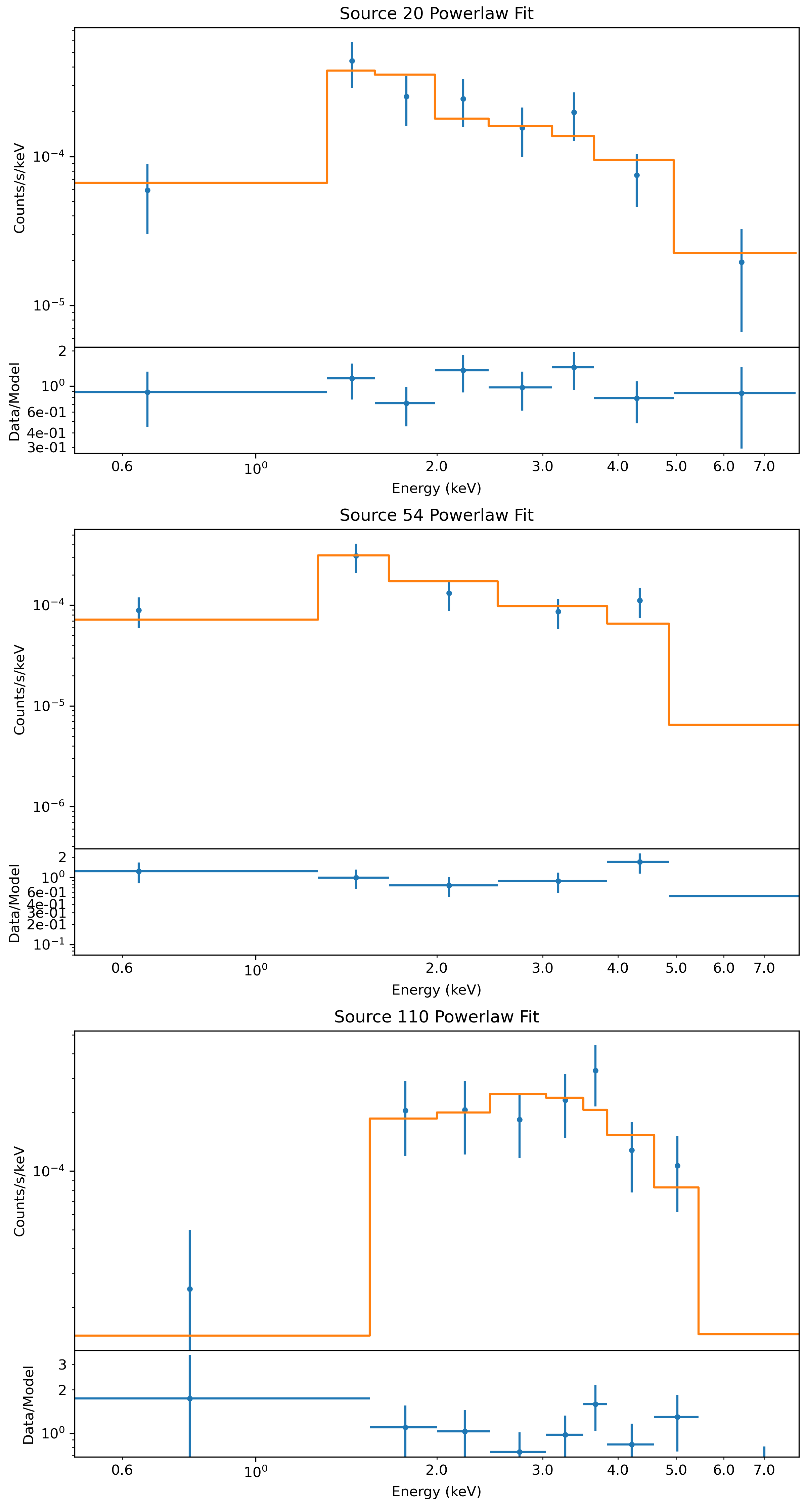}
    \includegraphics[width=0.37\textwidth,align=t]{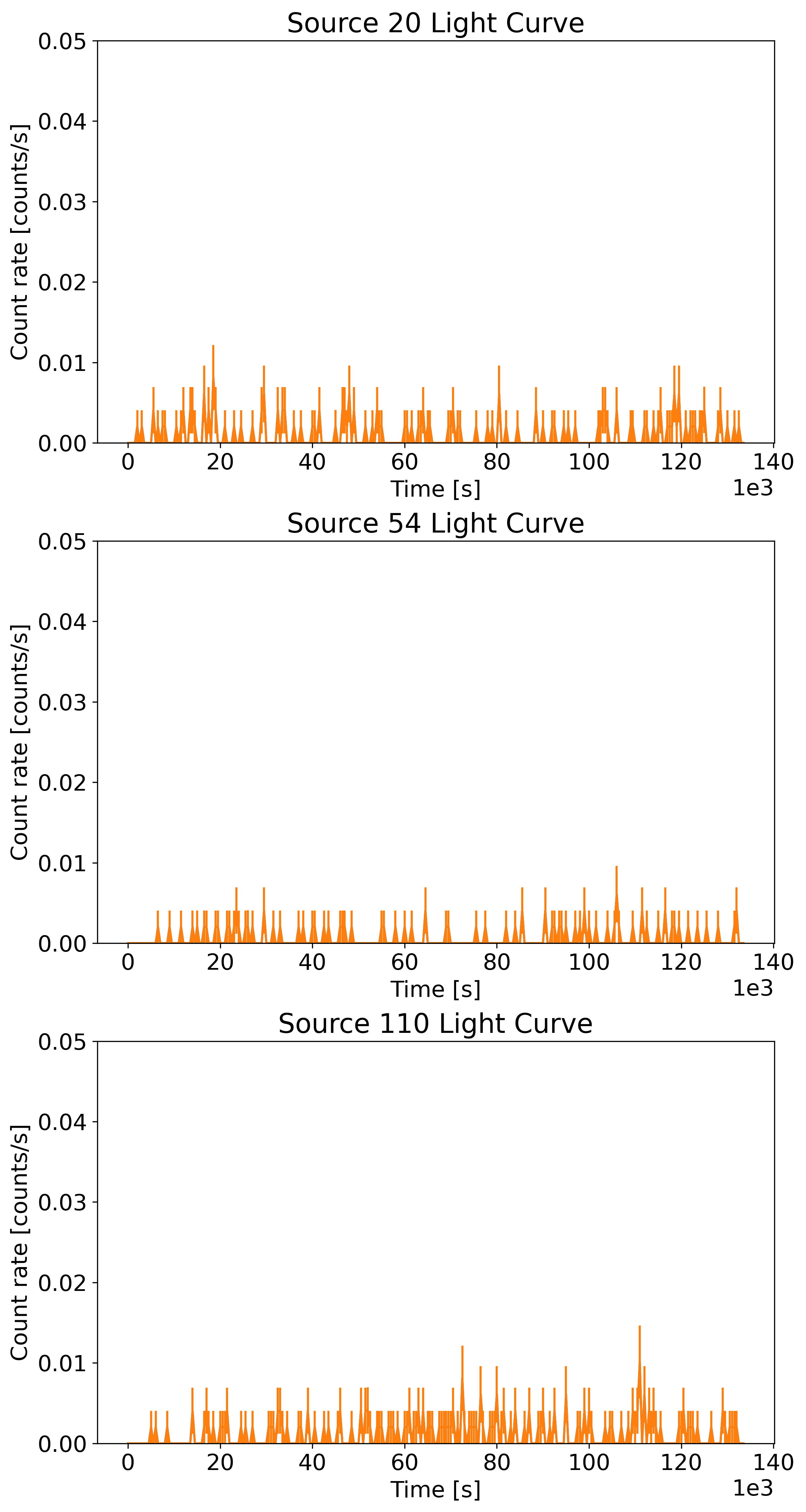}
    \caption{Spectra and lightcurves for CXO sources without MW counterparts. }
    \label{fig:ngc3532_no_counterparts}
\end{figure*}

Sources 20, 54, and 110 have no reliable MW counterparts, even in the DECaPS2 survey. Sources 20 and 54 exhibit X-ray properties similar to those of sources discussed in Section \ref{background_hard}, including location on the HR diagram, and hard or relatively hard spectra (see Figure \ref{fig:ngc3532_no_counterparts}) which are mostly well fitted by PL models with $\Gamma= 1.4-1.9$. 

Neither of these sources are confidently classified, but the most probable classes are LMXB and NS, as well as AGN for source 54. Given the relative brightness in X-rays, but the lack of counterparts down to the limiting magnitude of 21.7 (at 50\% recovery rate) in the z band of the DECaPS2 survey, we consider these CO classifications plausible.

Source 110 has an absorbed PL index with high uncertainty $\Gamma=3.1\pm0.8$, which may indicate a soft spectrum. Its X-ray spectrum resembles those of magnetars. At an assumed typical Galactic distance of $\sim4$ kpc, its X-ray luminosity would be $\sim2\times10^{31}$ erg s$^{-1}$. This absorbed luminosity is compatible with those of magnetars in quiescence \citep{olausen_mcgill_2014}. The corresponding unabsorbed luminosity of $\sim\SI{e32}{\erg\per\s}$ is too large for a non-flaring low mass star, while a higher mass star should be visible in DECaPS2. Source 110 is unconfidently classified by MUWCLASS as a NS at 56\% probability. 

Sources 20 and 110 have 1 ``bad'' detection $\sim 1''$ away in the DECaPS2 $g$-band, but without any reported fluxes. A deeper NIR observation would help to firmly establish the nature of these sources. 

\section{Summary}
\label{summary}

We performed multiwavelength analysis and classification of 131 X-ray sources detected in the field of the 300 Myr-old nearby cluster NGC 3532. Of these X-ray sources, 28\% are variable, and 95\% have multiwavelength counterparts in at least one of the surveys we used. We summarize the main results from our study below:

\begin{itemize}
    \item We confidently classified 40 CXO sources 
    to be low-mass stars or young low-mass stars, of which 31 belong to the cluster. Six flaring sources belong to the cluster, with the largest flare luminosity being $\SI{3.4e30}{\erg\per\s\per\square\cm}$.
    
    \item We confirm the previously reported inverse correlation between X-ray activity and rotation period in low-mass stars.
    
    \item Eight late B-type or early A-type cluster stars were detected in X-rays. While most of them likely have low-mass companions responsible for X-ray emission, Source 29$^{\ast}$ does not have reported evidence of binarity, and yet shows a strong, 5-ks long flare with an average flare luminosity of \SI{3.4e30}{\erg\per\s}.
    
    \item Detailed analysis of ML classification results confirms that the precision of LM-STAR and AGN classifications in the field of NGC 3532 are high, while completeness is lower. This could be due to biases and imbalances in the distribution of source classes in our TD. The classifications for CO classes are mostly unconfident, due to under-representation in the TD, and require additional observations/analysis to confirm.
    
    \item Among galactic background sources with MW counterparts, we found flaring sources (Sources 30, 42, 43, 49) showing symmetric flare profiles which differ from sharp-rise slow-decay profiles typical for flaring stars. Since such profiles are relatively rare for coronal stellar flares, these sources may have a different nature. Of these, Source 43 is the most interesting source, showing a strong flare distinct from typical coronal flares. Deeper CXO ACIS observations of these sources could uncover a possible compact object nature. 
    
    \item We identified several other background sources as candidate compact objects (Sources 4, 9, 12, 20, 54, 62, and 110), based on their spectral properties and higher X-ray luminosities at their fiducial distances. In particular, source 9 has a high tangential velocity of $\SI{340}{\km\per\s}$ which, combined with the hard X-ray spectrum, makes it likely to be a non-accreting neutron star in a binary system.
    
    \item The candidate compact objects are not likely to be cluster members of NGC 3532, because they lack reliable optical/IR counterparts. The CO remnants of the $\sim 20$ massive stars that have gone supernova at the cluster age have likely all escaped the cluster by this time. In theory, some types of COs (e.g., CVs or NSs from electron-capture SNe) could exist in NGC 3532. However, electron-capture SNe that form NSs are thought to be only a few percent of core collapse SNe \citep{wanajo_electron-capture_2010}, and thus may not have occurred in the cluster. Additionally, any companion stars of WDs may not hav had enough time to evolve to form CVs. The only two cluster members that could, in principle, harbour a CO are associated with the evolved star (Source 99) and the A0 star with a hard excess (Source 111).
    
\end{itemize}

\begin{deluxetable*}{cllllllllll}

\tablehead{ 
\colhead{Source} & \colhead{2CXO Name} & \colhead{Class} & \colhead{$P_\mathrm{Class}$} & \colhead{Can. CO} & \colhead{$C_\mathrm{net}$} & \colhead{$P_\mathrm{var}$} & \colhead{Gmag} & \colhead{Dist.} & \colhead{$\gamma$} & \colhead{$kT$
}
}

\startdata
4&J110522.5-585718&HMXB?&$0.36\pm0.10$&Y&131&0.47&16.8&$6090\substack{+2090\\-1020}$&$2.03\substack{+0.45\\-0.42}$&$5.40\substack{+9.72\\-1.99}$
\\
7&J110450.0-585559&LM-STAR&$1.00\pm0.01$&N&325&0.85&11.7&$286\substack{+1.03\\-1.06}$&$9.80\substack{+-\\-1.25}$&$0.56\substack{+0.04\\-0.04}$
\\
8&J110439.4-585550&YSO?&$0.55\pm0.15$&N&53&0.97&17.77&$616\substack{+44.7\\-35.8}$&$8.03\substack{+-\\-3.38}$&$0.39\substack{+0.10\\-0.09}$
\\
9&J110449.9-585549&CV?&$0.39\pm0.07$&Y&91&0.18&17.82&$3890\substack{+2130\\-1500}$&$1.63\substack{+0.49\\-0.45}$&$5.33\substack{+2.40\\-1.48}$
\\
10&J110455.3-585516&YSO?&$0.45\pm0.10$&N&56&1&18.92&$561\substack{+66.8\\-51.4}$&$2.43\substack{+0.75\\-0.39}$&$0.67\substack{+0.16\\-0.09}$
\\
12&J110423.1-585445&AGN?&$0.47\pm0.07$&N&382&0.96&&&$2.21\substack{+0.26\\-0.24}$&$4.70\substack{+2.79\\-1.17}$
\\
13&J110548.8-585438&NS?&$0.62\pm0.32$&N&104&0.44&&&$1.66\substack{+0.47\\-0.44}$&$14.36\substack{+29.66\\-7.03}$
\\
15&J110443.6-585425&HM-STAR&$0.61\pm0.10$&N&183&1&11.44&$1850\substack{+77.6\\-63.0}$&$2.67\substack{+0.32\\-0.30}$&$1.04\substack{+0.06\\-0.07}$
\\
18&J110428.3-585400&AGN?&$0.69\pm0.33$&N&114&0.42&22.37&&$1.90\substack{+0.41\\-0.38}$&$4.22\substack{+5.26\\-1.35}$
\\
20&J110605.6-585334&LMXB?&$0.53\pm0.23$&N&124&0.88&&&$1.78\substack{+0.51\\-0.45}$&$6.23\substack{+27.09\\-3.52}$
\\
22&J110414.8-585305&LM-STAR&$1.00\pm0.01$&N&222&0.99&12.47&$493\substack{+3.62\\-3.61}$&$5.12\substack{+0.68\\-0.60}$&$0.61\substack{+0.04\\-0.04}$
\\
25&J110535.7-585212&LM-STAR&$1.00\pm0.00$&N&153&0.82&12.04&$483\substack{+4.15\\-3.35}$&$9.63\substack{+-\\-1.33}$&$0.40\substack{+0.10\\-0.10}$
\\
26&J110507.7-585206&NS?&$0.65\pm0.18$&N&165&0.83&&&$1.72\substack{+0.34\\-0.32}$&$15.32\substack{+38.65\\-7.58}$
\\
27&J110610.7-585154&YSO?&$0.53\pm0.12$&N&139&1&18.13&$1110\substack{+170\\-118}$&$3.09\substack{+0.63\\-0.53}$&$0.75\substack{+0.08\\-0.18}$
\\
29&J110430.1-585147&LM-STAR?&$0.60\pm0.11$&N&330&1&9.84&$466\substack{+2.83\\-3.45}$&$2.93\substack{+0.29\\-0.26}$&$1.38\substack{+1.17\\-0.04}$
\\
30&J110443.6-585132&LMXB?&$0.38\pm0.08$&Y&53&1&19.87&$2920\substack{+1280\\-1530}$&$2.67\substack{+0.62\\-0.54}$&$0.59\substack{+0.08\\-0.10}$
\\
35&J110543.3-585053&LM-STAR?&$0.63\pm0.13$&N&220&0.64&15.55&$482\substack{+6.31\\-5.54}$&$3.75\substack{+0.46\\-0.42}$&$0.67\substack{+0.74\\-0.04}$
\\
41&J110456.3-585015&YSO?&$0.52\pm0.06$&N&219&1&16.74&$371\substack{+18.9\\-15.2}$&$3.75\substack{+0.43\\-0.39}$&$0.79\substack{+0.05\\-0.05}$
\\
42&J110420.3-585010&LMXB?&$0.36\pm0.08$&Y&131&1&16.6&$1480\substack{+113\\-83.3}$&$2.02\substack{+0.35\\-0.30}$&$0.65\substack{+-\\-0.05}$
\\
43&J110445.0-585009&LMXB&$0.70\pm0.11$&Y&151&1&&&$1.93\substack{+0.32\\-0.30}$&$4.27\substack{+1.60\\-0.99}$
\\
49&J110423.6-584935&LMXB?&$0.45\pm0.08$&N&51&1&19.61&$1220\substack{+1060\\-341}$&$3.17\substack{+1.12\\-0.79}$&$0.63\substack{+0.18\\-0.12}$
\\
51&J110438.6-584929&LM-STAR&$0.83\pm0.07$&N&71&0.91&9.73&$513\substack{+49.9\\-38.0}$&$8.78\substack{+-\\-1.93}$&$0.51\substack{+0.07\\-0.10}$
\\
52&J110524.4-584913&AGN?&$0.66\pm0.14$&N&101&0.97&&&$1.50\substack{+0.43\\-0.40}$&$31.10\substack{+-\\-22.36}$
\\
54&J110453.3-584900&NS?&$0.46\pm0.27$&N&89&0.52&&&$1.60\substack{+0.38\\-0.36}$&$24.56\substack{+-\\-17.68}$
\\
55&J110554.8-584859&LM-STAR&$0.87\pm0.07$&N&357&0.57&9.73&$434\substack{+26.7\\-23.9}$&$8.22\substack{+0.97\\-0.86}$&$0.45\substack{+0.04\\-0.07}$
\\
57&J110435.5-584824&LM-STAR&$1.00\pm0.01$&N&619&1&12.34&$475\substack{+3.00\\-3.11}$&$4.80\substack{+0.36\\-0.34}$&$0.60\substack{+0.03\\-0.03}$
\\
59&J110520.7-584757&LM-STAR&$0.78\pm0.09$&N&285&0.19&14.23&$406\substack{+2.63\\-2.40}$&$5.40\substack{+0.64\\-0.57}$&$0.52\substack{+0.11\\-0.05}$
\\
61&J110529.6-584720&LM-STAR?&$0.34\pm0.12$&N&74&1&18.54&$463\substack{+39.5\\-30.1}$&$3.06\substack{+0.65\\-0.56}$&$1.02\substack{+0.41\\-0.21}$
\\
62&J110439.8-584701&LMXB&$0.49\pm0.09$&Y&321&0.09&20.02&$3980\substack{+1990\\-1480}$&$1.54\substack{+0.25\\-0.24}$&$17.64\substack{+-\\-8.69}$
\\
64&J110518.4-584615&LM-STAR&$0.73\pm0.07$&N&66&0.06&9.65&$460\substack{+8.35\\-6.46}$&$10.00\substack{+-\\-1.82}$&$0.44\substack{+0.11\\-0.18}$
\\
65&J110535.8-584609&LM-STAR&$0.70\pm0.10$&N&66&0.17&8.97&$538\substack{+21.4\\-18.1}$&$10.00\substack{+-\\-1.78}$&$0.37\substack{+0.18\\-0.12}$
\\
66&J110535.5-584547&CV?&$0.42\pm0.09$&N&144&1&17.39&$450\substack{+16.6\\-15.6}$&$3.32\substack{+0.54\\-0.47}$&$0.67\substack{+0.06\\-0.06}$
\\
68&J110450.7-584543&LM-STAR?&$0.43\pm0.11$&N&179&1&16.46&$451\substack{+8.26\\-6.86}$&$3.19\substack{+0.49\\-0.43}$&$0.53\substack{+0.12\\-0.05}$
\\
70&J110542.6-584540&YSO?&$0.57\pm0.12$&N&209&0.14&15.95&$1810\substack{+130\\-111}$&$3.00\substack{+0.39\\-0.36}$&$0.99\substack{+0.08\\-0.08}$
\\
71&J110521.8-584528&LMXB?&$0.60\pm0.14$&N&135&1&17.95&&$3.19\substack{+0.51\\-0.45}$&$0.64\substack{+0.06\\-0.05}$
\\
77&J110441.4-584352&NS&$0.82\pm0.12$&Y&227&0.65&&&$1.29\substack{+0.31\\-0.29}$&$79.90\substack{+-\\-54.75}$
\\
90&J110621.8-585133&CV?&$0.35\pm0.08$&Y&109&0&19.62&$4560\substack{+1810\\-1200}$&$1.65\substack{+0.54\\-0.47}$&$6.52\substack{+20.21\\-3.36}$
\\
99&J110435.9-584520&LM-STAR&$0.90\pm0.07$&N&130&1&7.41&$429\substack{+6.09\\-6.21}$&$3.81\substack{+0.85\\-0.68}$&$0.24\substack{+0.06\\-0.04}$
\\
110&J110429.5-584406&NS?&$0.56\pm0.15$&N&97&0.9&&&$3.15\substack{+0.86\\-0.77}$&$1.88\substack{+0.85\\-0.69}$
\\
111&J110532.7-584349&LM-STAR?&$0.55\pm0.10$&N&56&0.93&8.91&$470\substack{+6.40\\-5.91}$&$10.00\substack{+-\\-1.75}$&$0.56\substack{+0.06\\-0.21}$
\\
118&J110515.6-585437&LMXB?&$0.43\pm0.10$&Y&36&0.94&20.78&$4690\substack{+1870\\-1870}$&$1.86\substack{+0.82\\-0.70}$&$5.05\substack{+-\\-2.69}$
\\
119&J110518.3-584842&LMXB?&$0.44\pm0.11$&Y&34&0.01&17.62&$1550\substack{+278\\-219}$&$3.28\substack{+0.95\\-0.79}$&$0.61\substack{+0.24\\-0.11}$
\\
131&J110457.9-584742&LM-STAR&$0.89\pm0.07$&N&&0.59&8.39&&&
\\
\enddata

\label{table:detailed}
\caption{Table of sources discussed in detail in Section \ref{detailed}. This table represents a subset of a larger  machine-readable table (MRT) which includes all 131 X-ray sources detected with $S/N>5$, available electronically. Columns shown in this table include: CSC2 name, most probable ML classification and probability, candidate CO status in 5-class scheme \textbf{(if CT$>2$ for CO class probability, see Equation \ref{eq:CT})}, net CXO counts, variability, Gaia eDR3 distance (pc), PL fit photon index $\Gamma$, and $kT$ (keV) from the {\tt mekal} fit. Unconfident classifications in 8-class scheme (as defined by Eq. \ref{eq:CT}) are marked with "?" Note that a source with the highest probability for a CO class in the 8-class scheme may still not be a candidate CO in the 5-class scheme, if its combined  probabilities for the CO-related classes (LMXB, NS, CV, and HMXB) are not high enough. }

\end{deluxetable*}

\section{Acknowledgement}

Support for this work was provided by the National Aeronautics and Space Administration through Chandra Award AR9-20005 issued by the Chandra X-ray Observatory Center, operated by the Smithsonian Astrophysical Observatory for the National Aeronautics Space Administration under contract NAS803060, and also by the NASA Astrophysics Data Analysis Program (ADAP) award 80NSSC19K0576. JH acknowledges support from an appointment to the NASA Postdoctoral Program at the Goddard Space Flight Center, administered by Oak Ridge Associated Universities through a contract with NASA.

Database: This work has made use of the Chandra Source Catalog, provided by the Chandra X-ray Center (CXC) as part of the Chandra Data Archive \citep{2020AAS...23515405E}; the SIMBAD database, operated at CDS, Strasbourg, France \citep{wenger_simbad_2000}; and the VizieR catalogue access tool, CDS, Strasbourg, France \citep{ochsenbein_vizier_2000}.

Software: 
Astropy \citep{collaboration_astropy_2013}, Astroquery \citep{ginsburg_astroquery_2019}, scikit-learn \citep{pedregosa_scikit-learn_2011}, imbalanced-learn \citep{lemaitre_imbalanced-learn_2017}, isochrones \citep{morton_isochrones_2015}, hvplot, and related holoviz packages.\footnote{\url{https://hvplot.holoviz.org/}}

Hardware: This work was completed in part with resources provided by the High Performance Computing Cluster at The George Washington University, Research Technology Services. 

\facilities{CXO, Gaia, CTIO:2MASS, WISE, NEOWISE, CTIO: DECAM, CTIO: VST}

\appendix

\section{Confusion Matrices}
\label{cm}

\begin{figure*}
\begin{center}
\includegraphics[width=200pt,trim=0 0 0 0]{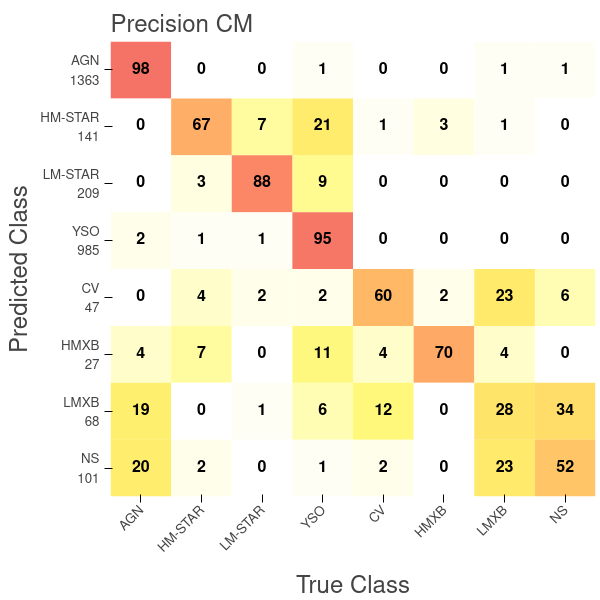}
\includegraphics[width=200pt,trim=0 0 0 0]{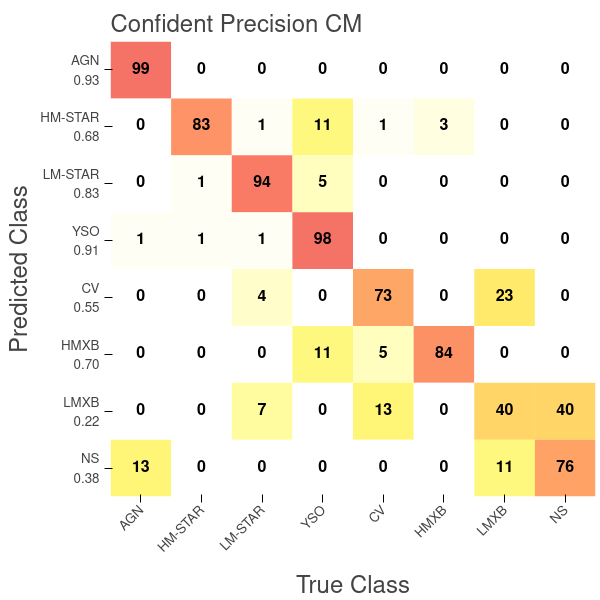}
\includegraphics[width=200pt,trim=0 0 0 0]{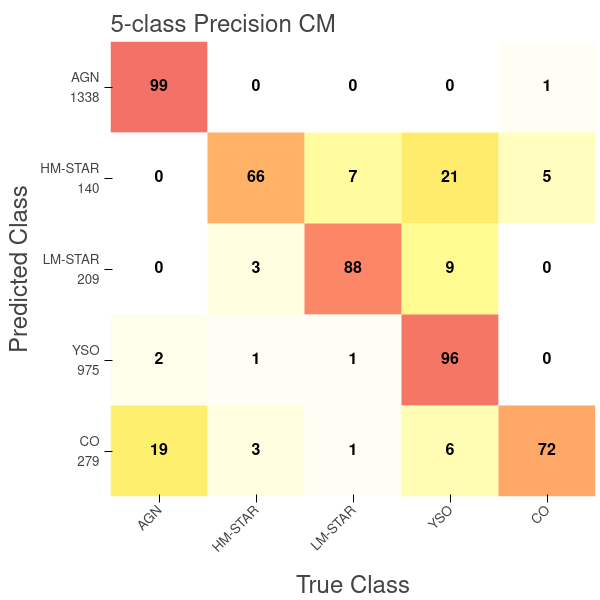}
\includegraphics[width=200pt,trim=0 0 0 0]{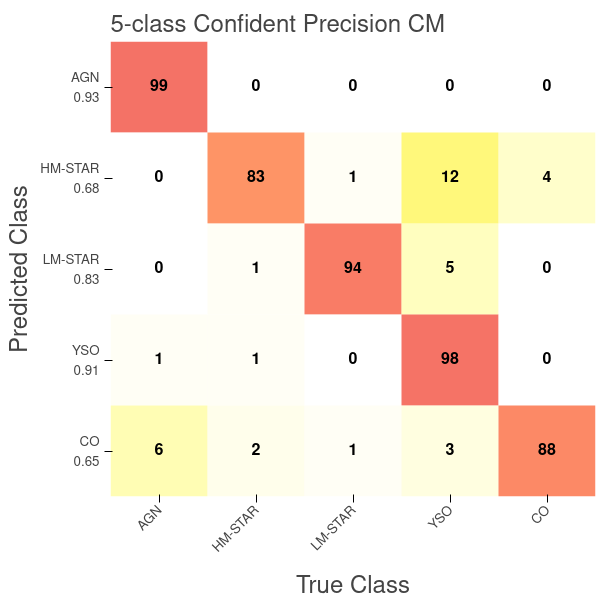}
\caption{Normalized precision confusion matrices (CMs) of the TD using the Leave-One-Out cross validation method where every AGN has been reddened using the extinction/absorption parameter from the NGC 3532 field. Compared to \citep{yang_classifying_2022}, Gaia eDR3 distances \citep{bailer-jones_estimating_2021} and associated luminosities are added, which improves performance slightly. 
The left panels shows the CMs of all classifications while the right panels shows the CMs for the confident classifications (CT$\ge2$). The upper panels show the CMs under the 8-class scheme and the lower panels show the CMs under the 5-class scheme. 
The value within each element of the CM is the percentage of sources in a true class, shown on the horizontal axis, that are from the predicted class, shown on the vertical axis.
The values under the class labels along the vertical axis in the left panels are the total numbers of the sources in the corresponding classes, while in the right panels these values are the fractions of the sources surviving the confidence cut (CT$\ge2$) for each class. 
Redder colors indicate higher classification percentage.} 
\label{fig:cm}
\end{center}
\end{figure*}

To validate the performance of MUWCLASS applied to the NGC 3532 field, we use the same TD (with additional distance and luminosity information) and leave-one-out-cross-validation (LOOCV) method as described in \cite{yang_classifying_2022}. Before running the LOOCV procedure, We apply reddening on AGNs in the TD using the extinction and absorption parameters ($E(B-V)=1.3$, \cite{ruiz_ruizcagdpyc_2018}, $n_{H}=\SI{9e21}{\cm^{-2}}$, \cite{guver_relation_2009}) through the Galactic plane in the direction of NGC 3532. The confusion matrices that summarize classification performance are shown in Figure \ref{fig:cm}. 

\section{Astrometric Correction}
\label{astrometric}

We apply astrometric corrections to CSC2 source coordinates by aligning the master level X-ray coordinates to Gaia eDR3 source coordinates. The Gaia eDR3 reference sources are built with a few filters applied to ensure the reliability of their astrometry ($G<23$, \textbf{Gaia position errors} e\_RA\_ICRS$<1$, e\_DE\_ICRS$<1$, \textbf{Gaia parallax and parallax error} $-2<\mathrm{Plx}<2$, $e\_\mathrm{Plx}<1$, \textbf{Gaia proper motion and proper motion error} PM$<20$, e\_PM$<1$, \texttt{RUWE}$<1.4$ and \textbf{Gaia astrometric excess noise} epsi$<1.898$, corresponding to the 90\% of the epsi distribution). Then, we propagate the Gaia coordinates to the X-ray observation epoch (MJD=54763 at 2008-10-24) using Gaia proper motions (if no proper motion value is available, we use the initial ICRS coordinates from Gaia eDR3 catalog at epoch MJD = 57388. at 2016-01-01). The X-ray sources are filtered on broadband significance $>5$ and broadband net counts (src\_cnts\_aper90\_b)$>$20 before they are matched to the proper-motion-corrected Gaia sources using the CIAO wcs\_match algorithm. For wcs\_match, we use ``trans" method with only translational correction, source match radius = 1.0, residtype=0, esidfac=0. The residlim is the residual limit used to eliminate the largest source pair position error, and we tested several different values of this parameter (0.1, 0.2, 0.3, 0.4). 

The astrometric (alignment) uncertainty (``PU\_astro\_68" column) is calculated using the following equation:

\begin{equation}
    {\rm PU_{\rm astro}} = (\sum_{i=1}^N (\frac{1}{\delta_{{\rm X},i}^2+\delta_{{\rm Gaia},i}^2}))^{-1/2}
\end{equation}

where $i$ goes through all matched pairs that remain after the final iteration of wcs\_match, $\delta_{\rm X}$ is the 1$\sigma$ X-ray PU calculated using the equation 14 from \cite{kim_chandra_2007}, and $\delta_{\rm Gaia}$ is the standard error in the Gaia coordinates. The final astrometric PUs are the arithmetic mean of the astrometric PUs in the RA and DEC directions. The astrometric solutions are summarized in Table \ref{tab:astro_solution} with different setting of residlims. We use residlim=0.2 since it is consistent with astrometric solutions calculated from residlim=0.1 and residlim=0.3 and the RMS residuals and the alignment uncertainties converge. 

We calculated the combined X-ray PU (``PU" column) by adding the 95\% level PU from \cite{kim_chandra_2007} (``PU\_kim95" column) and the alignment uncertainty (``PU\_astro\_68" column, multiplied by 2 to convert 1-$\sigma$ to 2-$\sigma$) in quadrature.

\begin{deluxetable*}{lrrrrr}
\tablehead{ 
\colhead{residlim} & \colhead{$\Delta$RA $\cos$(DEC)} & \colhead{$\Delta$DEC} & \colhead{${\rm PU_{\rm astro}}$} & \colhead{RMS Residuals$^{\rm a}$} & \colhead{\# of matched pairs} \\
\colhead{arcsec} & \colhead{arcsec} & \colhead{arcsec} & \colhead{arcsec} & \colhead{arcsec} & \colhead{} 
}
\startdata
0.1  & 0.32 & 0.12 & 0.124 & 0.041 & 5 \\
\textbf{0.2} & \textbf{0.23} & \textbf{0.15} & \textbf{0.092} & \textbf{0.092} & \textbf{9} \\
0.3 & 0.19 & 0.11 & 0.086 & 0.117 & 12 \\
0.4 & 0.23 & -0.03 & 0.062 & 0.190 & 22 \\ \hline
\enddata
\caption{Astrometric solutions of CXO observation of NGC 3532 (ObsID=8941) using a set of reslim parameter from \texttt{wcs\_match}. $^{\rm a}$ RMS Residuals is calculated from \texttt{wcs\_match}.
}
\label{tab:astro_solution}
\end{deluxetable*}

\section{Detailed Analysis for Additional Sources}
\label{detailed_appendix}

Here we present detailed analysis for additional sources not covered in Section \ref{detailed}.

\subsection{Cluster Members}

Source 71$^{\ast}$ misses 2MASS/Gaia counterparts by a tiny margin ($0.002''$ outside combined PU.), but is matched to a DECaPS2 counterpart. However, because DECaPS2 is not used in the ML pipeline, this source is unconfidently classified as an LMXB. The Gaia counterpart has proper motion ($-$9.981, 5.295) mas/yr and distance ($\approx 483$ pc) consistent with those of NGC 3532, is slightly above the main sequence on the binary track, and appears to be K-type. However, the \texttt{RUWE} value of 1.0 does not indicate binarity. Source 71 exhibited a large flare with luminosity of $\SI{3e30}{\erg\per\s}$ assuming a cluster distance. Since the spectrum and lightcurve of Source 71 resemble those of a relatively nearby coronally flaring low-mass star, we consider the 2MASS/Gaia counterpart to likely be the real match. 

\subsection{Cluster A-Type and B-Type Stars}

Sources 55 and 64, identified as \object{CPD-58 3086B}, \object{CPD-58 305} in SIMBAD, were seen in ROSAT \citep{franciosini_rosat_2000}. They exhibit evidence of binarity (\texttt{RUWE}=6.0, 1.5, and elevated positions above the solitary star track of the main sequence in the CMD). The X-ray spectra are soft and can be well-described by a {\tt mekal} model with $kT\approx0.4$ keV. They are non-variable, and have X-ray luminosities of $\sim\SI{e29}{\erg\per\s}$.

Source 99$^{\ast}$ has a Gaia DR3 counterpart coincident with the ``red clump'' region on the NGC 3532 isochrone shown in see Figure \ref{fig:ngc3532_mw_bp-rp_gmag}, and is known as \object{HD 96175} in SIMBAD. Its distance and proper motion are compatible with cluster membership. Using the isochrone fit, this star has initial mass $\approx \SI{3.3}{M_\sun}$, or spectral type $\approx$B8V. The source is variable in X-rays, displaying a small flare. The Gaia \texttt{RUWE} value is 1.79, consistent with previous identification as a spectroscopic binary. This source also appears in the Gaia DR3 ``Non-single stars catalog'' \citep{gaia_collaboration_gaia_2022} with a measured period of 240 days and \textbf{primary} semi-major axis of 0.286 AU. \textbf{This source is likely in binary with a lower-mass star responsible for the X-ray emissions.} 

Source 99 and 131 appear in the TD as LM-STARs, and were classified as such. As our manual analysis agrees with the classifications, we do not consider this to be of much concern.

\subsection{Foreground Stars}

\begin{figure*}
    \centering
    \includegraphics[width=0.37\textwidth,align=t]{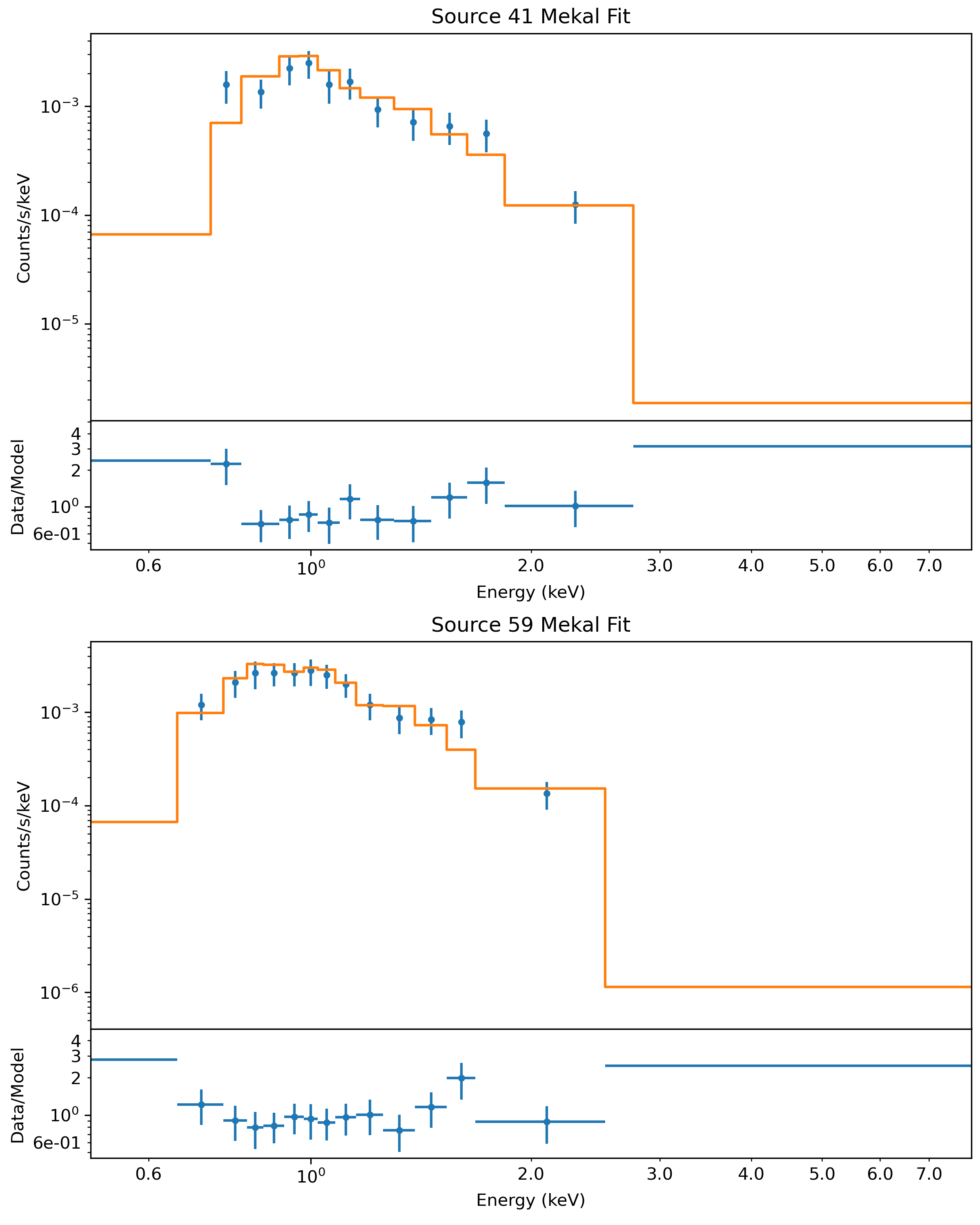}
    \includegraphics[width=0.37\textwidth,align=t]{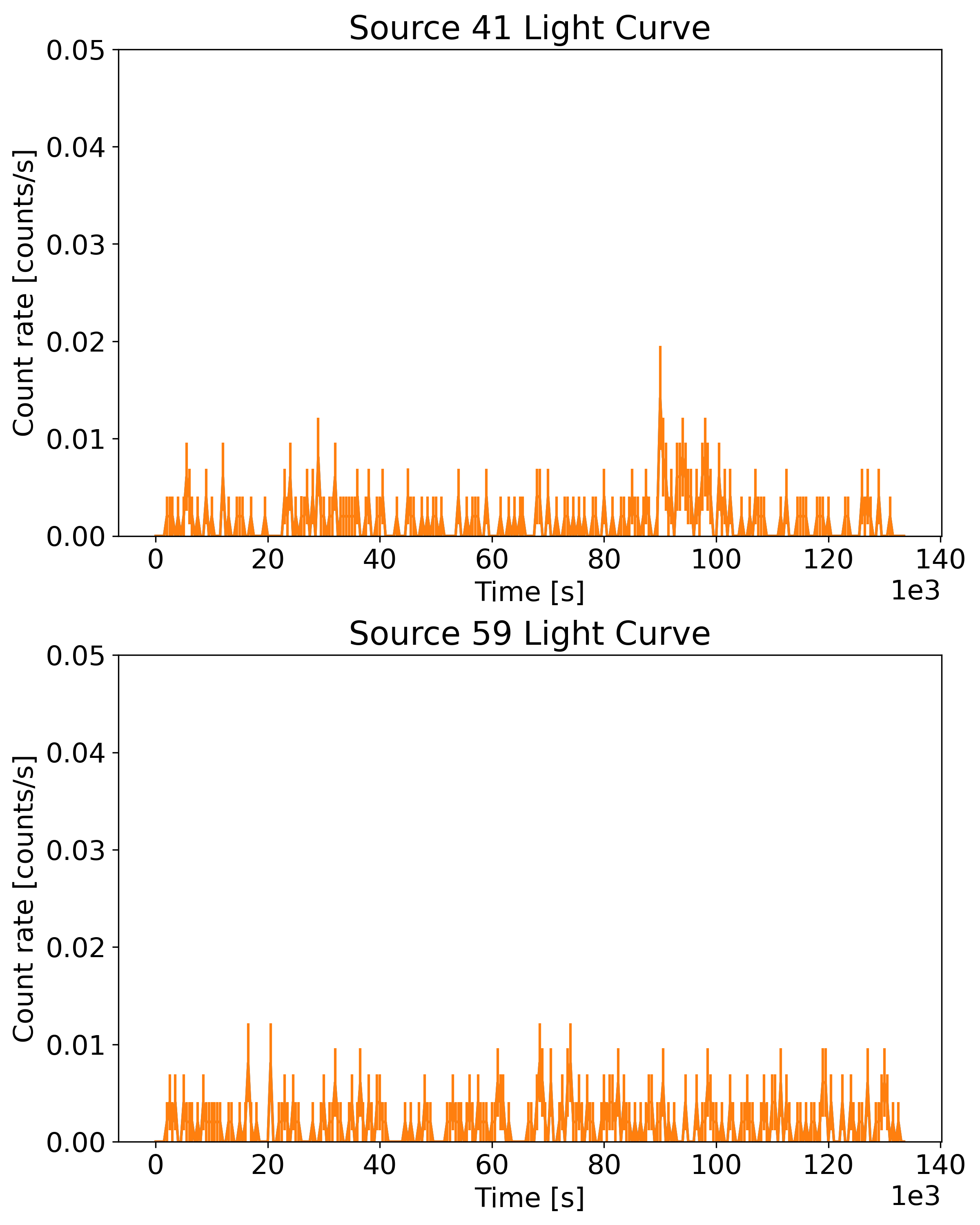}
    \caption{Spectra and lightcurves for CXO sources matched to cluster foreground stars. }
    \label{fig:ngc3532_foreground}
\end{figure*}

Sources 41$^{\ast}$ and 59 are coincident with foreground stars at $d\approx370$ and 400 pc, respectively, according to Gaia eDR3 distances \citep{bailer-jones_estimating_2021}. Their spectra and lightcurves are shown in Figure \ref{fig:ngc3532_foreground}. Both sources exhibit soft X-ray spectra which are adequately described by the {\tt mekal} model with $kT=0.78$ and 0.26 keV respectively. The former source is classified as 52\% YSO and 30\% LM-STAR while the latter is classified as 86\% LM-STAR.

The lightcurve of Source 41 shows a minor flare, while its \texttt{RUWE} value of 2.2 indicates binarity. Given the somewhat harder spectrum (compared to Source 59), it may be an active binary, which could be classified as a YSO by the ML pipeline. Source 59 is likely a coronally active low-mass star. 

\subsection{Background Sources}

Source 8, at $d\approx616\pm 40$ pc, is slightly beyond NGC 3532, although its Gaia PM ((-10.025, 5.026) mas/yr) is consistent with cluster membership. The source has a soft X-ray spectrum, which fits with the {\tt mekal} model, having $kT\approx 0.39$ keV. Its \texttt{RUWE} value of 1.3 may indicate binarity. The CXO lightcurve shows a small flare. The source is classified as 55\% YSO and 38\% LM-STAR, suggesting either a coronally active low-mass star or an active binary.

Source 15$^{\ast}$ was catalogued by \cite{fernandez_photometric_1980} and is listed as \object{Cl* NGC 3532 FERN 299} in SIMBAD. However, Gaia proper motion ($\mu_{\rm RA}$,$\mu_{\rm Dec}$)=(-6.137, 0.351) mas yr$^{-1}$, distance $d\approx 1850 \pm 75$ pc, as well as the position off the main sequence on the optical CMD are inconsistent with cluster membership. The source is significantly (but slowly) variable in X-rays with a relatively hard spectrum that's fit by an absorbed PL with $\Gamma\approx 2.7$. The X-ray luminosity is $\SI{5.7e30}{\erg\per\s}$. The \texttt{RUWE} value of 1.5 indicates binarity. Gaia DR3 astrophysical parameters are conflicting, with the ESP-ELS module suggesting a K-type star with $T\approx5,000$ K while the FLAME module gives a stellar mass of $3.4{M_\sun}$, implying a B-type star. The distance, brightness, and color suggests an evolved star, possibly of K-type. The X-ray source is classified by the pipeline as 74\% HM-STAR and 16\% YSO. The relatively bright X-ray emission may be from interaction with a companion. 

Source 27$^{\ast}$, at $d\approx1100$ pc, has UnWISE, 2MASS and Gaia counterparts and shows a relatively hard X-ray spectrum which can be described by {\tt mekal} with $kT\approx 0.7$ keV, with most of the photons detected during the flare. The flare has a sharp rise and slow decay profile typical for stellar flares. The source is classified as 53\% YSO, 26\% CV, and 18\% LMXB. The classifications are likely affected by the spectral hardening during the flare which dominates most of the spectral counts. 

Source 119 only has a Gaia counterpart, which is only detected in the G-band. This source is non-variable during the {\sl CXO} observation. Its spectrum fits the absorbed PL model with  $\Gamma\simeq 3.3$. The source is harder and more X-ray luminous ($L_X=\SI{6e29}{\erg\per\s}$) than most low-mass stars. The highest classification probabilities are 44\% LMXB, 18\% CV, and 18\% NS, and it's therefore classified as a candidate CO. It's possible that a lack of BP-RP color and NIR-IR counterparts disfavored it from being classified as a YSO. 

Source 118 is faint both in optical and X-rays, and has a negative parallax in Gaia DR3 and a rather uncertain proper motion ($\mu=(6.1\pm 1.6)$ mas yr$^{-1}$). The faintness of this source prevents us from drawing further conclusions. 

\subsection{Hard Sources with MW counterparts}

Sources 4 and 90 have Gaia and NIR counterparts, with Gaia distance beyond the cluster. The sources lie near the edge of the ACIS-I field-of-view, so the chance coincidence probability is larger. The spectra are relatively hard, and are well fit by both models, with PL photon indices $\Gamma\approx2.0,\ 1.6$ and {\tt mekal} $kT\approx 5.4,\ 6.5$ keV. Their X-ray luminosities ($L_X>\SI{e31}{\erg\per\s}$) are higher than a typical solitary low-mass star at their fiducial distances of 6 and 4.5 kpc, while their optical luminosity $L_{\rm O}\sim\SI{e32},\ \SI{e34}{\erg\per\s}$ are compatible with stellar luminosities. The \texttt{RUWE} values of $\sim 1$ do not provide evidence of binarity. Their total proper motion of ($\SI{6.4\pm 0.6}{mas/yr},\ \SI{7.0\pm 0.4}{mas/yr}$) translates to high velocities of $\approx \SI{180}{\km\per\s},\ \SI{150}{\km\per\s}$. However, these velocities may be mostly due to differential galactic rotation. These sources are classified as candidate COs in the 5-class scheme, which is supported by their hard spectra and high X-ray luminosities. 

\section{White Dwarfs}
\label{wd}

We cross-matched WDs and WD candidates in NGC 3532 in the literature to CXO sources. Only three WDs (None of them are the heavy WD VPHAS J110358.0-583709.2) are located within the field of view of the {\sl CXO} observation, and none of them had an X-ray counterpart. The list of WDs in NGC 3532 is given in Table \ref{table:wd}. Non-detection in X-rays is consistent with solitary WDs with temperatures of $\sim\SI{3e4}{K}$, derived in \cite{dobbie_new_2009} at the age of NGC 3532.

\begin{deluxetable*}{cllll}

\tablehead{ 
\colhead{Identifier} & \colhead{Object Type} & \colhead{RA} & \colhead{DEC} & \colhead{Reference
}
}

\startdata
Cl* NGC 3532     RK       8&WD*&168.2033&-58.8306&{[1]}
\\
 NGC 3532-WDC J1107-5848&Candidate\_WD*&166.8698074&-58.80675485&{[1]}
\\
 NGC 3532-WDC J1107-5842&Candidate\_WD*&166.8415686&-58.7034724&{[1]}
\\
 NGC 3532-WDC J1106-5847&Candidate\_WD*&166.7460896&-58.79267942&{[1]}
\\
 NGC 3532-WDC J1106-5843&Candidate\_WD*&166.7151416&-58.73028971&{[1]}
\\
 NGC 3532-WDC J1106-5905&Candidate\_WD*&166.5764723&-59.08813626&{[1]}
\\
 NGC 3532-WDC J1106-5856&Candidate\_WD*&166.5702729&-58.93469326&{[1]}
\\
Cl* NGC 3532     RK       5&WD*&166.5173497&-58.92221326&{[1]}
\\
Cl* NGC 3532     RK       6&WD*&166.4710669&-58.49197324&{[1]}
\\
\textbf{Cl* NGC 3532     RK       1}&WD*&166.3993072&-58.87401832&{[1]}
\\
 \textbf{NGC 3532-WDC J1105-5857}&Candidate\_WD*&166.3494859&-58.95636597&{[1]}
\\
Cl* NGC 3532     RK      10&WD*&165.8130725&-58.36229544&{[1]}
\\
VPHAS J110358.0-583709.2&WD&165.9916069&-58.6191961&{[2]}
\\
VPHAS J110434.5-583047.4&WD&166.14375&-58.51317&{[2]}
\\
\textbf{VPHAS J110547.2-584241.8}&WD&166.44667&-58.71161&{[2]}
\\
Cl* NGC 3532 RK 9&WD*&165.9054929&-58.31119815&{[3]}
\\
\enddata

\label{table:wd}
\caption{WDs and candidate WDs suggested to be cluster members of NGC 3532. WDs within the field of the CXO observation of NGC 3532 bolded. Some WDs have Gaia counterparts inconsistent with cluster membership, and are not shown in Fig. \ref{fig:ngc3532_mw_bp-rp_gmag}. References: [1]: \cite{dobbie_further_2012}, [2]: \cite{raddi_search_2016}, [3]: \cite{koester_spectroscopic_1993}}

\end{deluxetable*}
\clearpage

\bibliographystyle{aasjournal}
\bibliography{references.bib}

\begin{thebibliography}{}
\expandafter\ifx\csname natexlab\endcsname\relax\def\natexlab#1{#1}\fi
\providecommand{\url}[1]{\href{#1}{#1}}
\providecommand{\dodoi}[1]{doi:~\href{http://doi.org/#1}{\nolinkurl{#1}}}
\providecommand{\doeprint}[1]{\href{http://ascl.net/#1}{\nolinkurl{http://ascl.net/#1}}}
\providecommand{\doarXiv}[1]{\href{https://arxiv.org/abs/#1}{\nolinkurl{https://arxiv.org/abs/#1}}}

\bibitem[{Amard {et~al.}(2019)Amard, Palacios, Charbonnel, Gallet, Georgy,
  Lagarde, \& Siess}]{amard_first_2019}
Amard, L., Palacios, A., Charbonnel, C., {et~al.} 2019, Astronomy \&
  Astrophysics, 631, A77, \dodoi{10.1051/0004-6361/201935160}

\bibitem[{Bailer-Jones {et~al.}(2021)Bailer-Jones, Rybizki, Fouesneau,
  Demleitner, \& Andrae}]{bailer-jones_estimating_2021}
Bailer-Jones, C. A.~L., Rybizki, J., Fouesneau, M., Demleitner, M., \& Andrae,
  R. 2021, The Astronomical Journal, 161, 147, \dodoi{10.3847/1538-3881/abd806}

\bibitem[{Brown {et~al.}(2021)Brown, Vallenari, Prusti, Bruijne, Babusiaux,
  Biermann, Creevey, Evans, Eyer, Hutton, Jansen, Jordi, Klioner, Lammers,
  Lindegren, Luri, Mignard, Panem, Pourbaix, Randich, Sartoretti, Soubiran,
  Walton, Arenou, Bailer-Jones, Bastian, Cropper, Drimmel, Katz, Lattanzi,
  Leeuwen, Bakker, Cacciari, Castañeda, Angeli, Ducourant, Fabricius,
  Fouesneau, Frémat, Guerra, Guerrier, Guiraud, Piccolo, Masana, Messineo,
  Mowlavi, Nicolas, Nienartowicz, Pailler, Panuzzo, Riclet, Roux, Seabroke,
  Sordo, Tanga, Thévenin, Gracia-Abril, Portell, Teyssier, Altmann, Andrae,
  Bellas-Velidis, Benson, Berthier, Blomme, Brugaletta, Burgess, Busso, Carry,
  Cellino, Cheek, Clementini, Damerdji, Davidson, Delchambre, Dell’Oro,
  Fernández-Hernández, Galluccio, García-Lario, Garcia-Reinaldos,
  González-Núñez, Gosset, Haigron, Halbwachs, Hambly, Harrison,
  Hatzidimitriou, Heiter, Hernández, Hestroffer, Hodgkin, Holl, Janßen,
  Fombelle, Jordan, Krone-Martins, Lanzafame, Löffler, Lorca, Manteiga,
  Marchal, Marrese, Moitinho, Mora, Muinonen, Osborne, Pancino, Pauwels, Petit,
  Recio-Blanco, Richards, Riello, Rimoldini, Robin, Roegiers, Rybizki, Sarro,
  Siopis, Smith, Sozzetti, Ulla, Utrilla, Leeuwen, Reeven, Abbas, Aramburu,
  Accart, Aerts, Aguado, Ajaj, Altavilla, Álvarez, Cid-Fuentes, Alves,
  Anderson, Varela, Antoja, Audard, Baines, Baker, Balaguer-Núñez, Balbinot,
  Balog, Barache, Barbato, Barros, Barstow, Bartolomé, Bassilana, Bauchet,
  Baudesson-Stella, Becciani, Bellazzini, Bernet, Bertone, Bianchi,
  Blanco-Cuaresma, Boch, Bombrun, Bossini, Bouquillon, Bragaglia, Bramante,
  Breedt, Bressan, Brouillet, Bucciarelli, Burlacu, Busonero, Butkevich, Buzzi,
  Caffau, Cancelliere, Cánovas, Cantat-Gaudin, Carballo, Carlucci, Carnerero,
  Carrasco, Casamiquela, Castellani, Castro-Ginard, Sampol, Chaoul, Charlot,
  Chemin, Chiavassa, Cioni, Comoretto, Cooper, Cornez, Cowell, Crifo, Crosta,
  Crowley, Dafonte, Dapergolas, David, David, Laverny, Luise, March, Ridder,
  Souza, Teodoro, Torres, Peloso, Pozo, Delbo, Delgado, Delgado, Delisle,
  Matteo, Diakite, Diener, Distefano, Dolding, Eappachen, Edvardsson, Enke,
  Esquej, Fabre, Fabrizio, Faigler, Fedorets, Fernique, Fienga, Figueras,
  Fouron, Fragkoudi, Fraile, Franke, Gai, Garabato, Garcia-Gutierrez,
  García-Torres, Garofalo, Gavras, Gerlach, Geyer, Giacobbe, Gilmore, Girona,
  Giuffrida, Gomel, Gomez, Gonzalez-Santamaria, González-Vidal, Granvik,
  Gutiérrez-Sánchez, Guy, Hauser, Haywood, Helmi, Hidalgo, Hilger, Hładczuk,
  Hobbs, Holland, Huckle, Jasniewicz, Jonker, Campillo, Julbe, Karbevska,
  Kervella, Khanna, Kochoska, Kontizas, Kordopatis, Korn, Kostrzewa-Rutkowska,
  Kruszyńska, Lambert, Lanza, Lasne, Campion, Fustec, Lebreton, Lebzelter,
  Leccia, Leclerc, Lecoeur-Taibi, Liao, Licata, Lindstrøm, Lister, Livanou,
  Lobel, Pardo, Managau, Mann, Marchant, Marconi, Santos, Marinoni, Marocco,
  Marshall, Polo, Martín-Fleitas, Masip, Massari, Mastrobuono-Battisti, Mazeh,
  McMillan, Messina, Michalik, Millar, Mints, Molina, Molinaro, Molnár,
  Montegriffo, Mor, Morbidelli, Morel, Morris, Mulone, Munoz, Muraveva, Murphy,
  Musella, Noval, Ordénovic, Orrù, Osinde, Pagani, Pagano, Palaversa,
  Palicio, Panahi, Pawlak, Esteller, Penttilä, Piersimoni, Pineau, Plachy,
  Plum, Poggio, Poretti, Poujoulet, Prša, Pulone, Racero, Ragaini, Rainer,
  Raiteri, Rambaux, Ramos, Ramos-Lerate, Fiorentin, Regibo, Reylé, Ripepi,
  Riva, Rixon, Robichon, Robin, Roelens, Rohrbasser, Romero-Gómez, Rowell,
  Royer, Rybicki, Sadowski, Sellés, Sahlmann, Salgado, Salguero, Samaras,
  Gimenez, Sanna, Santoveña, Sarasso, Schultheis, Sciacca, Segol, Segovia,
  Ségransan, Semeux, Shahaf, Siddiqui, Siebert, Siltala, Slezak, Smart,
  Solano, Solitro, Souami, Souchay, Spagna, Spoto, Steele, Steidelmüller,
  Stephenson, Süveges, Szabados, Szegedi-Elek, Taris, Tauran, Taylor,
  Teixeira, Thuillot, Tonello, Torra, Torra, Turon, Unger, Vaillant, Dillen,
  Vanel, Vecchiato, Viala, Vicente, Voutsinas, Weiler, Wevers, Wyrzykowski,
  Yoldas, Yvard, Zhao, Zorec, Zucker, Zurbach, \& Zwitter}]{brown_gaia_2021}
Brown, A. G.~A., Vallenari, A., Prusti, T., {et~al.} 2021, Astronomy \&
  Astrophysics, 649, A1, \dodoi{10.1051/0004-6361/202039657}

\bibitem[{Clem {et~al.}(2011)Clem, Landolt, Hoard, \& Wachter}]{clem_deep_2011}
Clem, J.~L., Landolt, A.~U., Hoard, D.~W., \& Wachter, S. 2011, The
  Astronomical Journal, 141, 115, \dodoi{10.1088/0004-6256/141/4/115}

\bibitem[{Collaboration {et~al.}(2013)Collaboration, Robitaille, Tollerud,
  Greenfield, Droettboom, Bray, Aldcroft, Davis, Ginsburg, Price-Whelan,
  Kerzendorf, Conley, Crighton, Barbary, Muna, Ferguson, Grollier, Parikh,
  Nair, Unther, Deil, Woillez, Conseil, Kramer, Turner, Singer, Fox, Weaver,
  Zabalza, Edwards, Azalee~Bostroem, Burke, Casey, Crawford, Dencheva, Ely,
  Jenness, Labrie, Lim, Pierfederici, Pontzen, Ptak, Refsdal, Servillat, \&
  Streicher}]{collaboration_astropy_2013}
Collaboration, A., Robitaille, T.~P., Tollerud, E.~J., {et~al.} 2013, Astronomy
  and Astrophysics, 558, A33, \dodoi{10.1051/0004-6361/201322068}

\bibitem[{Collaboration {et~al.}(2022)Collaboration, Arenou, Babusiaux,
  Barstow, Faigler, Jorissen, Kervella, Mazeh, Mowlavi, Panuzzo, Sahlmann,
  Shahaf, Sozzetti, Bauchet, Damerdji, Gavras, Giacobbe, Gosset, Halbwachs,
  Holl, Lattanzi, Leclerc, Morel, Pourbaix, Fiorentin, Sadowski, Ségransan,
  Siopis, Teyssier, Zwitter, Planquart, Brown, Vallenari, Prusti, de~Bruijne,
  Biermann, Creevey, Ducourant, Evans, Eyer, Guerra, Hutton, Jordi, Klioner,
  Lammers, Lindegren, Luri, Mignard, Panem, Randich, Sartoretti, Soubiran,
  Tanga, Walton, Bailer-Jones, Bastian, Drimmel, Jansen, Katz, van Leeuwen,
  Bakker, Cacciari, Castañeda, De~Angeli, Fabricius, Fouesneau, Frémat,
  Galluccio, Guerrier, Heiter, Masana, Messineo, Nicolas, Nienartowicz,
  Pailler, Riclet, Roux, Seabroke, Sordo, Thévenin, Gracia-Abril, Portell,
  Altmann, Andrae, Audard, Bellas-Velidis, Benson, Berthier, Blomme, Burgess,
  Busonero, Busso, Cánovas, Carry, Cellino, Cheek, Clementini, Davidson,
  de~Teodoro, Campos, Delchambre, Dell'Oro, Esquej, Fernández-Hernández,
  Fraile, Garabato, García-Lario, Haigron, Hambly, Harrison, Hernández,
  Hestroffer, Hodgkin, Janßen, de~Fombelle, Jordan, Krone-Martins, Lanzafame,
  Löffler, Marchal, Marrese, Moitinho, Muinonen, Osborne, Pancino, Pauwels,
  Recio-Blanco, Reylé, Riello, Rimoldini, Roegiers, Rybizki, Sarro, Smith,
  Utrilla, van Leeuwen, Abbas, Ábrahám, Aramburu, Aerts, Aguado, Ajaj,
  Aldea-Montero, Altavilla, Álvarez, Alves, Anders, Anderson, Varela, Antoja,
  Baines, Baker, Balaguer-Núñez, Balbinot, Balog, Barache, Barbato, Barros,
  Bartolomé, Bassilana, Becciani, Bellazzini, Berihuete, Bernet, Bertone,
  Bianchi, Binnenfeld, Blanco-Cuaresma, Blazere, Boch, Bombrun, Bossini,
  Bouquillon, Bragaglia, Bramante, Breedt, Bressan, Brouillet, Brugaletta,
  Bucciarelli, Burlacu, Butkevich, Buzzi, Caffau, Cancelliere, Cantat-Gaudin,
  Carballo, Carlucci, Carnerero, Carrasco, Casamiquela, Castellani,
  Castro-Ginard, Chaoul, Charlot, Chemin, Chiaramida, Chiavassa, Chornay,
  Comoretto, Contursi, Cooper, Cornez, Cowell, Crifo, Cropper, Crosta, Crowley,
  Dafonte, Dapergolas, David, de~Laverny, De~Luise, De~March, De~Ridder,
  de~Souza, de~Torres, del Peloso, del Pozo, Delbo, Delgado, Delisle, Demouchy,
  Dharmawardena, Diakite, Diener, Distefano, Dolding, Enke, Fabre, Fabrizio,
  Fedorets, Fernique, Figueras, Fournier, Fouron, Fragkoudi, Gai,
  Garcia-Gutierrez, Garcia-Reinaldos, García-Torres, Garofalo, Gavel, Gerlach,
  Geyer, Gilmore, Girona, Giuffrida, Gomel, Gomez, González-Núñez,
  González-Santamaría, González-Vidal, Granvik, Guillout, Guiraud,
  Gutiérrez-Sánchez, Guy, Hatzidimitriou, Hauser, Haywood, Helmer, Helmi,
  Sarmiento, Hidalgo, Hładczuk, Hobbs, Holland, Huckle, Jardine, Jasniewicz,
  Piccolo, Jiménez-Arranz, Campillo, Julbe, Karbevska, Khanna, Kordopatis,
  Korn, Kóspál, Kostrzewa-Rutkowska, Kruszyńska, Kun, Laizeau, Lambert,
  Lanza, Lasne, Campion, Lebreton, Lebzelter, Leccia, Lecoeur-Taibi, Liao,
  Licata, Lindstrøm, Lister, Livanou, Lobel, Lorca, Loup, Pardo, Romeo,
  Managau, Mann, Manteiga, Marchant, Marconi, Marcos, Santos, Pina, Marinoni,
  Marocco, Marshall, Polo, Martín-Fleitas, Marton, Mary, Masip, Massari,
  Mastrobuono-Battisti, McMillan, Messina, Michalik, Millar, Mints, Molina,
  Molinaro, Molnár, Monari, Monguió, Montegriffo, Montero, Mor, Mora,
  Morbidelli, Morris, Muraveva, Murphy, Musella, Nagy, Noval, Ocaña, Ogden,
  Ordenovic, Osinde, Pagani, Pagano, Palaversa, Palicio, Pallas-Quintela,
  Panahi, Payne-Wardenaar, Esteller, Penttilä, Pichon, Piersimoni, Pineau,
  Plachy, Plum, Poggio, Prša, Pulone, Racero, Ragaini, Rainer, Raiteri, Ramos,
  Ramos-Lerate, Regibo, Richards, Diaz, Ripepi, Riva, Rix, Rixon, Robichon,
  Robin, Robin, Roelens, Rogues, Rohrbasser, Romero-Gómez, Rowell, Royer,
  Mieres, Rybicki, Núñez, Sellés, Salguero, Samaras, Gimenez, Sanna,
  Santoveña, Sarasso, Schultheis, Sciacca, Segol, Segovia, Semeux, Siddiqui,
  Siebert, Siltala, Silvelo, Slezak, Slezak, Smart, Snaith, Solano, Solitro,
  Souami, Souchay, Spagna, Spina, Spoto, Steele, Steidelmüller, Stephenson,
  Süveges, Surdej, Szabados, Szegedi-Elek, Taris, Taylor, Teixeira, Tolomei,
  Tonello, Torra, Torra, Elipe, Trabucchi, Tsounis, Turon, Ulla, Unger,
  Vaillant, van Dillen, van Reeven, Vanel, Vecchiato, Viala, Vicente,
  Voutsinas, Weiler, Wevers, Wyrzykowski, Yoldas, Yvard, Zhao, Zorec, \&
  Zucker}]{gaia_collaboration_gaia_2022}
Collaboration, G., Arenou, F., Babusiaux, C., {et~al.} 2022,
  \dodoi{10.48550/ARXIV.2206.05595}

\bibitem[{Cutri {et~al.}(2021)Cutri, Wright, Conrow, Fowler, Eisenhardt,
  Grillmair, Kirkpatrick, Masci, McCallon, Wheelock, Fajardo-Acosta, Yan,
  Benford, Harbut, Jarrett, Lake, Leisawitz, Ressler, Stanford, Tsai, Liu,
  Helou, Mainzer, Gettngs, Gonzalez, Hoffman, Marsh, Padgett, Skrutskie, Beck,
  Papin, \& Wittman}]{cutri_vizier_2021}
Cutri, R.~M., Wright, E.~L., Conrow, T., {et~al.} 2021, VizieR Online Data
  Catalog, II/328.
\newblock \url{https://ui.adsabs.harvard.edu/abs/2014yCat.2328....0C}

\bibitem[{Davenport {et~al.}(2019)Davenport, Covey, Clarke, Boeck, Cornet, \&
  Hawley}]{davenport_evolution_2019}
Davenport, J. R.~A., Covey, K.~R., Clarke, R.~W., {et~al.} 2019, The
  Astrophysical Journal, 871, 241, \dodoi{10.3847/1538-4357/aafb76}

\bibitem[{Dobbie {et~al.}(2012)Dobbie, Day-Jones, Williams, Casewell, Burleigh,
  Lodieu, Parker, \& Baxter}]{dobbie_further_2012}
Dobbie, P.~D., Day-Jones, A., Williams, K.~A., {et~al.} 2012, Monthly Notices
  of the Royal Astronomical Society, 423, 2815,
  \dodoi{10.1111/j.1365-2966.2012.21090.x}

\bibitem[{Dobbie {et~al.}(2009)Dobbie, Napiwotzki, Burleigh, Williams, Sharp,
  Barstow, Casewell, \& Hubeny}]{dobbie_new_2009}
Dobbie, P.~D., Napiwotzki, R., Burleigh, M.~R., {et~al.} 2009, Monthly Notices
  of the Royal Astronomical Society, 395, 2248,
  \dodoi{10.1111/j.1365-2966.2009.14688.x}

\bibitem[{Drew {et~al.}(2014)Drew, Gonzalez-Solares, Greimel, Irwin,
  Küpcü~Yoldas, Lewis, Barentsen, Eislöffel, Farnhill, Martin, Walsh,
  Walton, Mohr-Smith, Raddi, Sale, Wright, Groot, Barlow, Corradi, Drake,
  Fabregat, Frew, Gänsicke, Knigge, Mampaso, Morris, Naylor, Parker,
  Phillipps, Ruhland, Steeghs, Unruh, Vink, Wesson, \&
  Zijlstra}]{drew_vst_2014}
Drew, J.~E., Gonzalez-Solares, E., Greimel, R., {et~al.} 2014, Monthly Notices
  of the Royal Astronomical Society, 440, 2036, \dodoi{10.1093/mnras/stu394}

\bibitem[{Eggen(1981)}]{eggen_open_1981}
Eggen, O.~J. 1981, The Astrophysical Journal, 246, 817, \dodoi{10.1086/158977}

\bibitem[{Evans {et~al.}(2020)Evans, Primini, Miller, Evans, Allen, Anderson,
  Becker, Budynkiewicz, Burke, Chen, Civano, D'Abrusco, Doe, Fabbiano,
  Martinez~Galarza, Gibbs, Glotfelty, Graessle, Grier, Hain, Hall, Harbo,
  Houck, Lauer, Laurino, Lee, McCollough, McDowell, McLaughlin, Morgan,
  Mossman, Nguyen, Nichols, Nowak, Paxson, Perdikeas, Plummer, Rots,
  Siemiginowska, Sundheim, Thong, Tibbetts, Van~Stone, Winkelman, \&
  Zografou}]{2020AAS...23515405E}
Evans, I.~N., Primini, F.~A., Miller, J.~B., {et~al.} 2020, 235, 154.05.
\newblock \url{https://ui.adsabs.harvard.edu/abs/2020AAS...23515405E}

\bibitem[{Farias {et~al.}(2015)Farias, Smith, Fellhauer, Goodwin, Candlish,
  Blaña, \& Dominguez}]{farias_difficult_2015}
Farias, J.~P., Smith, R., Fellhauer, M., {et~al.} 2015, Monthly Notices of the
  Royal Astronomical Society, 450, 2451, \dodoi{10.1093/mnras/stv790}

\bibitem[{Fernandez \& Salgado(1980)}]{fernandez_photometric_1980}
Fernandez, J.~A., \& Salgado, C.~W. 1980, Astronomy and Astrophysics Supplement
  Series, 39, 11.
\newblock \url{https://ui.adsabs.harvard.edu/abs/1980A&AS...39...11F/abstract}

\bibitem[{Fouesneau {et~al.}(2022)Fouesneau, Frémat, Andrae, Korn, Soubiran,
  Kordopatis, Vallenari, Heiter, Creevey, Sarro, de~Laverny, Lanzafame, Lobel,
  Sordo, Rybizki, Slezak, Álvarez, Drimmel, Garabato, Delchambre,
  Bailer-Jones, Hatzidimitriou, Lorca, Fustec, Pailler, Mary, Robin, Utrilla,
  Aramburu, Bakker, Bellas-Velidis, Bijaoui, Blomme, Bouret, Brouillet,
  Brugaletta, Burlacu, Carballo, Casamiquela, Chaoul, Chiavassa, Contursi,
  Cooper, Dafonte, Demouchy, Dharmawardena, García-Lario, García-Torres,
  Gomez, González-Santamaría, Piccolo, Kontizas, Lebreton, Licata,
  Lindstrøm, Livanou, Romeo, Manteiga, Marocco, Martayan, Marshall, Nicolas,
  Ordenovic, Palicio, Pallas-Quintela, Pichon, Poggio, Recio-Blanco, Riclet,
  Santoveña, Schultheis, Segol, Silvelo, Smart, Süveges, Thévenin, Elipe,
  Ulla, van Dillen, Zhao, \& Zorec}]{fouesneau_gaia_2022}
Fouesneau, M., Frémat, Y., Andrae, R., {et~al.} 2022,
  \dodoi{10.48550/ARXIV.2206.05992}

\bibitem[{Franciosini {et~al.}(2000)Franciosini, Randich, \&
  Pallavicini}]{franciosini_rosat_2000}
Franciosini, E., Randich, S., \& Pallavicini, R. 2000, Astronomy and
  Astrophysics, 357, 139.
\newblock \url{http://adsabs.harvard.edu/abs/2000A%26A...357..139F}

\bibitem[{Fritzewski {et~al.}(2019)Fritzewski, Barnes, James, Geller, Meibom,
  \& Strassmeier}]{fritzewski_spectroscopic_2019}
Fritzewski, D.~J., Barnes, S.~A., James, D.~J., {et~al.} 2019, Astronomy \&
  Astrophysics, 622, A110, \dodoi{10.1051/0004-6361/201833587}

\bibitem[{Fritzewski {et~al.}(2021)Fritzewski, Barnes, James, \&
  Strassmeier}]{fritzewski_rotation_2021}
Fritzewski, D.~J., Barnes, S.~A., James, D.~J., \& Strassmeier, K.~G. 2021,
  Astronomy \& Astrophysics, 652, A60, \dodoi{10.1051/0004-6361/202140894}

\bibitem[{Fruscione {et~al.}(2006)Fruscione, McDowell, Allen, Brickhouse,
  Burke, Davis, Durham, Elvis, Galle, Harris, Huenemoerder, Houck, Ishibashi,
  Karovska, Nicastro, Noble, Nowak, Primini, Siemiginowska, Smith, \&
  Wise}]{2006SPIE.6270E..1VF}
Fruscione, A., McDowell, J.~C., Allen, G.~E., {et~al.} 2006, 6270, 62701V,
  \dodoi{10.1117/12.671760}

\bibitem[{Garmire {et~al.}(2003)Garmire, Bautz, Ford, Nousek, \&
  Ricker}]{garmire_advanced_2003}
Garmire, G.~P., Bautz, M.~W., Ford, P.~G., Nousek, J.~A., \& Ricker, Jr., G.~R.
  2003, 4851, 28, \dodoi{10.1117/12.461599}

\bibitem[{Gessner \& Janka(2018)}]{gessner_hydrodynamical_2018}
Gessner, A., \& Janka, H.-T. 2018, The Astrophysical Journal, 865, 61,
  \dodoi{10.3847/1538-4357/aadbae}

\bibitem[{Ginsburg {et~al.}(2019)Ginsburg, Sipőcz, Brasseur, Cowperthwaite,
  Craig, Deil, Guillochon, Guzman, Liedtke, Lian~Lim, Lockhart, Mommert,
  Morris, Norman, Parikh, Persson, Robitaille, Segovia, Singer, Tollerud,
  de~Val-Borro, Valtchanov, Woillez, {Astroquery Collaboration}, \& {a subset
  of astropy Collaboration}}]{ginsburg_astroquery_2019}
Ginsburg, A., Sipőcz, B.~M., Brasseur, C.~E., {et~al.} 2019, The Astronomical
  Journal, 157, 98, \dodoi{10.3847/1538-3881/aafc33}

\bibitem[{Güdel \& Nazé(2009)}]{gudel_x-ray_2009}
Güdel, M., \& Nazé, Y. 2009, Astronomy and Astrophysics Review, 17, 309,
  \dodoi{10.1007/s00159-009-0022-4}

\bibitem[{Günther {et~al.}(2022)Günther, Melis, Robrade, Schneider, Wolk, \&
  Yadav}]{gunther_coronal_2022}
Günther, H.~M., Melis, C., Robrade, J., {et~al.} 2022, The Astronomical
  Journal, 164, 8, \dodoi{10.3847/1538-3881/ac6ef6}

\bibitem[{Güver \& Özel(2009)}]{guver_relation_2009}
Güver, T., \& Özel, F. 2009, Monthly Notices of the Royal Astronomical
  Society, 400, 2050, \dodoi{10.1111/j.1365-2966.2009.15598.x}

\bibitem[{Igoshev {et~al.}(2021)Igoshev, Chruslinska, Dorozsmai, \&
  Toonen}]{igoshev_combined_2021}
Igoshev, A.~P., Chruslinska, M., Dorozsmai, A., \& Toonen, S. 2021,
  arXiv:2109.10362 [astro-ph].
\newblock \url{http://arxiv.org/abs/2109.10362}

\bibitem[{Jaehnig {et~al.}(2021)Jaehnig, Bird, \&
  Holley-Bockelmann}]{jaehnig_membership_2021}
Jaehnig, K., Bird, J., \& Holley-Bockelmann, K. 2021, The Astrophysical
  Journal, 923, 129, \dodoi{10.3847/1538-4357/ac1d51}

\bibitem[{Jennings {et~al.}(2018)Jennings, Kaplan, Chatterjee, Cordes, \&
  Deller}]{jennings_binary_2018}
Jennings, R.~J., Kaplan, D.~L., Chatterjee, S., Cordes, J.~M., \& Deller, A.~T.
  2018, The Astrophysical Journal, 864, 26, \dodoi{10.3847/1538-4357/aad084}

\bibitem[{Judge {et~al.}(2008)Judge, Solomon, \& Ayres}]{judge_estimate_2008}
Judge, P., Solomon, S., \& Ayres, a. 2008, The Astrophysical Journal, 593, 534,
  \dodoi{10.1086/376405}

\bibitem[{Kim {et~al.}(2007)Kim, Kim, Wilkes, Green, Kim, Anderson, Barkhouse,
  Evans, Ivezić, Karovska, Kashyap, Lee, Maksym, Mossman, Silverman, \&
  Tananbaum}]{kim_chandra_2007}
Kim, M., Kim, D.-W., Wilkes, B.~J., {et~al.} 2007, The Astrophysical Journal
  Supplement Series, 169, 401, \dodoi{10.1086/511634}

\bibitem[{Koester \& Reimers(1993)}]{koester_spectroscopic_1993}
Koester, D., \& Reimers, D. 1993, Astronomy and Astrophysics, 275, 479.
\newblock \url{https://ui.adsabs.harvard.edu/abs/1993A&A...275..479K/abstract}

\bibitem[{Lada \& Lada(2003)}]{lada_embedded_2003}
Lada, C.~J., \& Lada, E.~A. 2003, Annual Review of Astronomy and Astrophysics,
  41, 57, \dodoi{10.1146/annurev.astro.41.011802.094844}

\bibitem[{Larsen(2010)}]{larsen_young_2010}
Larsen, S.~S. 2010, Philosophical Transactions of the Royal Society A:
  Mathematical, Physical and Engineering Sciences, 368, 867,
  \dodoi{10.1098/rsta.2009.0255}

\bibitem[{Lemaître {et~al.}(2017)Lemaître, Nogueira, \&
  Aridas}]{lemaitre_imbalanced-learn_2017}
Lemaître, G., Nogueira, F., \& Aridas, C.~K. 2017, Journal of Machine Learning
  Research, 18, 1.
\newblock \url{http://jmlr.org/papers/v18/16-365.html}

\bibitem[{Marocco {et~al.}(2021)Marocco, Eisenhardt, Fowler, Kirkpatrick,
  Meisner, Schlafly, Stanford, Garcia, Caselden, Cushing, Cutri, Faherty,
  Gelino, Gonzalez, Jarrett, Koontz, Mainzer, Marchese, Mobasher, Schlegel,
  Stern, Teplitz, \& Wright}]{marocco_catwise2020_2021}
Marocco, F., Eisenhardt, P. R.~M., Fowler, J.~W., {et~al.} 2021, The
  Astrophysical Journal Supplement Series, 253, 8,
  \dodoi{10.3847/1538-4365/abd805}

\bibitem[{Marrese {et~al.}(2021)Marrese, Marinoni, Fabrizio, \&
  Altavilla}]{2021gdr3.reptE...9M}
Marrese, P.~M., Marinoni, S., Fabrizio, M., \& Altavilla, G. 2021, Gaia {EDR3}
  documentation {Chapter} 9: {Cross}-match with external catalogues, Tech. rep.
\newblock \url{https://ui.adsabs.harvard.edu/abs/2021gdr3.reptE...9M}

\bibitem[{McGale {et~al.}(1996)McGale, Pye, \& Hodgkin}]{mcgale_rosat_1996}
McGale, P.~A., Pye, J.~P., \& Hodgkin, S.~T. 1996, Monthly Notices of the Royal
  Astronomical Society, 280, 627, \dodoi{10.1093/mnras/280.3.627}

\bibitem[{Morton(2015)}]{morton_isochrones_2015}
Morton, T.~D. 2015, Astrophysics Source Code Library, ascl:1503.010.
\newblock \url{https://ui.adsabs.harvard.edu/abs/2015ascl.soft03010M}

\bibitem[{Mowlavi {et~al.}(2021)Mowlavi, Rimoldini, Evans, Riello, De~Angeli,
  Palaversa, Audard, Eyer, Garcia-Lario, Gavras, Holl, Jevardat~de Fombelle,
  Lecœur-Taïbi, \& Nienartowicz}]{mowlavi_large-amplitude_2021}
Mowlavi, N., Rimoldini, L., Evans, D.~W., {et~al.} 2021, Astronomy and
  Astrophysics, 648, A44, \dodoi{10.1051/0004-6361/202039450}

\bibitem[{Notsu {et~al.}(2019)Notsu, Maehara, Honda, Hawley, Davenport,
  Namekata, Notsu, Ikuta, Nogami, \& Shibata}]{notsu_kepler_2019}
Notsu, Y., Maehara, H., Honda, S., {et~al.} 2019, The Astrophysical Journal,
  876, 58, \dodoi{10.3847/1538-4357/ab14e6}

\bibitem[{Ochsenbein {et~al.}(2000)Ochsenbein, Bauer, \&
  Marcout}]{ochsenbein_vizier_2000}
Ochsenbein, F., Bauer, P., \& Marcout, J. 2000, Astronomy and Astrophysics
  Supplement Series, 143, 23, \dodoi{10.1051/aas:2000169}

\bibitem[{Olausen \& Kaspi(2014)}]{olausen_mcgill_2014}
Olausen, S.~A., \& Kaspi, V.~M. 2014, The Astrophysical Journal Supplement
  Series, 212, 6, \dodoi{10.1088/0067-0049/212/1/6}

\bibitem[{Pedregosa {et~al.}(2011)Pedregosa, Varoquaux, Gramfort, Michel,
  Thirion, Grisel, Blondel, Prettenhofer, Weiss, Dubourg, Vanderplas, Passos,
  Cournapeau, Brucher, Perrot, \& Duchesnay}]{pedregosa_scikit-learn_2011}
Pedregosa, F., Varoquaux, G., Gramfort, A., {et~al.} 2011, Journal of Machine
  Learning Research, 12, 2825.
\newblock \url{http://jmlr.org/papers/v12/pedregosa11a.html}

\bibitem[{Pizzocaro {et~al.}(2019)Pizzocaro, Stelzer, Poretti, Raetz, Micela,
  Belfiore, Marelli, Salvetti, \& De~Luca}]{pizzocaro_activity_2019}
Pizzocaro, D., Stelzer, B., Poretti, E., {et~al.} 2019, Astronomy and
  Astrophysics, 628, A41, \dodoi{10.1051/0004-6361/201731674}

\bibitem[{Possolo {et~al.}(2019)Possolo, Merkatas, \&
  Bodnar}]{possolo_asymmetrical_2019}
Possolo, A., Merkatas, C., \& Bodnar, O. 2019, Metrologia, 56, 045009,
  \dodoi{10.1088/1681-7575/ab2a8d}

\bibitem[{Pye {et~al.}(2015)Pye, Rosen, Fyfe, \& Schröder}]{pye_survey_2015}
Pye, J.~P., Rosen, S., Fyfe, D., \& Schröder, A.~C. 2015, Astronomy \&
  Astrophysics, 581, A28, \dodoi{10.1051/0004-6361/201526217}

\bibitem[{Raddi {et~al.}(2016)Raddi, Catalán, Gänsicke, Hermes, Napiwotzki,
  Koester, Tremblay, Barentsen, Farnhill, Mohr-Smith, Drew, Groot,
  Guzman-Ramirez, Parker, Steeghs, \& Zijlstra}]{raddi_search_2016}
Raddi, R., Catalán, S., Gänsicke, B.~T., {et~al.} 2016, Monthly Notices of
  the Royal Astronomical Society, 457, 1988, \dodoi{10.1093/mnras/stw042}

\bibitem[{Ruiz(2018)}]{ruiz_ruizcagdpyc_2018}
Ruiz, A. 2018, ruizca/gdpyc v1.0,  Zenodo, \dodoi{10.5281/zenodo.1482888}

\bibitem[{Saydjari {et~al.}(2022)Saydjari, Schlafly, Lang, Meisner, Green,
  Zucker, Zelko, Speagle, Daylan, Lee, Valdes, Schlegel, \&
  Finkbeiner}]{saydjari_dark_2022}
Saydjari, A.~K., Schlafly, E.~F., Lang, D., {et~al.} 2022, The {Dark} {Energy}
  {Camera} {Plane} {Survey} 2 ({DECaPS2}): {More} {Sky}, {Less} {Bias}, and
  {Better} {Uncertainties}, Tech. rep.
\newblock \url{https://ui.adsabs.harvard.edu/abs/2022arXiv220611909S}

\bibitem[{Schlafly {et~al.}(2019)Schlafly, Meisner, \&
  Green}]{schlafly_unwise_2019}
Schlafly, E.~F., Meisner, A.~M., \& Green, G.~M. 2019, The Astrophysical
  Journal Supplement Series, 240, 30, \dodoi{10.3847/1538-4365/aafbea}

\bibitem[{Simon(2000)}]{simon_x-ray_2000}
Simon, T. 2000, Publications of the Astronomical Society of the Pacific, 112,
  599, \dodoi{10.1086/316563}

\bibitem[{Skrutskie {et~al.}(2006)Skrutskie, Cutri, Stiening, Weinberg,
  Schneider, Carpenter, Beichman, Capps, Chester, Elias, Huchra, Liebert,
  Lonsdale, Monet, Price, Seitzer, Jarrett, Kirkpatrick, Gizis, Howard, Evans,
  Fowler, Fullmer, Hurt, Light, Kopan, Marsh, McCallon, Tam, Dyk, \&
  Wheelock}]{skrutskie_two_2006}
Skrutskie, M.~F., Cutri, R.~M., Stiening, R., {et~al.} 2006, The Astronomical
  Journal, 131, 1163, \dodoi{10.1086/498708}

\bibitem[{Stevenson {et~al.}(2022)Stevenson, Willcox, Vigna-Gomez, \&
  Broekgaarden}]{stevenson_wide_2022}
Stevenson, S., Willcox, R., Vigna-Gomez, A., \& Broekgaarden, F. 2022,
  arXiv:2205.03989 [astro-ph].
\newblock \url{http://arxiv.org/abs/2205.03989}

\bibitem[{van~der Meij {et~al.}(2021)van~der Meij, Guo, Kaper, \&
  Renzo}]{van_der_meij_confirming_2021}
van~der Meij, V., Guo, D., Kaper, L., \& Renzo, M. 2021, Astronomy \&
  Astrophysics, 655, A31, \dodoi{10.1051/0004-6361/202040114}

\bibitem[{Wanajo {et~al.}(2010)Wanajo, Janka, \&
  Müller}]{wanajo_electron-capture_2010}
Wanajo, S., Janka, H.-T., \& Müller, B. 2010, The Astrophysical Journal, 726,
  L15, \dodoi{10.1088/2041-8205/726/2/L15}

\bibitem[{Wenger {et~al.}(2000)Wenger, Ochsenbein, Egret, Dubois, Bonnarel,
  Borde, Genova, Jasniewicz, Laloë, Lesteven, \& Monier}]{wenger_simbad_2000}
Wenger, M., Ochsenbein, F., Egret, D., {et~al.} 2000, Astronomy and
  Astrophysics Supplement Series, 143, 9, \dodoi{10.1051/aas:2000332}

\bibitem[{Wilms {et~al.}(2000)Wilms, Allen, \& McCray}]{wilms_absorption_2000}
Wilms, J., Allen, A., \& McCray, R. 2000, The Astrophysical Journal, 542, 914,
  \dodoi{10.1086/317016}

\bibitem[{Wright {et~al.}(2010)Wright, Eisenhardt, Mainzer, Ressler, Cutri,
  Jarrett, Kirkpatrick, Padgett, McMillan, Skrutskie, Stanford, Cohen, Walker,
  Mather, Leisawitz, Gautier, McLean, Benford, Lonsdale, Blain, Mendez, Irace,
  Duval, Liu, Royer, Heinrichsen, Howard, Shannon, Kendall, Walsh, Larsen,
  Cardon, Schick, Schwalm, Abid, Fabinsky, Naes, \&
  Tsai}]{wright_wide-field_2010}
Wright, E.~L., Eisenhardt, P. R.~M., Mainzer, A.~K., {et~al.} 2010, The
  Astronomical Journal, 140, 1868, \dodoi{10.1088/0004-6256/140/6/1868}

\bibitem[{Yang {et~al.}(2022)Yang, Hare, Kargaltsev, Volkov, Chen, \&
  Rangelov}]{yang_classifying_2022}
Yang, H., Hare, J., Kargaltsev, O., {et~al.} 2022,
  \dodoi{10.48550/ARXIV.2206.13656}

\end{thebibliography}

\end{document}